\documentclass[
aps,
pra,
twocolumn,
superscriptaddress,
floatfix
]{revtex4-1}
\usepackage[dvipdfmx]{graphicx}
\usepackage{dcolumn}
\usepackage{bm}
\usepackage{ulem}
\usepackage{amsmath}
\usepackage{amssymb}
\usepackage{txfonts}
\usepackage{hyperref}
\usepackage{color} 
\usepackage{algorithm}
\usepackage{algpseudocode}
\newcommand{\bs}   {\boldsymbol}
\newcommand{\mb}   {\mathbf}

\newcommand{\Tr}{{\rm Tr}}
\newcommand{\e}{{\rm e}}
\newcommand{\imag}{{\rm i}}

\hypersetup{
  colorlinks=true,
  linkcolor=[rgb]{0.60,0.00,0.0},
  citecolor=[rgb]{0.0,0.0,0.6},
  urlcolor=[rgb]{0.0,0.0,0.6},
  setpagesize=false
}

\begin{document}

\title{
  Thermodynamic properties of 
  an $S=1/2$
  ring-exchange model on the triangular lattice
}

\author{Kazuhiro~Seki}
\affiliation{Computational Quantum Matter Research Team, RIKEN, Center for Emergent Matter Science (CEMS), Saitama 351-0198, Japan}

\author{Seiji~Yunoki}
\affiliation{Computational Quantum Matter Research Team, RIKEN, Center for Emergent Matter Science (CEMS), Saitama 351-0198, Japan}
\affiliation{Computational Condensed Matter Physics Laboratory, RIKEN Cluster for Pioneering Research (CPR), Saitama 351-0198, Japan}
\affiliation{Computational Materials Science Research Team, RIKEN Center for Computational Science (R-CCS),  Hyogo 650-0047,  Japan}

\begin{abstract}
  By using a numerically exact diagonalization technique and
  a block-extended version of the finite-temperature Lanczos method, 
  we study thermodynamic properties of an $S=1/2$ Heisenberg model 
  on the triangular lattice with 
  an antiferromagnetic nearest-neighbor interaction $J$ and
  a four-spin ring-exchange interaction $J_{\rm c}$. 
  Calculations are performed on small clusters 
  under the periodic-boundary conditions. 
  In contrast to the purely triangular case with $J_{\rm c}=0$, 
  the specific heat exhibits a characteristic double-peak structure
  for $J_{\rm c}/J \gtrsim 0.04$.   
  From the calculations of the entropy and the uniform magnetic susceptibility,
  it is shown that nonmagnetic excitations exist below
  the magnetic excitation for $J_{\rm c}/J \gtrsim 0.04$.
\end{abstract}

\date{\today}

\maketitle
\section{Introduction}
The $S=1/2$ Heisenberg antiferromagnet on the triangular lattice is
a prototypical frustrated quantum system
and has been a candidate of a resonating-valence-bond (RVB)
or a spin-liquid ground state~\cite{Anderson1973,Fazekas1974}.  
Although the ground state is likely to be 
the conventional 120$^\circ$ N\'{e}el state according to
the recent numerical and theoretical studies
~\cite{Bernu1992,Bernu1994,Misguich1999,Capriotti1999,Yunoki2006,White2007},  
the quest for a spin-liquid state in the same lattice is still continuing 
by incorporating additional terms to stabilize a spin-liquid state,
such as the next-nearest-neighbor exchange interaction 
~\cite{Lecheminant1995,Manuel1999,Kaneko2014,Iqbal2016,Ferrari2019} and 
the four-spin ring-exchange interaction $J_{\rm c}$~\cite{Misguich1998,Misguich1999,Motrunich2005,Mishmash2013}.
These exchange interactions can be considered as an introduction of
the charge fluctuation~\cite{Calzado2004,Tanaka2018} 
and thus become more relevant for describing magnetic properties of
Mott insulators in proximity of the metal-insulator transition
~\cite{Morita2002,Koretsune2007,Sahebsara2008,Yoshioka2009,
  Tocchio2013,Yamada2014,Laubach2015,Misumi2017,Shirakawa2017,Szasz2018,Skolimowski2019}.
While the ring-exchange interaction itself has long been considered 
for describing the magnetism in the three-dimensional 
solid $^3$He~\cite{McMahan1975,Hetherington1975,Roger1980,Yosida1980,Ceperley1987,Roger2011,Candido2011},
NiS$_2$~\cite{Yosida1981}, and
the parent compounds of high-$T_{\rm c}$ cuprate superconductor such as 
La$_2$CuO$_4$~\cite{Lorenzana1999,Coldea2001,Katanin2002,Headings2010,Rutonjski2016,Yamamoto2019}, 
its importance in triangular-lattice systems near the Mott transition 
is attracting a renewed attention recently~\cite{Motrunich2005,Mishmash2013,Law2017,He2018}  
in organic Mott insulators  $\kappa$-(ET)$_2$Cu$_2$(CN)$_3$~\cite{Shimizu2003,Kurosaki2005,Manna2010} and 
$\mathrm{Et}{\mathrm{Me}}_{3}\mathrm{Sb}{[\mathrm{Pd}{(\text{dmit})}_{2}]}_{2}$~\cite{Itou2008,Yamashita2010}, and 
a charge-density-wave Mott insulator $1T$-TaS$_2$~\cite{FazekasTosatti1979,Klanjsek2017}.

\begin{figure}
  \begin{center}
    \includegraphics[width=0.9\columnwidth]{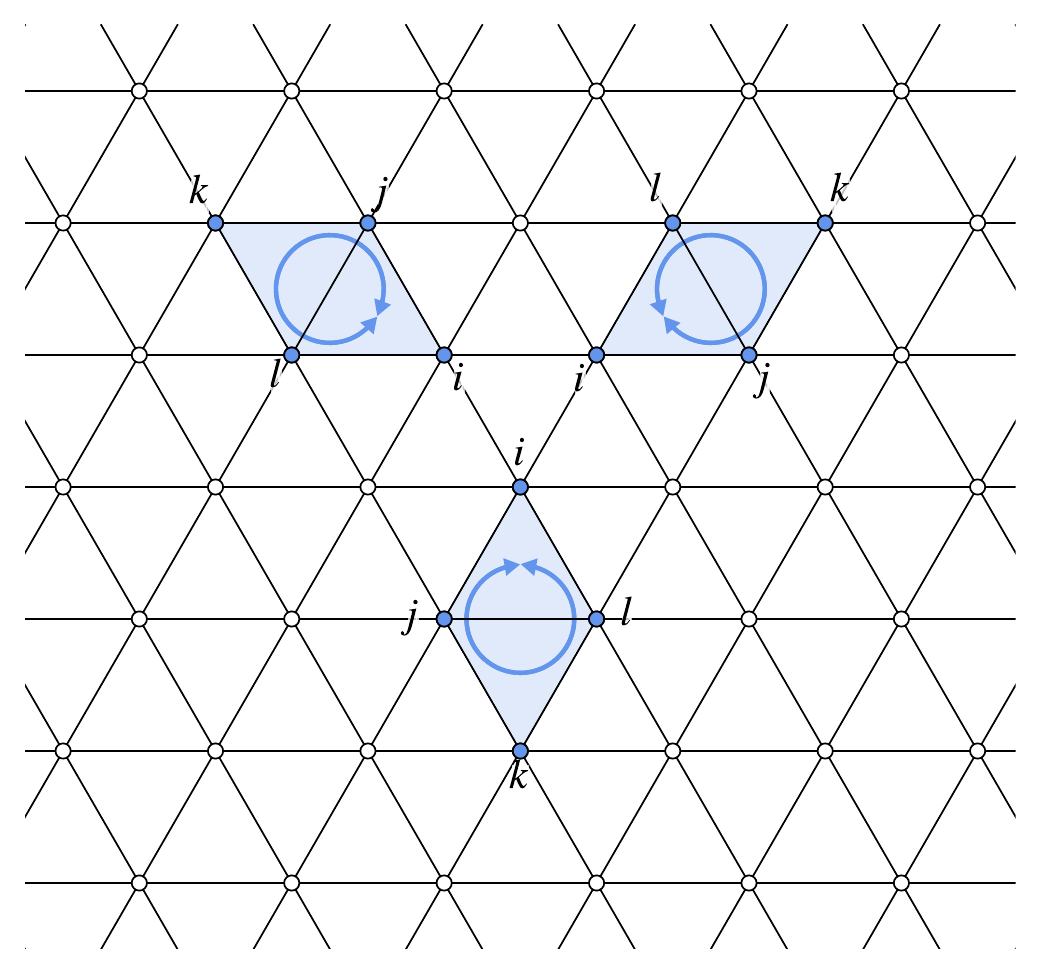}
    \caption{
      Schematic of the model described in Eq.~(\ref{ham}) on the triangular lattice. 
      Elementary parallelograms, where $\hat{P}_{ijkl}$ and $\hat{P}_{ijkl}^\dag$ act 
      (indicated by circular arrows), are indicated. 
      \label{fig:model}}
  \end{center}
\end{figure}

As an effective model for the triangular-lattice materials
near the Mott transition but with frozen charge degrees of freedom,
the ring-exchange model on the triangular lattice
has been proposed~\cite{Motrunich2005,Mishmash2013,Law2017,He2018}. 
The model is described by the following Hamiltonian: 
\begin{equation}
  \hat{H} = J \sum_{\langle ij\rangle} \hat{\mathbf{S}}_i \cdot \hat{\mathbf{S}}_j
  + J_{\rm c} \sum_{\langle ijkl \rangle} \left(\hat{P}_{ijkl} + \hat{P}_{ijkl}^{\dag} \right),
  \label{ham}
\end{equation}
where
$J$ is the nearest-neighbor exchange coupling, 
$J_{\rm c}$ is the four-spin ring-exchange coupling,
$\hat{\mathbf{S}}_{i}=(\hat{S}_i^x,\hat{S}_i^y,\hat{S}_i^z)$ is the spin $S=1/2$ operator, and 
$\hat{P}_{ijkl}$ permutes four spins 
at sites $i,j,k,$ and $l$ on
an elementary parallelogram
cyclically connected as $i$-$j$-$k$-$l$-$i$ (see Fig.~\ref{fig:model}). 
More specifically, we define that
$i$-$k$ and $j$-$l$ are diagonals of the parallelogram, and
$k$ is the next-nearest neighbor of $i$ on the triangular lattice, as indicated in Fig.~\ref{fig:model}. 
The ring-exchange operator $\hat{P}_{ijkl}$ can be expressed 
by a product of permutation operators as 
\begin{equation}
  \hat{P}_{ijkl}=\hat{P}_{il}\hat{P}_{ik}\hat{P}_{ij},    
\end{equation}
where $\hat{P}_{ij} = \hat{P}_{ij}^\dag 
=2 \hat{\mathbf{S}}_i \cdot \hat{\mathbf{S}}_j + \frac{1}{2}$ is the permutation operator exchanging spins at site $i$ and $j$. 
It follows that $\hat{P}_{ijkl}^\dag= \hat{P}_{lkji}=\hat{P}_{ijkl}^{-1}$.
The sum indicated by $\langle ij\rangle$ in the first term of $\hat{H}$ runs over all pairs of nearest-neighbor 
sites $i$ and $j$, and the second sum indicated by $\langle ijkl\rangle$ runs over all elementary parallelograms 
(denoted by shaded blue in Fig.~\ref{fig:model}) formed by sites $i,j,k$, and $l$. 

In terms of the $t/U$ expansion of the half-filled Hubbard model 
with the nearest-neighbor hopping $t$ and the on-site interaction $U$,     
the ring-exchange term appears in the fourth-order expansion with
$J_{\rm c}=20t^4/U^3$, although 
there are additional correction terms in the expansion 
with the fourth order~\cite{Calzado2004,Tanaka2018}. 
Note that the Hamiltonian in Eq.~(\ref{ham}) has been 
considered as a model for the nuclear magnetism of a $^3$He film
adsorbed on graphite preplated with $^4$He
at a particular commensurate density but with a ferromagnetic $J<0$
~\cite{Roger1984,Roger1990,Ishida1997,Roger1998,
  Misguich1999,Momoi1999,Momoi2006,Fukuyama2008,Fuseya2009,Seki2009,Momoi2012},
although a recent fixed-node diffusion Monte Carlo calculation 
poses a question on the realization of such a commensurate
crystalline state~\cite{Moroni2019}. 

Since $J>0$ and $J_{\rm c}\geqslant 0$ are relevant for 
the magnetism near Mott transition,  
let us briefly summarize the ground-state phase diagram of the model in Eq.~(\ref{ham})
for $J>0$ and $J_{\rm c}\geqslant 0$ discussed in the previous literature. 
For $J>0$ and $J_{\rm c}=0$, the ground state is the three-sublattice (120$^\circ$) N{\'e}el ordered state
~\cite{Bernu1992,Bernu1994,Misguich1999,Capriotti1999,Motrunich2005,Yunoki2006,White2007}. 
For $J=0$ and $J_{\rm c}>0$, the ground state is
a spin-liquid state (SL-I), which corresponds to
the RVB state, with no spin gap~\cite{Misguich1999,LiMing2000,Fuseya2009}. 
Another spin-liquid state (SL-II) appears
for moderate $J_{\rm c}/J$~\cite{Misguich1999,LiMing2000,Motrunich2005,Fuseya2009},
where the SL-II phase has many singlet excitations in the spin gap (i.e., below the lowest magnetic excitation).

Besides exploring the spin-liquid ground states, 
it is also crucial to study 
excitation properties such as thermodynamics
as they can be measured experimentally~\cite{Manna2010,Kato2014,Rawl2017,Cui2018}. 
In this paper, we examine 
the effect of the ring-exchange interaction 
on the thermodynamic properties such as
the specific heat, entropy, uniform magnetic susceptibility,
and generalized Wilson ratio.
Recently, these thermodynamic properties,  
except for the specific heat, 
of a model similar to Eq.~(\ref{ham}) but without the terms corresponding to Eq.~(\ref{eq.JcSS})
on a $28$-site cluster 
has been reported~\cite{prelovsek2019vanishing} 
using
an improved version~\cite{Morita2019} of
the finite-temperature Lanczos method~\cite{Jaklic1994,Jaklic2000,Prelovsek}.  
Here,  we propose 
an extended version of the finite-temperature Lanczos method
with the block Lanczos algorithm,  
and adopt it for small-cluster calculations up to $36$ sites.  
The block-Lanczos extension allows
for an efficient sampling over
random states that is required for
approximate evaluation of the trace over a
basis set of the Hilbert space.

The rest of this paper is organized as follows. 
The finite-temperature Lanczos method with the 
extension to the block-Lanczos algorithm 
is described in Sec.~\ref{sec.Method}. 
The method is applied in Sec.~\ref{sec.Results}
to calculate the entropy, the specific heat, the uniform magnetic susceptibility,
and the generalized Wilson ratio of the model 
for various values of $J_{\rm c}/J$. 
The results are summarized and discussed in Sec.~\ref{sec.Conclusions}. 
An algorithm to find a spin configuration from a given state label
in a Hilbert space of a fixed magnetization $S^z=\sum_{i}S_i^z$ is
described in Appendix~\ref{App.A}. 
The effect of the ring-exchange
interaction $J_{\rm c}$ on the spin-wave excitation in 
the $120^{\circ}$ N\'{e}el ordered state is studied
within the linear spin-wave theory in Appendix~\ref{App.B}. 
Throughout the paper, we set $\hbar=k_{\rm B}=1$.

\section{Method}\label{sec.Method}

In this section, we describe the finite-temperature
Lanczos method, which allows us to evaluate the
partition function and thermal averages of physical
observables approximately, without full numerical
diagonalization of the Hamiltonian. 
Before entering the details, let us first briefly summarize the procedure of 
the finite-temperature Lanczos method. 
The key approximations made in the finite-temperature Lanczos method are
(i) stochastic evaluation of the trace of operator $\hat{O}$
and
(ii) approximate evaluation of Boltzmann factor $\e^{-\beta \hat{H}}$
by the Lanczos method, where $\beta$ is the inverse temperature. 
We use the random-phase states for stochastic samplings in (i) and   
adopt the block Lanczos method for (ii).

\subsection{Exact partition function}
The partition function $Z$ at temperature $T=1/\beta$ is defined by
\begin{equation}
  Z = \Tr \left[\e^{-\beta \hat{H}} \right] =
  \sum_{n=1}^{N_{\rm st}} \e^{-\beta{E}_n},    
\end{equation}
where $E_n$ is an eigenvalue of $\hat{H}$
associated with an eigenstate $|E_n \rangle$, 
i.e.,
\begin{equation}
  \hat{H} |E_n \rangle = E_n |E_n \rangle, 
\end{equation}
and $N_{\rm st}$ is the number of eigenstates.
The thermal average of operator $\hat{A}$
is given by 
\begin{equation}
\left\langle \hat{A} \right\rangle =
\frac{1}{Z}\Tr \left[\e^{-\beta \hat{H}} \hat{A} \right] =
\frac{1}{Z}\Tr \left[\e^{-\beta \hat{H}/2}\hat{A} \e^{-\beta \hat{H}/2}\right].
\label{eq.Thermalave}
\end{equation}

In practice, one can make use of symmetries of the Hamiltonian  
to reduce the computational cost for 
numerical diagonalization or Lanczos iterations as 
\begin{equation}
  Z = 
  \sum_{\alpha=1}^{N_{\rm sym}} Z^{(\alpha)}
\end{equation}
with
\begin{equation}
  Z^{(\alpha)} = \sum_{n=1}^{N_{\rm st}^{(\alpha)}} \e^{-\beta{E}_n^{(\alpha)}}   
\end{equation}
and
\begin{equation}  
  \hat{H}^{(\alpha)} |E_n^{(\alpha)} \rangle = E_n^{(\alpha)} |E_n^{(\alpha)} \rangle,   
\end{equation}
where
$\alpha$ labels symmetry sectors of the Hamiltonian, 
$N_{\rm sym}$ is the number of symmetry sectors,
$N_{\rm st}^{(\alpha)}$ is the number of states in a given symmetry sector $\alpha$ satisfying
$N_{\rm st} = \sum_{\alpha=1}^{N_{\rm sym}} {N_{\rm st}^{(\alpha)}}$, and
$\hat{H}^{(\alpha)}$ is the block-diagonalized Hamiltonian, i.e.,  
$\hat{H} = \oplus_{\alpha=1}^{N_{\rm sym}} \hat{H}^{(\alpha)}$.

\begin{figure}
  \begin{center}
    \includegraphics[width=0.95\columnwidth]{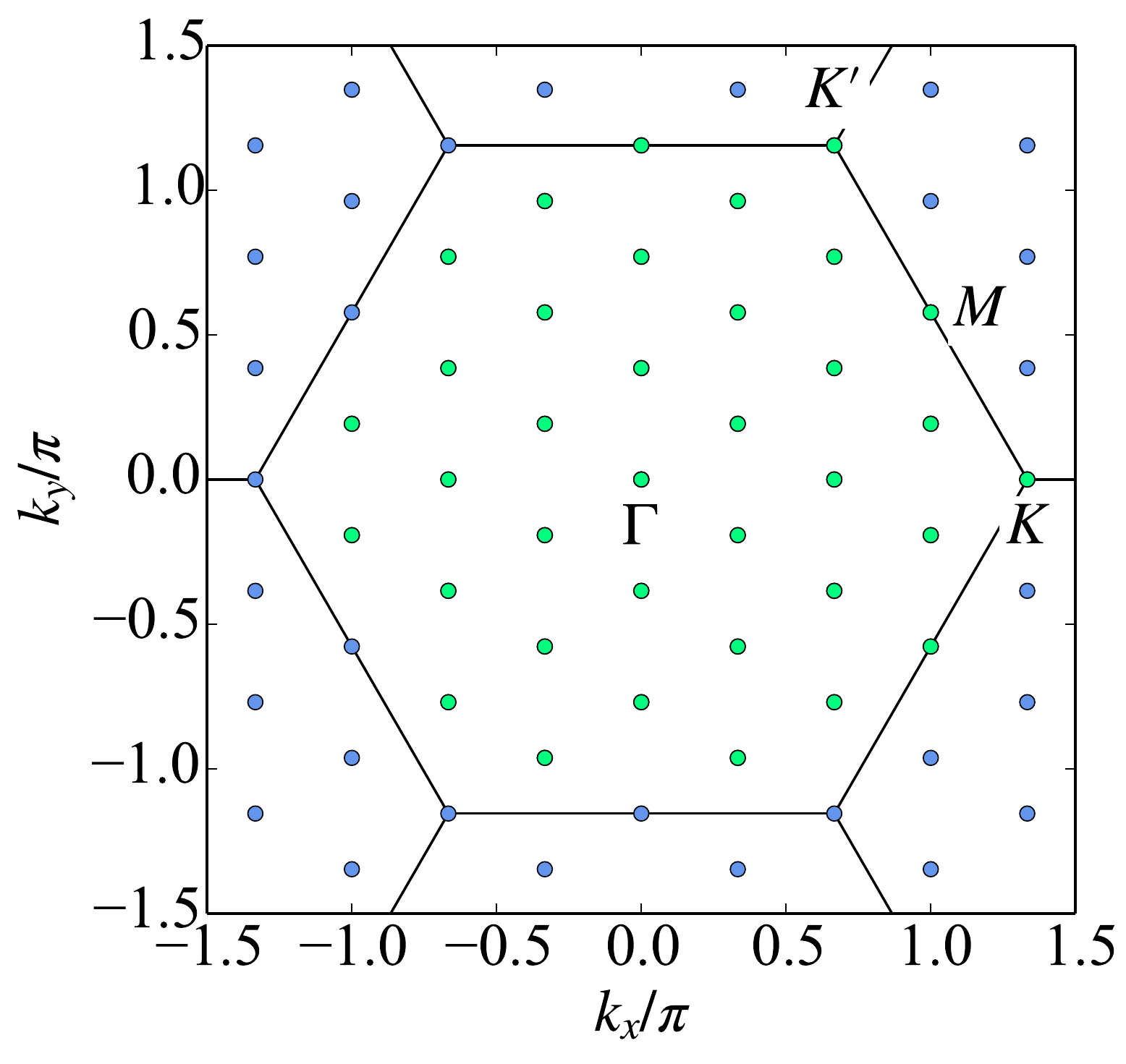}
    \caption{
      Available momenta $\mathbf{k}=(k_x,k_y)$ for the $L=6\times 6$ cluster under the periodic 
      boundary conditions. 
      Solid lines denote the Brillouin-zone boundaries and 
      light green circles indicate the 36 momenta inside the first Brillouin zone. 
      High symmetric momenta, $\Gamma$:~$(0,0)$,
        $K$:~$(4\pi/3,0)$, $M$:~$(\pi,\pi/\sqrt{3})$, and 
        $K'$:~$(2\pi/3,2\pi/\sqrt{3})$, are also indicated. 
      \label{fig.kpoints}}
  \end{center}
\end{figure}

We consider the Hamiltonian in Eq.~(\ref{ham}) on small clusters
under the periodic-boundary conditions.  
The symmetry sectors are labeled as 
$(\alpha)=(\mathbf{k},S^z)$, where
$\mathbf{k}$ is the momentum and $S^{z}$ is the eigenvalue of 
$\hat{S}^z = \sum_{i=1}^{L} \hat{S}^z_i$,
and $L$ is the number of sites.
This labeling of the symmetry sectors results in $N_{\rm sym} = L (L+1)$.
Figure~\ref{fig.kpoints} shows the available momenta for the $L=6 \times 6$ cluster, 
which is the largest size used in the present study. 
In Appendix~\ref{App.A}, we describe
an algorithm to find a spin configuration for a given state label
in the fixed-magnetization Hilbert space.

We evaluate $Z^{(\alpha)}$ numerically exactly 
if $N_{\rm st}^{(\alpha)} \leqslant 10^4$. 
For evaluation of $Z^{(\alpha)}$ with larger $N_{\rm st}^{(\alpha)}$,
we employ the 
finite-temperature Lanczos method~\cite{Jaklic1994,Jaklic2000,Prelovsek}
combined with the block-Lanczos algorithm described in the following sections. 
Below we drop the superscript $(\alpha)$ labeling the symmetry sectors for brevity.

\subsection{Random-phase state}\label{sec.rps}
Following Refs.~\cite{Iitaka2004,Weisse2006}, 
here we review some properties of the random-phase states,
which is relevant to the stochastic evaluation
of the trace. 
Consider a state $|r\rangle$ such that 
\begin{equation}
  |r\rangle = 
  \sum_{x=1}^{N_{\rm st}} \e^{{\rm i} \theta_x^r} |x \rangle,
  \label{rps}
\end{equation}
where $\{|x\rangle\}$ is an arbitrary complete orthonormal set
satisfying $\hat{1}=\sum_{x=1}^{N_{\rm st}}|x\rangle \langle x|$ and $\langle x | x' \rangle=\delta_{xx'}$,
and $\theta_x^r$ are random variables
distributing uniformly in $[0,2\pi)$~\cite{Drabold1993}. 
Notice that $|r\rangle$ is {\it not} normalized because $\langle r | r \rangle = N_{\rm st}$. 
  
We now define a statistical average as
\begin{equation}
  \langle \langle \cdots \rangle \rangle =\lim_{R\to\infty} \frac{1}{R} \sum_{r=1}^R \cdots, 
  \label{statave}
\end{equation}
where $r$ denotes a different set of the random variables.
Since 
$
\left\langle \left \langle \e^{\imag \theta_x^r} \right\rangle \right \rangle= 0  
$
and 
$
\left \langle \left\langle \left(\e^{\imag \theta_{x'}^r}\right)^* \e^{\imag \theta_x^r}\right\rangle \right \rangle
= \left \langle \left\langle \e^{\imag (\theta_{x}^r-\theta_{x'}^r)}  \right\rangle \right \rangle
= \delta_{xx^\prime}
$, 
we can easily show that $|r\rangle$'s are statistically complete
\begin{equation}
  \left \langle \left \langle | r \right \rangle \right \langle r | \rangle  \rangle = 
  \sum_{x=1}^{N_{\rm st}} | x \rangle  \langle x | = \hat{1}.
  \label{random_complete}
\end{equation}
The expectation value of operator $\hat{O}$ with respect to $|r\rangle$
is given by
\begin{equation}
  \langle r| \hat{O}|r \rangle =
  \sum_{x=1}^{N_{\rm st}}
  \langle x |\hat{O} |x\rangle 
+ \sum_{x=1}^{N_{\rm st}}
\sum_{x'=1}^{N_{\rm st}}
\left(
\e^{\imag 
(\theta_{x}^r - \theta_{x'}^r) 
}
-\delta_{xx'}
\right)
\langle x'| \hat{O} | x \rangle.
\label{eq.random_matelm}
\end{equation}
Therefore, the trace can be evaluated stochastically as 
\begin{equation}
  \Tr \left[\hat{O}\right] 
  = \sum_{x=1}^{N_{\rm st}} \langle x | \hat{O} | x \rangle 
  = \langle \langle \langle r | \hat{O} | r \rangle \rangle \rangle.  
  \label{random_trace}
\end{equation}

Finally, if the statistical average is truncated at a finite number $R$
of the random-phase states in Eq.~(\ref{random_trace}), 
the leading error $|\delta O|$,
where $\delta O$ is the second term of
the right-hand side of Eq.~(\ref{eq.random_matelm}), 
is estimated as~\cite{Iitaka2004,Weisse2006}
\begin{eqnarray}
  |\delta O|^2
  &=& \frac{1}{R} \sum_{x \not= x'} |\langle x'|\hat{O}|x \rangle|^2 \notag \\
  &=& \frac{1}{R} \left(\Tr\left[\hat{O}^2\right] - \sum_{x=1}^{N_{\rm st}} \langle x|\hat{O}|x \rangle^2 \right).
  \label{error_rps}
\end{eqnarray}
Here, $\hat{O}$ is assumed to be a Hermitian operator. 
Note, however, that $\hat{O}=\e^{-\beta \hat{H}} \hat{A}$ is not Hermitian  
if $\hat{A}$ does not commute with $\hat{H}$, even if $\hat{A}$ itself is Hermitian.  
In such a case, $\hat{O}$ can still be chosen Hermitian if the symmetric form 
\begin{equation}
  \hat{O}= \e^{-\beta \hat{H}/2} \hat{A} \e^{-\beta \hat{H}/2}   
\end{equation}
is used as in Eq.~(\ref{eq.Thermalave}).

\subsection{Finite-temperature Lanczos method}
From Eqs.~(\ref{statave}) and ~(\ref{random_trace}) we obtain
\begin{equation}
  Z
  =
  \lim_{R\to\infty}
  \frac{1}{R} \sum_{r=1}^{R} \langle r | \e^{-\beta \hat{H}} | r \rangle. 
\end{equation}
Now the matrix element $\langle r| \e^{-\beta \hat{H}} |r \rangle$
has to be evaluated. 
If the full diagonalization of $\hat{H}$ were possible,
the matrix element could be evaluated exactly  
by inserting the identity with the eigenstates 
$\hat{P}_{\rm Eig}=\sum_{n=1}^{N_{\rm st}} |E_n \rangle \langle E_n |=\hat{1}$. 
In the finite-temperature Lanczos method,  $\hat{P}_{\rm Eig}$
is approximated by the projection onto the Ritz states  
$\hat{P}_{\rm Ritz}=\sum_{l=1}^{N_{\rm L}} |\epsilon_l^r \rangle \langle \epsilon_l^r |$,
where 
$|\epsilon_{l}^{r}\rangle$ is the $l$-th Ritz state
associated with the Ritz value $\epsilon_{l}^{r}$ 
obtained by the Lanczos algorithm terminated at the $N_{\rm L}$th
step of the Lanczos iteration started with the initial state $|r\rangle$. 
The partition function is thus approximated as 
\begin{eqnarray}
  Z
  \approx
  \frac{1}{R} \sum_{r=1}^{R} \langle r | \e^{-\beta \hat{H}} | r \rangle  
  &\approx&
  \frac{1}{R} \sum_{r=1}^{R} \sum_{l=1}^{N_{\rm L}}
  \e^{-\beta \epsilon_l^r} |\langle \epsilon_l^r | r \rangle |^2,  
  \label{Zapprox}
\end{eqnarray}
where 
the first approximation is made by
truncating the number of the random states at a finite value $R$, and
the second approximation is made by approximating
the Boltzmann factor as
$\e^{-\beta \hat{H}}\approx \e^{-\beta \hat{H}} \hat{P}_{\rm Ritz}
= \sum_{l=1}^{N_{\rm L}}
\e^{-\beta \epsilon_{l}^r} | \epsilon_{l}^r \rangle \langle \epsilon_{l}^r|$. 
Equation~(\ref{Zapprox}) is the approximate partition function
calculated in the finite-temperature Lanczos method~\cite{Jaklic1994,Jaklic2000,Prelovsek}.
Notice that since $|r\rangle$ defined in Eq.~(\ref{rps}) is not normalized,
differently from Refs.~\cite{Jaklic1994,Jaklic2000,Prelovsek},  
the factor $N_{\rm st}$ does not appear in Eq.~(\ref{Zapprox}). 
Such a factor is taken into account in $|\langle \epsilon_l^r | r \rangle |^2$
in our formulation.

\subsection{Block Lanczos algorithm}
Here,   
we describe the block Lanczos algorithm~\cite{Chatelin,Shirakawa2014,Allerdt2015,Seki2018}
to adopt it for the finite-temperature Lanczos method. 
As the initial states, we first generate $M_{\rm B}$ random-phase states
\begin{equation}
  \left|r_1 \right\rangle, \ \left|r_2 \right \rangle, \quad \cdots, \quad
  \left|r_{M_{\rm B}} \right\rangle.   
  \label{initialstates2}
\end{equation}
To describe the algorithm, it is convenient to move to the matrix notation.
Let 
$\boldsymbol{Y}\in \mathbb{C}^{N_{\rm st} \times M_{\rm B}}$
be a matrix representation of the set of random-phase states in Eq.~(\ref{initialstates2}) 
in the orthonormal basis $\{|x \rangle\}$ used in Eq.~(\ref{rps}), i.e., 
\begin{equation}
  \left[\boldsymbol{Y}\right]_{xb} = \langle x | r_b \rangle =
  \e^{\imag \theta_{x}^{r_b}}.   
\end{equation}
Namely, $\bs{Y}$ contains $M_{\rm B}$ random-phase vectors as column vectors.

Since the $M_{\rm B}$ random-phase vectors are not orthonormalized to each other,
$\bs{Y}$ itself cannot be used as the initial vectors for the block Lanczos algorithm.
Instead, $M_{\rm B}$ orthonormalized vectors
can be obtained from a QR factorization of $\boldsymbol{Y}$ as 
\begin{equation}
  \boldsymbol{Y}=\boldsymbol{Q}_1\boldsymbol{B}_0, 
  \label{SQR}
\end{equation}
where 
$\boldsymbol{Q}_1 \in \mathbb{C}^{N_{\rm st} \times M_{\rm B}}$ 
satisfies $\boldsymbol{Q}_1^\dag \boldsymbol{Q}_1 = \boldsymbol{I}$ and 
$\boldsymbol{B}_0 \in \mathbb{C}^{M_{\rm B}  \times M_{\rm B}}$ 
is an upper triangular matrix satisfying
$\boldsymbol{Y}^\dag \boldsymbol{Y}=\boldsymbol{B}_0^\dag \boldsymbol{B}_0$.
Now $\bs{Q}_1$ can be used as the initial vectors for the block-Lanczos algorithm.
Block-Lanczos vectors $\bs{Q}_2, \bs{Q}_3, \cdots, \bs{Q}_{k_{\rm max}}$ 
with $k_{\rm max} = N_{\rm L}/M_{\rm B}$
are constructed successively by iterating the following procedures
for $k=1$ to  $k_{\rm max}$: 
\begin{eqnarray}
  \label{BL1}
  \bs{A}_k &:=& \bs{Q}_{k}^\dag \bs{H} \bs{Q}_{k} \\
  \label{BL2}
  \bs{X}_k &:=& \bs{H} \bs{Q}_{k} - \bs{Q}_{k} \bs{A}_k  - \bs{Q}_{k-1} \bs{B}_{k-1}^\dag  \\
  \label{BL3}
  \bs{X}_k &=:& \bs{Q}_{k+1}\bs{B}_k, 
\end{eqnarray} 
where $\bs{Q}_0 := \bs{0}$ and  
$\left[ \bs{H} \right]_{xx'} = \langle x | \hat{H} | x' \rangle$ is 
the matrix representation of $\hat{H}$.
The procedure in Eq.~(\ref{BL3}) should be read as the QR factorization 
of $\bs{X}_k \in \mathbb{C}^{N_{\rm st} \times M_{\rm B}}$ yielding the $(k+1)$st block-Lanczos vectors 
$\bs{Q}_{k+1}\in \mathbb{C}^{N_{\rm st} \times M_{\rm B}}$ 
with $\bs{Q}_{k'}^\dag \bs{Q}_{k} = \delta_{k',k} \bs{I}$
and an upper-triangular matrix 
$\bs{B}_k \in \mathbb{C}^{M_{\rm B} \times M_{\rm B}}$. 
The procedure in Eq.~(\ref{BL1}) requires 
$M_{\rm B}$ matrix-vector multiplications. 
Note that $N_{\rm L}$ is assumed to be a multiple of $M_{\rm B}$ for simplicity.   
However, if $N_{\rm L}$ is not a multiple of $M_{\rm B}$,  
$k_{\rm max}$ should be read as 
${\rm nint}(N_{\rm L}/M_{\rm B})$ for example and  
$N_{\rm L}$ below as $k_{\rm max} M_{\rm B}$,
where ${\rm nint(\cdot)}$ denotes the nearest-integer function.

Defining  
$\tilde{\bs{Q}}_{k} = \left(\bs{Q}_1,\cdots,\bs{Q}_{k} \right)
\in \mathbb{C}^{N_{\rm st} \times k M_{\rm B}}$,   
$\bs{T}_{k} = \tilde{\bs{Q}}_k^\dag \bs{H} \tilde{\bs{Q}}_{k} \in \mathbb{C}^{kM_{\rm B} \times kM_{\rm B}}$
can be constructed after the procedure (\ref{BL1}) of the $k$th block-Lanczos iteration. 
It follows from Eqs.~(\ref{BL1})--(\ref{BL3}) that
$\bs{Q}_{j'}^\dag \bs{H} \bs{Q}_{j} = \bs{A}_j \delta_{j',j} + \bs{B}_j \delta_{j',j+1} + \bs{B}_{j'}^\dag \delta_{j',j-1}$.
Therefore, $\bs{T}_k$ is a Hermitian-band matrix of the form 
\begin{equation}
  \label{Bandmatrix}
  \bs{T}_k 
  =
  \left[
    \begin{array}{ccccc}
      \bs{A}_1 &  \bs{B}^\dag_1 & 0 & \cdots & 0      \\
      \bs{B}_1 &  \bs{A}_2      & \bs{B}^\dag_2 & \ddots & \vdots \\
      0 & \ddots        & \ddots & \ddots & 0 \\
      \vdots & \ddots & \bs{B}_{k-2} & \bs{A}_{k-1} & \bs{B}^\dag_{k-1} \\
      0 & \cdots & 0 & \bs{B}_{k-1} & \bs{A}_k 
    \end{array}
  \right]. 
\end{equation}
A diagonalization of $\bs{T}_{k_{\rm max}}$ gives 
$N_{\rm L}$ Ritz values as its eigenvalues, i.e., 
\begin{equation}
  \bs{D} = \bs{U}^\dag \bs{T}_{k_{\rm max}} \bs{U}
  = {\rm diag}(\epsilon_1^{\{r\}}, \cdots, \epsilon_{N_{\rm L}}^{\{r\}}),
  \label{eq.Tdiag}
\end{equation}
where $\bs{U}$ is a unitary matrix. 
Here, the superscript $\{r\}$ denotes that
the Ritz values are obtained by the block-Lanczos method
with the initial states
$\{r\}=\{r_1,r_2,\cdots,r_{M_{\rm B}}\}$.
It follows from Eq.~(\ref{eq.Tdiag}) and 
$\bs{T}_{k_{\rm max}} = \tilde{\bs{Q}}_{k_{\rm max}}^\dag \bs{H} \tilde{\bs{Q}}_{k_{\rm max}}$   
that 
$\bs{D}=\left(\tilde{\bs{Q}}_{k_{\rm max}} \bs{U} \right)^\dag \bs{H} \left(\tilde{\bs{Q}}_{k_{\rm max}} \bs{U} \right)$.
Therefore, the Ritz state $|\epsilon_l^{\{r\}}\rangle$ which satisfies 
$\hat{H}|\epsilon_l^{\{r\}} \rangle = \epsilon_l^{\{r\}} |\epsilon_l^{\{r\}} \rangle$ and 
$\langle \epsilon_l^{\{r\}}| \epsilon_{l'}^{\{r\}}\rangle=\delta_{ll'}$ 
is given by 
\begin{equation} 
  \langle x | \epsilon_l^{\{r\}} \rangle = [\tilde{\bs{Q}}_{k_{\rm max}}\bs{U}]_{xl}. 
  \label{Ritzvector}
\end{equation}

Finally, the overlap between the initial state and the $l$-th Ritz state 
is given by 
\begin{eqnarray}
  \langle \epsilon_l^{\{r\}} | r_b \rangle 
  &=&\left[\bs{U}^\dag \tilde{\bs{Q}}_{k_{\rm max}}^\dag \bs{Y} \right]_{lb} 
  = \left[\bs{U}^\dag \tilde{\bs{Q}}_{k_{\rm max}}^\dag \bs{Q}_1 \bs{B}_0 \right]_{lb} \notag \\
  &=& \sum_{m=1}^{M_{\rm B}}\left[\bs{U}^\dag\right]_{lm}\left[\bs{B}_{0}\right]_{mb},   
  \label{eq.overlap} 
\end{eqnarray}
where $\bs{Q}_{j'}^\dag \bs{Q}_{j} = \delta_{j',j} \bs{I}$ is used in the last equality.

\subsection{Block-extended finite-temperature Lanczos method}
Now the block-extended version of 
the finite-temperature Lanczos method can be formulated.   
For simplicity, we assume that the number $R$
of the random-phase states is a multiple of
the number $M_{\rm B}$ of the block size.
Introducing 
\begin{equation}
  R_{\rm B}=\frac{R}{M_{\rm B}}, 
\end{equation}
the approximate partition function in Eq.~(\ref{Zapprox})
can be expressed as 
\begin{eqnarray}
  Z
  &\approx&
  \frac{1}{R} \sum_{r=1}^{R} \langle r | \e^{-\beta \hat{H}} | r \rangle 
  =
  \frac{1}{R_{\rm B} M_{\rm B}} \sum_{r=1}^{R_{\rm B}} \sum_{b=1}^{M_{\rm B}}
  \langle r_b | \e^{-\beta \hat{H}} | r_b \rangle \notag \\
  &\approx&
  \frac{1}{R_{\rm B}}
  \sum_{r=1}^{R_{\rm B}}
  \sum_{l=1}^{N_{\rm L}}
  \e^{-\beta \epsilon_l^{\{r\}}}
  \left(
  \frac{1}{M_{\rm B}}
  \sum_{b=1}^{M_{\rm B}}  
  |\langle \epsilon_l^{\{r\}} | r_b \rangle |^2
  \right). 
  \label{block_ftlm}
\end{eqnarray}
On the equality of the first line, the $R\,(=R_{\rm B}M_{\rm B})$ random-phase states
are simply relabeled by a combination of the subscripts $r$ and $b$. 
To obtain the second line, the projection operator 
$\hat{P}_{\rm Ritz}=\sum_{l=1}^{N_{\rm L}} |\epsilon_l^{\{r\}} \rangle\langle \epsilon_l^{\{r\}}|$ 
is inserted. 
A formal difference from the standard finite-temperature Lanczos method
is that the
overlap squared,
$|\langle \epsilon_l^r | r \rangle |^2$, 
in Eq.~(\ref{Zapprox})
is replaced by the averaged one over the $M_{\rm B}$ random-phase states,
$\sum_{b=1}^{M_{\rm B}}|\langle \epsilon_{l}^{\{r\}} | r_{b} \rangle|^2/M_{\rm B}$, 
in Eq.~(\ref{block_ftlm}).
Here, the overlap
$\langle \epsilon_{l}^{\{r\}} | r_{b} \rangle$  
can be calculated through Eq.~(\ref{eq.overlap}). 
Obviously, 
Eq.~(\ref{block_ftlm}) reproduces Eq.~(\ref{Zapprox}) 
when $M_{\rm B}=1$.

Similarly to the partition function,
the numerator of Eq.~(\ref{eq.Thermalave})
is approximated as
\begin{eqnarray}
  &&\Tr \left[\e^{-\beta \hat{H}/2}\hat{A} \e^{-\beta \hat{H}/2}\right]
  \approx \frac{1}{R}\sum_{r=1}^R \langle r|\e^{-\beta \hat{H}/2} \hat{A} \e^{-\beta \hat{H}/2} |r\rangle 
  \notag \\
  &\approx&
  \frac{1}{R_{\rm B}}
  \sum_{r=1}^{R_{\rm B}}
  \sum_{l=1}^{N_{\rm L}}
  \sum_{l'=1}^{N_{\rm L}}
  \e^{-\beta \left(\epsilon_l^{\{r\}} + \epsilon_{l'}^{\{r\}} \right)/2} \notag \\
&\times&
\left(
  \frac{1}{M_{\rm B}}
  \sum_{b=1}^{M_{\rm B}}
  \langle r_b | \epsilon_{l'}^{\{r\}} \rangle
  \langle \epsilon_{l'}^{\{r\}} | \hat{A} |  \epsilon_l^{\{r\}} \rangle 
  \langle \epsilon_l^{\{r\}} | r_b \rangle 
  \right). 
  \label{block_ftlm_A}
\end{eqnarray}
Here, the right-most expression of Eq.~(\ref{eq.Thermalave})
is adopted as in the low-temperature Lanczos method~\cite{Aichhorn2003}. 
If $\hat{A}$ commutes with $\hat{H}$,
then $|\epsilon_{l}^{\{r\}} \rangle$ are simultaneous eigenstates of
$\hat{A}$ and $\hat{H}$. 
In this case, Eq.~(\ref{block_ftlm_A}) can be further simplified 
because 
$
\langle \epsilon_{l'}^{\{r\}} | \hat{A}   |  \epsilon_l^{\{r\}} \rangle 
=A_{l}^{\{r\}} \delta_{ll'}$,
where $A_{l}^{\{r\}}$ is an eigenvalue of $\hat{A}$.

A nice property of the block-extended version of the finite-temperature Lanczos method 
[Eqs.~(\ref{block_ftlm}) and (\ref{block_ftlm_A})]
is that one can flexibly choose 
$R_{\rm B}$ and $M_{\rm B}$ 
to exploit the computational resource efficiently.  
For example, the summation 
$\sum_{r=1}^{R_{\rm B}}\cdots$ 
can be done independently for each $r$, 
while a block size of $M_{\rm B}>1$ allows for 
the better performance in a single process as compared 
to the case of $M_{\rm B}=1$. 
To be more specific, 
let us consider an on-the-fly Hamiltonian multiplication to
the block-Lanczos vectors.  
In that case, the dominant computational costs are
generating Hamiltonian matrix elements   
rather than performing simple multiply-add operations. 
Since the block Lanczos method multiplies 
the Hamiltonian matrix to $M_{\rm B}$ vectors simultaneously,  
$M_{\rm B}$ times less operations for generating the matrix elements are required 
to achieve the same number of Hamiltonian-vector multiplications,
as compared to the standard Lanczos method.
We remark that such simultaneous Hamiltonian multiplication
to vectors can be employed also in the polynomial expansion technique~\cite{Weisse2006}.

In the block Lanczos method, 
at least $2M_{\rm B}$ vectors (of $N_{\rm st}$ dimension)
have to be stored.  
When the required memory for storing the $2M_{\rm B}$ vectors 
exceeds the limit of the available resource,
one can simply reduce the number $M_{\rm B}$ of the block size,
or even switch to the standard finite-temperature Lanczos method 
merely by setting $M_{\rm B}=1$.
Fortunately, 
the smaller number $R$ of samplings 
is required for the larger $N_{\rm st}$
to maintain a statistical accuracy  
(see for example Refs.~\cite{Schnack2018,Schnack2020}
and Sec.~\ref{sec.cTPQ}). 

Now we have three parameters $R_{\rm B}$, $M_{\rm B}$, and $N_{\rm L}$
for controlling the accuracy of the block-extended version of the 
finite-temperature Lanczos method. 
Values of these parameters
will be specified for each result in Sec.~\ref{sec.Results}.

\subsection{Connection with the canonical thermal-pure-quantum state}\label{sec.cTPQ} 
The finite-temperature Lanczos method
for observables commuting with $\hat{H}$~\cite{Jaklic1994,Jaklic2000,Prelovsek}, 
the low-temperature Lanczos method for 
observables not commuting with $\hat{H}$~\cite{Aichhorn2003},   
and the block-extended version of the finite-temperature Lanczos method 
for observables not commuting with $\hat{H}$ 
described in the previous section,  
can all be regarded as a method that makes use of 
the canonical thermal-pure-quantum (CTPQ) state~\cite{Sugiura2013},
as recently demonstrated 
with the standard finite-temperature Lanczos method in Ref.~\cite{nishida2019typicalitybased}. 
For example, 
the matrix element $\langle r_b | \e^{-\beta \hat{H}} | r_b \rangle$ 
appearing in Eq.~(\ref{block_ftlm}) is 
the inner product of the (unnormalized) CTPQ state $\e^{-\beta \hat{H}/2} |r_b \rangle$. 
There are several ways to evaluate matrix functions
operated to vectors
without full diagonalization,  
such as polynomial expansion techniques~\cite{TalEzer1984,
  Wang1994,Wang1994PRL,Iitaka1994,Vijay2002,Iitaka2003,
  Machida2005,Weisse_book,Seki2019Dirac}. 
With the Lanczos method used here, 
the CTPQ state is approximated by 
a linear combination of the $N_{\rm L}$
Ritz states $|\epsilon_l^{\{r\}} \rangle$ as
\begin{equation}
  \e^{-\beta \hat{H}/2} |r_b\rangle \approx  
  \hat{P}_{\rm Ritz} \e^{-\beta \hat{H}/2} |r_b\rangle  =
  \sum_{l=1}^{N_{\rm L}}
  \e^{-\beta \epsilon_{l}^{\{r\}}/2} \langle \epsilon_l^{\{r\}} | r_b \rangle
  | \epsilon_l^{\{r\}} \rangle. 
  \label{cTPQlanczos}
\end{equation}
In this sense, although it is difficult to estimate
the systematic error associated with the approximation
made in Eq.~(\ref{cTPQlanczos}), 
one can still refer to the 
convergence analysis of CTPQ states~\cite{Sugiura2013}.
For instance, the better convergence in probability
to the ensemble average is expected
for the larger $\mathcal{D}(T)=\e^{Ls(T)}$ with 
$s(T)$ being the entropy density. 
Here $\mathcal{D}(T)$ can be interpreted as a temperature-dependent
effective dimension of the Hilbert space, because it satisfies 
$\lim_{T\to \infty} \mathcal{D}(T) = N_{\rm st}$ and 
$\lim_{T\to 0} \mathcal{D}(T) = g$,
where $g$ is the ground-state degeneracy.

Note that the (block) Lanczos method
approximates well 
the extremal eigenvalues and eigenstates within
a few hundreds of the Lanczos steps $N_{\rm L}$,
almost independently of the realization of
the initial random-phase state $|r_b\rangle$.
Therefore, the (block) Lanczos approach
to the matrix exponential, as in Eq.~(\ref{cTPQlanczos}),  
complements the CTPQ approach
at low temperatures by its fast convergence
to the ground state and low-lying
excited states for each symmetry sector.
In particular, the block Lanczos method can better approximate
the low-lying excited states, especially within the block size, 
as compared to the standard Lanczos method~\cite{Chatelin}. 
On the other hand, empirically, 
the convergence of the (block) Lanczos method 
to the inner (i.e., non extremal) eigenpairs with dense spectra seems ``random'', 
in the sense that the convergence depends on 
the realization of $|r_b\rangle$ for fixed $N_{\rm L}$, 
as observed in spectra of dynamical correlation functions~\cite{Prelovsek}.
This implies that relatively large error bars are expected
at temperatures 
where the specific heat exhibits a peak, 
because the larger specific
heat indicates the larger fluctuation of the
internal energy $\langle \hat{H} \rangle$,
thus implying the denser eigen spectra of $\hat{H}$.   
Finally, we remark that a connection between the finite-temperature Lanczos 
method and the eigenstate-thermalization hypothesis~\cite{Deutsch1991,Srednicki1994}
has been discussed recently in Ref.~\cite{Rousochatzakis2019}.

\section{Results}\label{sec.Results}

\begin{figure}
  \begin{center}
    \includegraphics[width=0.95\columnwidth]{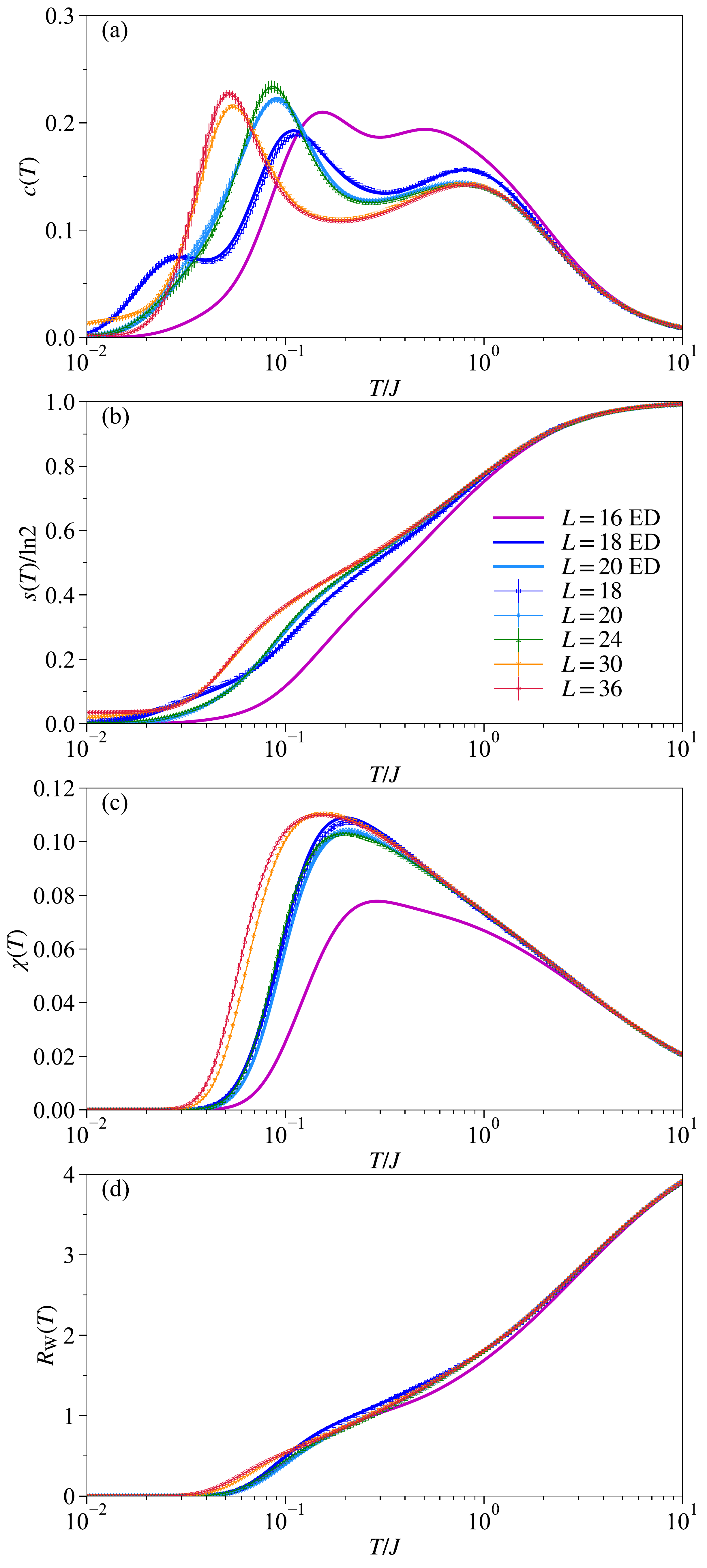}
    \caption{
      (a) Specific heat $c(T)$,
      (b) entropy density $s(T)$, 
      (c) uniform susceptibility $\chi(T)$, and 
      (d) Wilson ratio $R_{\rm W}(T)$    
      at $J_{\rm c}/J=0.07$ for $L=4\times4$, $18$,
      $5\times 4$, $6\times 4$, $30$, and $6\times 6$ 
      clusters (see Fig.~\ref{fig:cluster}). 
      Solid lines are results obtained by the full exact diagonalization. 
      Block-Lanczos parameters are
      $R_{\rm B}=24$, $M_{\rm B}=6$, and $N_{\rm L}=120$ for $L=4\times 5$,
      $R_{\rm B}=24$, $M_{\rm B}=8$, and $N_{\rm L}=160$ for $L=18$,  
      $R_{\rm B}=8$,  $M_{\rm B}=8$, and $N_{\rm L}=200$ for $L=6\times 4$, 
      $R_{\rm B}=6$,  $M_{\rm B}=8$, and $N_{\rm L}=320$ for $L=30$, and 
      $R_{\rm B}=6$,
      $M_{\rm B}=4$ for $0 \leqslant |S_z| \leqslant 1$,
      $M_{\rm B}=6$ for $2 \leqslant |S_z| \leqslant 5$, 
      $M_{\rm B}=6$ for $6 \leqslant |S_z|$,       
      and $N_{\rm L}=720$ for $L=6\times 6$. 
      \label{Ldep}}
  \end{center}
\end{figure}

\begin{figure}
  \begin{center}
    \includegraphics[width=0.95\columnwidth]{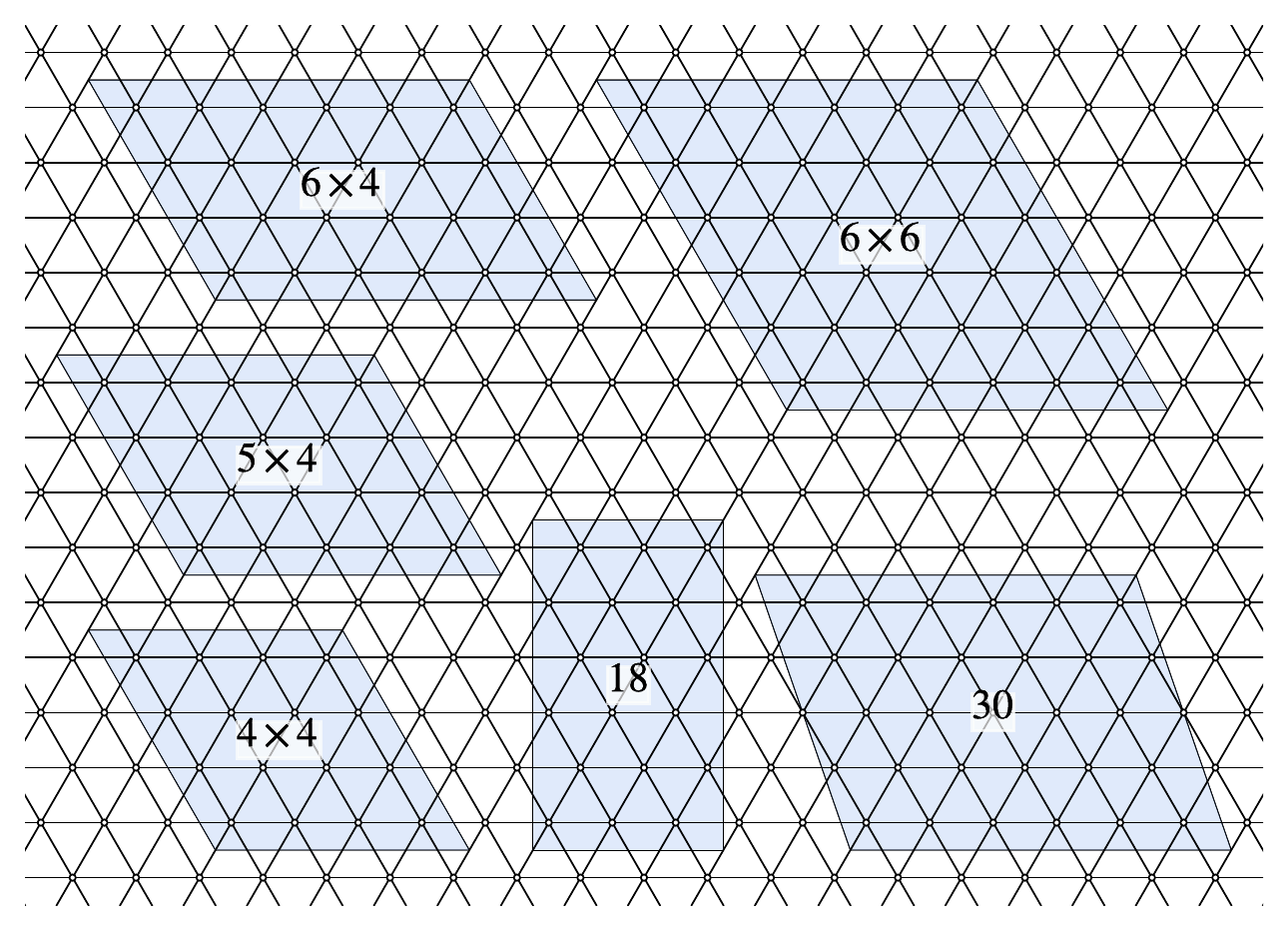}
    \caption{
    Cluster structures used for the calculations. 
    The periodic boundary conditions are imposed.
    \label{fig:cluster}}
  \end{center}
\end{figure}

Figure~\ref{Ldep} shows the specific heat
\begin{equation}
  c(T) = \frac{1}{LT^2} \left[
    \left \langle \hat{H}^2 \right \rangle -
    \left \langle \hat{H}   \right \rangle^2  \right],
\end{equation}
the entropy density 
\begin{equation}
  s(T) = \frac{1}{LT}
  \left[ \left \langle \hat{H} \right \rangle + T\ln Z
    \right], 
\end{equation}
the uniform magnetic susceptibility
\begin{equation}
  \chi(T) = \frac{1}{LT} \left[
  \left \langle \left( \hat{S}^z \right)^2 \right \rangle -
  \left \langle \hat{S}^z     \right \rangle^2 \right],  
\end{equation}
and the generalized temperature-dependent Wilson ratio~\cite{Prelovsek2019} 
\begin{equation}
  R_{\rm W}(T) = \frac{4\pi^2 T \chi(T)}{3 s(T)}   
\end{equation}
at $J_{\rm c}/J=0.07$ for $L=4\times4$, $18$, $5\times 4$, $6\times 4$, $30$, and $6\times 6$ 
(see Fig.~\ref{fig:cluster}).  
Notice that the entropy density $s(T)$ is normalized with respect to $\lim_{T\to\infty}s(T)=\ln{2}$ in the figure.  
Since these quantities involve only the thermal average of the quantities
that commute with $\hat{H}$, the calculations
are particularly efficient as compared to 
the quantities that do not commute with $\hat{H}$. 
Each of the error bars represents 
the standard error of the mean 
$\tilde{\sigma}/\sqrt{R_{\rm B}}$ 
with $\tilde{\sigma}$ being 
the estimated standard deviation defined by
\begin{equation}
  \tilde{\sigma} = \sqrt{ \frac{1}{R_{\rm B}-1}\sum_{r=1}^{R_{\rm B}} \left(X_r - \bar{X} \right)^2 },
\end{equation}
where 
$X_r$ is calculated $c(T)$, $s(T)$, $\chi(T)$, or $R_{\rm W}(T)$
but for a given $r$ (without averaging over $r$), and  
$\bar{X}$ is $c(T)$, $s(T)$, $\chi(T)$, or $R_{\rm W}(T)$ itself.
For comparison, the full-diagonalization results are also shown in Fig.~\ref{Ldep} for $L \leqslant 20$. 
It is confirmed for $L=18$ and $L=5\times 4$ 
that the results obtained by the block-extended version of the finite-temperature Lanczos method 
mostly coincide with the full-diagonalization results within error bars~\cite{SM}.

\begin{figure*}
  \begin{center}
    \includegraphics[width=1.9\columnwidth]{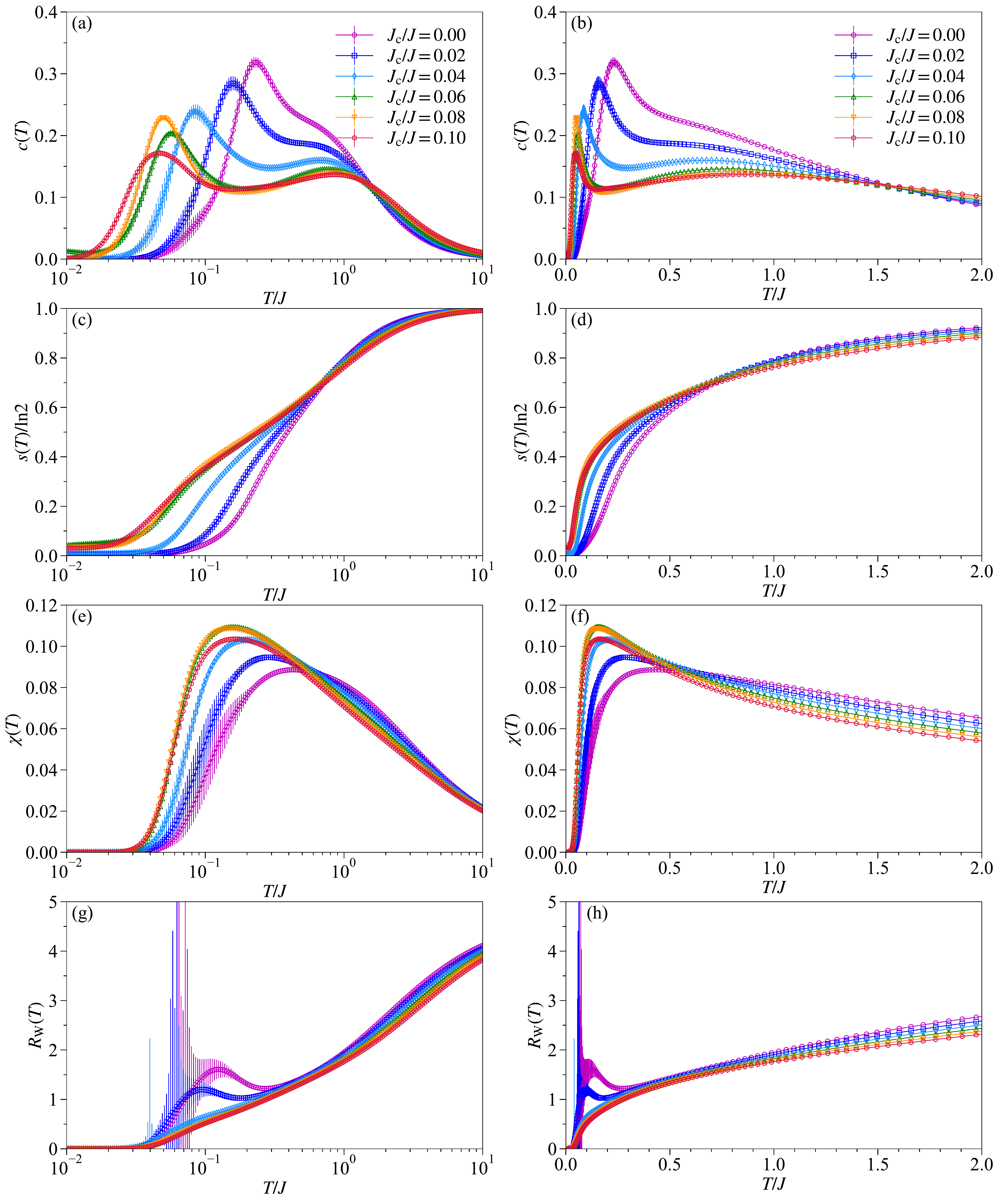}
    \caption{
      Semilog (left) and linear (right) plots of 
      (a),(b) specific heat $c(T)$,
      (c),(d) entropy density $s(T)$, 
      (e),(f) uniform susceptibility $\chi(T)$, and 
      (g),(h) Wilson ratio $R_{\rm W}(T)$      
      for several values of $J_{\rm c}/J$, indicated in the figures, and $L=6\times 6$. 
      Block-Lanczos parameters are 
      $R_{\rm B}=6$,
      $M_{\rm B}=4$ for $0 \leqslant |S_z| \leqslant 1$,
      $M_{\rm B}=6$ for $2 \leqslant |S_z| \leqslant 5$, 
      $M_{\rm B}=6$ for $6 \leqslant |S_z|$,       
      and $N_{\rm L}=720$. 
      \label{Kdep}}
  \end{center}
\end{figure*}

Figure~\ref{Kdep} shows
the $J_{\rm c}$ dependence of
$c(T)$, $s(T)$, $\chi(T)$, and $R_{\rm W}(T)$
for $L=6\times6$, which is the largest cluster available
and preserves all the symmetries of the triangular lattice. 
Without the ring-exchange interaction ($J_{\rm c}=0$),
$c(T)$ exhibits a peak around $T/J=0.2$ and a broad shoulder
for $T/J\gtrsim 0.5$, while no significant structure can be found in $s(T)$.  
This is in good agreement with the previous results
calculated by the finite-temperature Lanczos and
the exponential tensor-renormalization-group 
methods~\cite{Prelovsek2018,Chen2018,Chen2019TRG,prelovsek2019vanishing}. 
At low temperatures, 
a power-law dependence of $c(T)\sim T^2$
is expected with the N\'eel order~\cite{Bernu2001}.
However, such a power-law dependence is 
not found here due to the 
energy gap intrinsic to the finite-size calculation.  

For $J_{\rm c}/J \geqslant 0.04$, the specific heat $c(T)$
shows a double-peak structure with
a broad high-temperature peak at $T=T_{\rm high}\sim J$ and
a sharp low-temperature peak at $T=T_{\rm low} \ll J$.
Moreover,
it is observed that the high-temperature peak shifts towards
higher temperature with increasing $J_{\rm c}/J$ like
$T_{\rm high}\sim J+5J_{\rm c}$. 
Such a behavior of the high-temperature peak
can be expected from Eq.~(\ref{eq.Hfull}),
where the effective nearest-neighbor exchange $J+5J_{\rm c}$
becomes a dominant energy scale at high temperatures.

As shown in Fig.~\ref{Ldep}(a), 
the low-temperature peak position for $J_{\rm c}/J=0.07$ 
tends to be lowered for the larger clusters,
while the high-temperature peak is insensitive to the system size.
For example, for the $L=6\times 6$ cluster, 
the high-temperature peak appears at $T_{\rm high}/J\approx 0.8$ 
and the low-temperature peak is found at $T_{\rm low}/J\approx 0.05$.  
At the highest temperature around $T/J \sim 10$,
the entropy density reaches $s=\ln 2 \approx 0.693$, indicating that
the system is in the paramagnetic state.
In the temperature regime where $c(T)$ shows a dip between the two peaks, 
$s(T)$ exhibits a shoulder-like structure which is visible
in the semilog plot shown in Fig.~\ref{Kdep}(c).  
Interestingly, about the half of the total entropy 
$s=\frac{1}{2}\ln 2 \approx 0.347$ remains at such a temperature regime. 
The shoulder-like structure of $s(T)$ becomes more prominent
for the larger system size [see Fig.~\ref{Ldep}(b)].

As shown in Figs.~\ref{Kdep}(e) and \ref{Kdep}(f), 
the uniform magnetic susceptibility $\chi(T)$ decreases quickly
below temperature $T_{\chi}$ at which $\chi(T)$ takes a maximum. 
The peak position $T_{\chi}$ varies
from $T_\chi \approx 0.3J$ for $J_{\rm c}/J=0$
to $T_\chi \approx 0.15J$ for $J_{\rm c}/J=0.1$.
In particular, a rapid decrease of $T_{\chi}$ can be observed 
for $0 \leqslant J_{\rm c}/J \leqslant 0.06$. 
It is also found that for $J_{\rm c}/J \geqslant 0.04$ there exists a temperature region where 
the entropy and specific heat are finite
while $\chi(T)$ is almost zero.   
This implies that many nonmagnetic excitations exist below
the first magnetic excitation,
which is consistent with
the strong-coupling expansion of the Hubbard model~\cite{Yang2010}. 
Such low-lying nonmagnetic excited states
are thus essential for forming the low-temperature peak
in the specific heat.

These characteristic low-lying excitations can be better seen in the 
temperature-dependent Wilson ratio $R_{\rm W}(T)$~\cite{Prelovsek2019} shown in Figs.~\ref{Kdep}(g) and \ref{Kdep}(h). 
If this quantity tends to zero, 
it is indicative that the magnetic excitations are inactive while nonmagnetic ones are active.
Although the error bars are too large to discuss
its behavior for $T/J<0.1$ and $J_{\rm c}/J \lesssim 0.02$, 
the slight upturn of $R(T)$ for $J_{\rm c}/J=0$ at $T/J\sim 0.2$ 
is consistent with the result for the pure-triangular case
reported in Ref.~\cite{Prelovsek2019}. 
Despite the large error bars, one can still observe 
a clear change of behavior in $R_{\rm W}(T)$
for $T/J<0.4$ between the parameter regions
$J_{\rm c}/J \leqslant 0.02$ and 
$J_{\rm c}/J \geqslant 0.04$.

Finally, it is observed in Fig.~\ref{Kdep} that the error bars become larger below
the temperature at which the specific heat takes the maximum
(the low-temperature maximum for $J\geqslant 0.04$).
This behavior is expected from the discussion in Sec.~\ref{sec.cTPQ}.

\section{Summary and Discussion}\label{sec.Conclusions}
The thermodynamic properties of an $S=1/2$ antiferromagnetic Heisenberg model 
on the triangular lattice with the ring-exchange interaction 
have been studied by the block-extended version of 
the finite-temperature Lanczos method. 
The results for entropy $s(T)$, uniform magnetic susceptibility $\chi(T)$,
and Wilson ratio $R_{\rm W}(T)$ have shown that 
there exist low-energy nonmagnetic excitations 
for $J_{\rm c}/J \geqslant 0.04$.
The specific heat $c(T)$
exhibits a characteristic double-peak structure
for $J_{\rm c}/J \geqslant 0.04$,  
with the low-temperature peak being 
caused by these nonmagnetic excitations.

As it is apparent from $s(T)$, $\chi(T)$, and $R_{\rm W}(T)$, there is a 
great deal of similarity in the low-lying excitations 
between the ring-exchange model studied here and
the $J_{1}-J_{2}$ model on the triangular lattice or
the kagome-lattice antiferromagnet~\cite{Lecheminant1997,Waldtmann1998,Prelovsek2019,prelovsek2019vanishing}. 
However, the double-peak structure found here in $c(T)$ for $J_{\rm c}/J \geqslant 0.04$ 
distinguishes the ring-exchange model from the other models. 
Indeed, such a double-peak structure has not been 
observed in the $J_1-J_2$ model on the triangular lattice 
for $J_2/J_1=0.1$ and $0.2$~\cite{Prelovsek2018}. 
Moreover, the separation of these two peaks for the ring-exchange model is 
found to be more pronounced with increasing the system size. 
Such a system-size dependence 
of the low-temperature peak positions is
in contrast to that in the kagome-lattice antiferromagnet 
where the lower-temperature peak moves towards
higher temperatures with increasing the system size~\cite{Shimokawa2016,Schnack2018}.  
Instead, a system-size dependence similar to the ring-exchange model found here has also been observed
in the Kitaev model~\cite{Yamaji2016}.
This implies that the excitations corresponding to the high-temperature
peak are spatially local, while those corresponding to the low-temperature
peak are not.

It is interesting to compare the present results with
the recent experiments on Ba$_2$CoNb$_6$O$_{24}$, 
which is considered to be the $S=1/2$ two-dimensional 
triangular-lattice Heisenberg antiferromagnet 
with a nearest-neighbor coupling
$J=0.144$~meV~\cite{Rawl2017} or $J=1.66\pm0.06$~K~\cite{Cui2018}. 
In this material, no indication of the magnetic
order has been found in the thermodynamic measurements
down to $T=80~{\rm mK}$.  
After subtraction of the phonon contribution ($\propto T^3$),
the specific heat takes a single-peak structure.
Considering the absence of the double-peak structure in the specific heat, 
the case without the ring-exchange interaction (i.e., $J_{\rm c}/J=0$) is
rather more relevant to Ba$_2$CoNb$_6$O$_{24}$ than the ring-exchange model. 
In the literature~\cite{Rawl2017,Cui2018},
the absence of the 120$^\circ$ long-range order 
at finite temperatures is attributed to 
a realization of the Mermin-Wagner theorem~\cite{Mermin1966}
on the real material. 

Thermodynamic properties of $1T$-TaS$_2$ have also been 
measured experimentally~\cite{Kratochvilova2017,Ribak2017,Maruyama2020}. 
So far, no indication of a double-peak structure 
in the magnetic heat capacity has been reported.   
For example, only a single broad hump in the
magnetic heat capacity has been observed in Ref.~\cite{Kratochvilova2017}. 
However, the entropy at high temperature, 
obtained by integrating the magnetic heat capacity over the whole temperature region
measured, 
reaches only $\approx 40\%$ of $\ln2$~\cite{Kratochvilova2017}. 
If we assume that there exists a sharp peak in the magnetic heat capacity
at temperature lower than the experimental reach,
such a missing entropy is not inconsistent with our results, 
because our result implies that $s\approx\frac{1}{2}\ln 2$
remains at the temperature where $c(T)$ exhibits a dip.
Therefore, a further study on the missing entropy in $1T$-TaS$_2$ is highly desirable. 
We should note that a similar scenario on
the missing entropy and the double-peak structure in the heat capacity  
had been discussed in the context of nuclear magnetism of $^3$He film~\cite{Greywall1989},
which was resolved by the lower-temperature measurement of the heat capacity~\cite{Ishida1997}.

In Appendix~\ref{App.B}, we study 
the effect of the ring-exchange interaction $J_{\rm c}$
on the spin-wave dispersion in the 120$^\circ$ N\'{e}el 
ordered state, within the linear spin-wave theory.
It is found that the spin-excitation energies
near the $M$ point and symmetrically equivalent points 
are decreased drastically with $J_{\rm c}$. 
However, the spin-wave analysis,
which takes into account only the magnon excitation, 
was not able to capture 
the characteristic thermodynamic features, including
the double-peak structure of $c(T)$, found in our numerical calculations.
In particular, the microscopic understanding
of the double-peak structure in $c(T)$ found here 
requires a rather systematic analysis for larger clusters and 
is left for the future study.

\acknowledgments
The authors are grateful to
Tao Li for useful discussions and
Yusuke Nomura and Shohei Miyakoshi for helpful comments. 
The numerical computations have been done
on HOKUSAI GreatWave and HOKUSAI BigWaterfall supercomputers
at RIKEN under Project Nos.~G19011 and G20015. 
This work was supported by Grant-in-Aid for Research Activity start-up (No.~19K23433)
and Grant-in-Aid for Scientific Research (B) (No.~18H01183) 
from MEXT, Japan.

\appendix

\section{Algorithm to find a spin configuration for a given state label in a fixed-magnetization Hilbert space}\label{App.A}
The two-dimensional search technique introduced by Lin~\cite{Lin1990}
is an efficient method to find a state label $j$ 
for a given spin configuration $i$, i.e.,
$j(i)$, with a relatively small 
amount of storage, whose dimension is $~ 2 \times 2^{L/2}$.
Here, a set of the binary digits $\{b_l\}$ that represents $i$ with 
\begin{equation}
  i = \sum_{l=1}^{L} b_{l} 2^{l-1} \equiv (b_L b_{L-1} \ldots b_{1})_2
\end{equation}
is assigned to a spin configuration, 
by identifying $b_{l}=0$ ($b_{l}=1$) with 
the presence of a spin-$\downarrow$ (spin-$\uparrow$) at the $l$th site.

The inverse table,
which returns a spin configuration $i$
for a given state label $j$, i.e., $i(j)$,
is often stored. 
For a fixed-magnetization Hilbert space,  
the length of the inverse table is 
given by the binomial coefficient 
\begin{equation}
\binom{L}{N_\uparrow}
=\frac{L!}{N_{\uparrow}! (L-N_{\uparrow})!}
=\binom{L}{L-N_\uparrow},
\label{eq.length}
\end{equation}
where $N_{\sigma}$ is the number of spins with spin $\sigma$, 
$N_{\uparrow} + N_{\downarrow}=L$, and 
the magnetization is given by
$S^z = (N_{\uparrow} - N_{\downarrow})/2$.  
The range of the state label $j$ can be chosen as 
\begin{equation}
  1 \leqslant j \leqslant \binom{L}{N_{\uparrow}}. 
  \label{eq.jrange}
\end{equation}
For a concrete example of the correspondence
between $j$ and $i$, see Table~\ref{tab:j_to_i}. 
Since the range of $i$ is given by 
\begin{equation}
  2^{N_{\uparrow}}-1 \leqslant i \leqslant 2^{L}-2^{L-N_{\uparrow}}, 
\end{equation}
$i$ might be 64 bit integer for $L\geqslant32$. 
An algorithm that returns
a spin configuration $i$ for a given state label $j$ 
may be useful when spin configurations $i$
do not appear sequentially 
during the calculation of 
matrix elements of the Hamiltonian,
due to, for example, a parallelization
of the on-the-fly matrix-vector multiplication. 

\begin{table}
  \caption{
    Correspondence between state label $j$
    and spin configuration $i$ for $L=6$ and $N_{\uparrow}=3$.
    Both $j$ and $i$ are assumed to be in ascending order.  
    \label{tab:j_to_i}}
  {
  \begin{tabular}{ccccccc}
    \hline
    \hline
    $j$ & $i(L,N_{\uparrow},j)$ & & \text{\qquad} & $j$  & $i(L,N_{\uparrow},j)$    &   \\ 
    \hline
    $1$ & $(000111)_2=$&$7$   &               & $11$     & $(100011)_2=$&$35$        \\
    $2$ & $(001011)_2=$&$11$  &               & $12$     & $(100101)_2=$&$37$        \\
    $3$ & $(001101)_2=$&$13$  &               & $13$     & $(100110)_2=$&$38$        \\
    $4$ & $(001110)_2=$&$14$  &               & $14$     & $(101001)_2=$&$41$        \\
    $5$ & $(010011)_2=$&$19$  &               & $15$     & $(101010)_2=$&$42$        \\
    $6$ & $(010101)_2=$&$21$  &               & $16$     & $(101100)_2=$&$44$        \\
    $7$ & $(010110)_2=$&$22$  &               & $17$     & $(110001)_2=$&$49$        \\
    $8$ & $(011001)_2=$&$25$  &               & $18$     & $(110010)_2=$&$50$        \\
    $9$ & $(011010)_2=$&$26$  &               & $19$     & $(110100)_2=$&$52$        \\
   $10$ & $(011100)_2=$&$28$  &               & $20$     & $(111000)_2=$&$56$      \\
    \hline
    \hline
  \end{tabular}
  }
\end{table}

\begin{figure}
  \begin{center}
    \includegraphics[width=0.95\columnwidth]{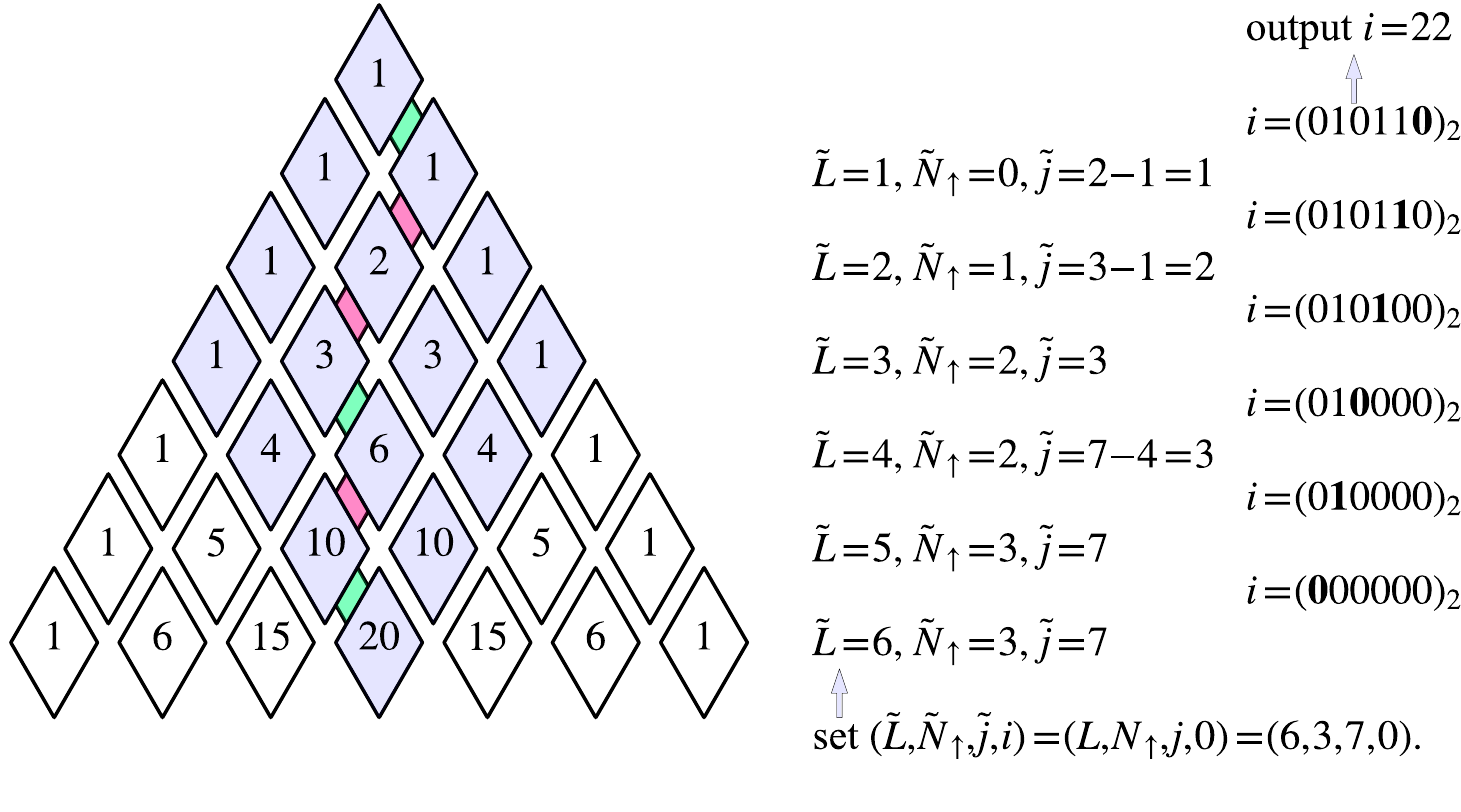} \\
    \caption{      
      Schematic figure of the algorithm 
      to find a spin configuration $i$ for a given state label $j$.
      The figure should be read from bottom to top
      to compare with Algorithm~\ref{findconfig}. 
      For a given set of $L$, $N_{\uparrow}$, and $j$,  
      one of the shortest paths from the vertex $\binom{L}{L-N_{\uparrow}}$
      to the topmost vertex $\binom{0}{0}$ of Pascal's triangle
      is assigned, and the path determines the spin configuration
      $i = \sum_{\tilde{L}=1}^{L} b_{\tilde{L}}2^{\tilde{L}-1}=(b_{L} b_{L-1}\ldots b_{1})_2$.
      The path goes rightward if 
      $\tilde{j} > \binom{\tilde{L}-1}{\tilde{N}_{\uparrow}}$ (indicated by magenta),
      or else leftward (indicated by green).
      The rightward (leftward) path from $\tilde{L}$th row to $\tilde{L}-1$th row  
      implies that $b_{\tilde{L}}=1$ ($b_{\tilde{L}}=0$). 
      The figure refers to the input $(L,N_{\uparrow},j) = (6,3,7)$,
      which results in the output $i=(010110)_2=22$. 
      The $(N_{\uparrow}+1)\times (N_{\downarrow}+1)=16$
      vertices on the possible $\binom{6}{3}=20$ shortest paths are highlighted
      with shaded blue color,
      and $b_{\tilde{L}}$ in $i$ is highlighted with boldface.
      Although a quick return is possible at $\tilde{L}=3$ in this example, 
      according to the lines 9--11 of Algorithm~\ref{findconfig},
      the remaining processes 
      corresponding to the lines 12--15 of Algorithm~\ref{findconfig} 
      for $\tilde{L}\leqslant 3$ are also shown here in this figure. 
      \label{pascal}}
  \end{center}
\end{figure}

Here we introduce such a function $i(j)$
by assuming that both $i$ and $j$
are in the ascending order, as in Table~\ref{tab:j_to_i}. 
The basic idea is to assign a state label $j$ to one of  
the shortest paths from the vertex $\binom{L}{L-N_{\uparrow}}$
to the topmost vertex $\binom{0}{0}$ on Pascal's triangle 
(see Fig.~\ref{pascal}).
Since there are $\binom{L}{L-N_{\uparrow}}$ different paths,
a one-to-one correspondence between the shortest paths and $\{j\}$ should exist.

To find a correspondence between binary numbers and
the shortest paths on Pascal's triangle, 
the following combinatorial recursion formula should be reminded;
\begin{equation}
  \binom{L}{L-N_{\uparrow}} = 
  \binom{L-1}{L-N_{\uparrow}-1} +
  \binom{L-1}{L-N_{\uparrow}}. 
  \label{eq.recursion}
\end{equation}
In terms of Pascal's triangle,
Eq.~(\ref{eq.recursion}) relates the current vertex (left-hand side) with its
upper left vertex (first term of the right-hand side) and
upper right vertex (second term of the right-hand side). 
More specifically, among the total 
$\binom{L}{L-N_{\uparrow}}$ spin configurations, 
$\binom{L-1}{L-N_{\uparrow}-1}=\binom{L-1}{N_{\uparrow}}$ spin configurations have
``0'' at the $L$th binary digit, and 
$\binom{L-1}{L-N_{\uparrow}}=\binom{L-1}{N_{\uparrow}-1}$ spin configurations have
``1'' at the $L$th binary digit, 
assuming that the number of 1's is $N_{\uparrow}$.
By taking into account also the assumption 
that both $i$ and $j$ are in the ascending order with
Eq.~(\ref{eq.jrange}), 
the $L$th binary digit $b_{L}$ of $i$ for a given $j$ is determined as 
\begin{eqnarray}
  b_{L}=
  \left\{
  \begin{array}{ll}
    0 & {\text{if }} j \leqslant \binom{L-1}{N_{\uparrow}}, \\
    1 & {\text{otherwise}}. 
  \end{array}
  \right.
  \label{eq.bL}
\end{eqnarray}
This property holds for any $(L,N_{\uparrow},j)$, 
implying that $i$ can be determined by
repeatedly evaluating the above for 
the remaining binary digits $\{b_{l}\}_{l=1}^{L-1}$
with a proper manipulation (decrement) of $(L,N_{\uparrow},j)$. 
A proposed function of 
finding a spin configuration $i$ for a given set of $(L,N_{\uparrow},j)$
is summarized in Algorithm~\ref{findconfig}.

\begin{algorithm}[H]
  \begin{algorithmic}[1]
    \Require
    \Statex {\bf Input:} integer $L,N_{\uparrow}$, and $j$
    \Statex $0 \leqslant N_{\uparrow} \leqslant L$ 
    \Statex $1\leqslant j \leqslant \binom{L}{N_{\uparrow}}$ 
    \Statex Temporal integer variables $\tilde{L},\tilde{N}_{\uparrow}$, and $\tilde{j}$ 
    \Ensure
    \Statex {\bf Output:} integer $i$ 
    \Statex $2^{N_{\uparrow}}-1 \leqslant i \leqslant 2^{L}-2^{L-N_{\uparrow}}$ 
    \Function{Find-Configuration}{$L,N_{\uparrow},j$}    
    \State $i=0$ \Comment{initialization}
    \State $\tilde{j}=j$  \Comment{initialization}
    \State $\tilde{N}_{\uparrow}=N_{\uparrow}$ \Comment{initialization}
    \For{$\tilde{L} = L, \ L-1, \ \ldots, \ 1$} \Comment{sweep all binary digits of $i$}
    \If{$\tilde{j} = 1$} 
    \State $i=i+2^{\tilde{N}_{\uparrow}}-1$ \Comment{Eq.~(\ref{eq.fill_1})}
    \State{\Return $i$} \Comment{$i$ is determined}
    \ElsIf{$\tilde{j} = \binom{\tilde{L}}{\tilde{N}_{\uparrow}}$} 
    \State $i = i+2^{\tilde{L}}-2^{\tilde{L}-\tilde{N}_{\uparrow}}$ \Comment{Eq.~(\ref{eq.fill_2})}
    \State{\Return $i$} \Comment{$i$ is determined}
    \ElsIf{$\tilde{j} > \binom{\tilde{L}-1}{\tilde{N}_{\uparrow}}$} 
    \State $i=i+2^{\tilde{L}-1}$ \Comment{$\tilde{L}$th binary digit of $i$ is 1} 
    \State $\tilde{j}=\tilde{j}-\binom{\tilde{L}-1}{\tilde{N}_{\uparrow}}$
    \Comment{to satisfy Eq.~(\ref{eq.jcond})}
    \State $\tilde{N}_{\uparrow}=\tilde{N}_{\uparrow}-1$ \Comment{decrement ``\# of $\uparrow$ spins'' by 1}
    \EndIf
    \EndFor
    \EndFunction
  \end{algorithmic}
  \caption{
    A function that returns a spin configuration $i$ for given number $L$ of sites, 
    number $N_{\uparrow}$ of up spins, and state label $j$.
    Comments are given in the right-most side. 
    \label{findconfig}
  }
\end{algorithm}

Several remarks on Algorithm~\ref{findconfig} are in order.
\begin{enumerate}

\item Binomial coefficients 
should be calculated and stored in advance for the better performance. 

\item Regarding the lines 3--5 of Algorithm~\ref{findconfig},  
the temporal variables
$\tilde{j}$, $\tilde{N}_{\uparrow}$, and  $\tilde{L}$ can be considered as
temporal state label, 
temporal number of $\uparrow$ spins, and 
temporal system size, respectively.
In terms of the shortest paths on Pascal's triangle, 
the decrementing loop of $\tilde{L}$ 
means that the shortest path is determined 
by climbing up Pascal's triangle from its $L$th row,  
and $\tilde{N}_{\uparrow}$ is the remaining number of rightward paths. 
$\tilde{N}_{\uparrow}$ and $\tilde{j}$ also have to be decremented 
properly in the loop (lines 14--15 of Algorithm~\ref{findconfig}), 
as it will be described in remark 5 below. 

\item Regarding the lines 6--8 of Algorithm~\ref{findconfig}, 
  the condition $\tilde{j}=1$ indicates that,  
  among the remaining $\tilde{L}$ binary digits of $i$, 
  the lowest $\tilde{N}_{\uparrow}$ digits should be filled with 1's,
  i.e.,
  \begin{equation}
    i=
    (
    b_{L}b_{L-1}\ldots b_{\tilde{L}+1} \
    \overbrace{
      \underbrace{00\ldots0}_{\tilde{L}-\tilde{N}_{\uparrow}} \
      \underbrace{11\ldots1}_{\tilde{N}_{\uparrow}} 
    }^{\tilde{L}} 
    )_2.
    \label{eq.fill_1}
  \end{equation}  
  In terms of the shortest paths on Pascal's triangle, 
  this implies that 
  the rest of the path goes first in the 
  upper left direction $\tilde{L}-\tilde{N}_{\uparrow}$ times and then 
  in the upper right direction $\tilde{N}_{\uparrow}$ times.

\item Regarding the lines 9--11 of Algorithm~\ref{findconfig}, 
  the condition $\tilde{j}=\binom{\tilde{L}}{\tilde{N}_\uparrow}$
  indicates that,  
  among the remaining $\tilde{L}$ binary digits of $i$, 
  the highest $\tilde{N}_{\uparrow}$ digits should be filled with 1's, i.e., 
  \begin{equation}
    i=
    (
    b_{L}b_{L-1}\ldots b_{\tilde{L}+1} \
    \overbrace{
    \underbrace{11\ldots1}_{\tilde{N}_{\uparrow}} \
    \underbrace{00\ldots0}_{\tilde{L}-\tilde{N}_{\uparrow}}
    }^{\tilde{L}} 
    )_2.
    \label{eq.fill_2}
  \end{equation}  
  In terms of the shortest paths on Pascal's triangle, 
  this implies that 
  the rest of the path goes first in 
  the upper right direction $\tilde{N}_{\uparrow}$ times and then 
  in the upper left direction $\tilde{L}-\tilde{N}_{\uparrow}$ times. 

\item Regarding the lines 12--15 of Algorithm~\ref{findconfig}, 
the condition $\tilde{j}>\binom{\tilde{L}-1}{\tilde{N}_\uparrow}$
indicates that the $\tilde{L}$th binary digit of $i$ is 1, 
as discussed around Eqs.~(\ref{eq.recursion}) and (\ref{eq.bL}). 
In terms of the shortest paths on Pascal's triangle,
this implies that the rightward path is chosen
to go from the $\tilde{L}$th row to the ($\tilde{L}-1$)th row.  
As in line 15, $\tilde{N}_{\uparrow}$ is decreased by 1 
because the remaining rightward paths have to be decreased by 1. 
As in line 14, $\tilde{j}$ has to be decreased in order to satisfy 
\begin{equation}
  1 \leqslant \tilde{j} \leqslant \binom{\tilde{L}}{\tilde{N}_{\uparrow}}
  \label{eq.jcond}
\end{equation}
for the next loop. This allows us to make use of 
the relation between the combinatorial recursion and 
the binary digits for $(\tilde{L},\tilde{N}_{\uparrow}, \tilde{j})$.  

\item Although it is not implemented in Algorithm~\ref{findconfig},
  at some $\tilde{L}$ 
  one can switch to refer to a ``small'' table $i(\tilde{L},\tilde{N}_{\uparrow},\tilde{j})$
  stored in advance in the memory to determine the remaining $\tilde{L}$ binary digits of $i$,
  instead of fully performing the loop over $\tilde{L}$.
  One can also implement a quick return
  when $\tilde{N}_{\uparrow}=1$ (when the current vertex is on the line next to the right edge) 
  or $\tilde{L}-\tilde{N}_{\uparrow}=1$ (when the current vertex is on the line next to the left edge) 
  is satisfied.
  
\end{enumerate}

Figure~\ref{pascal} shows 
a concrete example of
the algorithm for $L=6$, $N_{\uparrow}=3$, and $j=7$.
The path from the vertex $\binom{6}{3}=20$ 
to the topmost vertex is uniquely determined, 
and accordingly the algorithm returns
the corresponding spin configuration  
$i(L=6,N_{\uparrow}=3,j=7) = (010110)_2 = 22$.

The algorithm is applicable also to other models
such as the Hubbard model
where the total electron configuration can be given
as a tensor product of up-spin and down-spin electron configurations,
and the $t$-$J$ model where the
total electron configuration can be given 
as a tensor product of hole and spin configurations,
if the Hilbert space is constructed for
fixed magnetization and number of electrons.

\section{Linear spin-wave theory}\label{App.B}
Here we study the effect of the cyclic exchange interaction $J_{\rm c}$ 
on the spin-wave dispersion in the 120$^\circ$ N\'{e}el ordered state
within the linear spin-wave theory.
A comparison of the spin-wave dispersion of
the Heisenberg model on the triangular lattice
with the nearest and the next-nearest-neighbor interactions
($J$-$J'$ model) is also be made.

\subsection{Full Hamiltonian}
Before starting the linear spin-wave approximation,
it is convenient to rewrite the full Hamiltonian $\hat{H}$
in terms of the sum of inner products of spin operators.  
The four-spin exchange term can be written as 
\begin{eqnarray}
  \hat{P}_{ijkl} + \hat{P}_{ijkl}^\dag
  &=& \frac{1}{4} + \sum_{i'<j'\in \langle ijkl \rangle}
  \hat{\mb{S}}_{i'} \cdot \hat{\mb{S}}_{j'} \notag \\
  &+&
  4\left(\hat{Q}_{ijkl} + \hat{Q}_{iljk} - \hat{Q}_{ikjl}\right), 
  \label{eq:P4}
\end{eqnarray}
where
\begin{equation}
  \hat{Q}_{ijkl}=
  \left(\hat{\mb{S}}_i \cdot \hat{\mb{S}}_j \right)
  \left(\hat{\mb{S}}_k \cdot \hat{\mb{S}}_l \right). 
\end{equation}
If the sum over all plaquettes $\sum_{\langle ijkl \rangle}$
is performed, the first term (multiplied by $J_{\rm c}$) results in
\begin{equation}
  J_{\rm c} \sum_{\langle ijkl \rangle} \frac{1}{4} = \frac{3J_{\rm c}}{4}L, 
\end{equation}
because there exist $3L$ plaquettes for the $L$-site system under
periodic-boundary conditions (see Fig.~\ref{fig:model}).
Similarly, the second term results in
\begin{equation}
  J_{\rm c} \sum_{\langle ijkl \rangle}
  \sum_{i'<j' \in \langle ijkl \rangle}
  \hat{\mb{S}}_{i'} \cdot \hat{\mb{S}}_{j'}
  =
  5J_{\rm c} \sum_{\langle ij \rangle} \hat{\mb{S}}_i \cdot \hat{\mb{S}}_j +
  J_{\rm c} \sum_{\langle \langle ij \rangle \rangle} \hat{\mb{S}}_i \cdot \hat{\mb{S}}_j, \label{eq.JcSS}
\end{equation}
where $\langle \langle ij \rangle \rangle$ denotes a pair of spins 
on the next-nearest-neighbor sites $i$ and $j$ on the triangular lattice.
The factor 5 in the first term is
because the nearest-neighbor bonds
($(i',j')=\{(i,j),(j,k),(k,l),(l,i),(j,l)\}$) appear five times 
in the sum over the plaquettes for the ring-exchange term.
Similarly, the factor 1 in the second term is
because the next-nearest-neighbor bond 
($(i',j')=\{(i,k)\}$) appears once for each plaquette 
and is distinct for different plaquettes. 
Now the full Hamiltonian is written as
\begin{eqnarray}
  \hat{H} &=& (J+5J_{\rm c})
  \sum_{\langle ij \rangle} \hat{\mb{S}}_i \cdot \hat{\mb{S}}_j +
  J_{\rm c} \sum_{\langle \langle ij \rangle \rangle} \hat{\mb{S}}_i \cdot \hat{\mb{S}}_j \notag \\
  &+& 4J_{\rm c} \sum_{\langle ijkl \rangle} \left(
  \hat{Q}_{ijkl} + \hat{Q}_{iljk} - \hat{Q}_{ikjl}
  \right)
  +\frac{3J_{\rm c}L}{4}.
  \label{eq.Hfull}
\end{eqnarray}

\subsection{Rotating frame}
\begin{figure}
  \begin{center}
    \includegraphics[width=0.95\columnwidth]{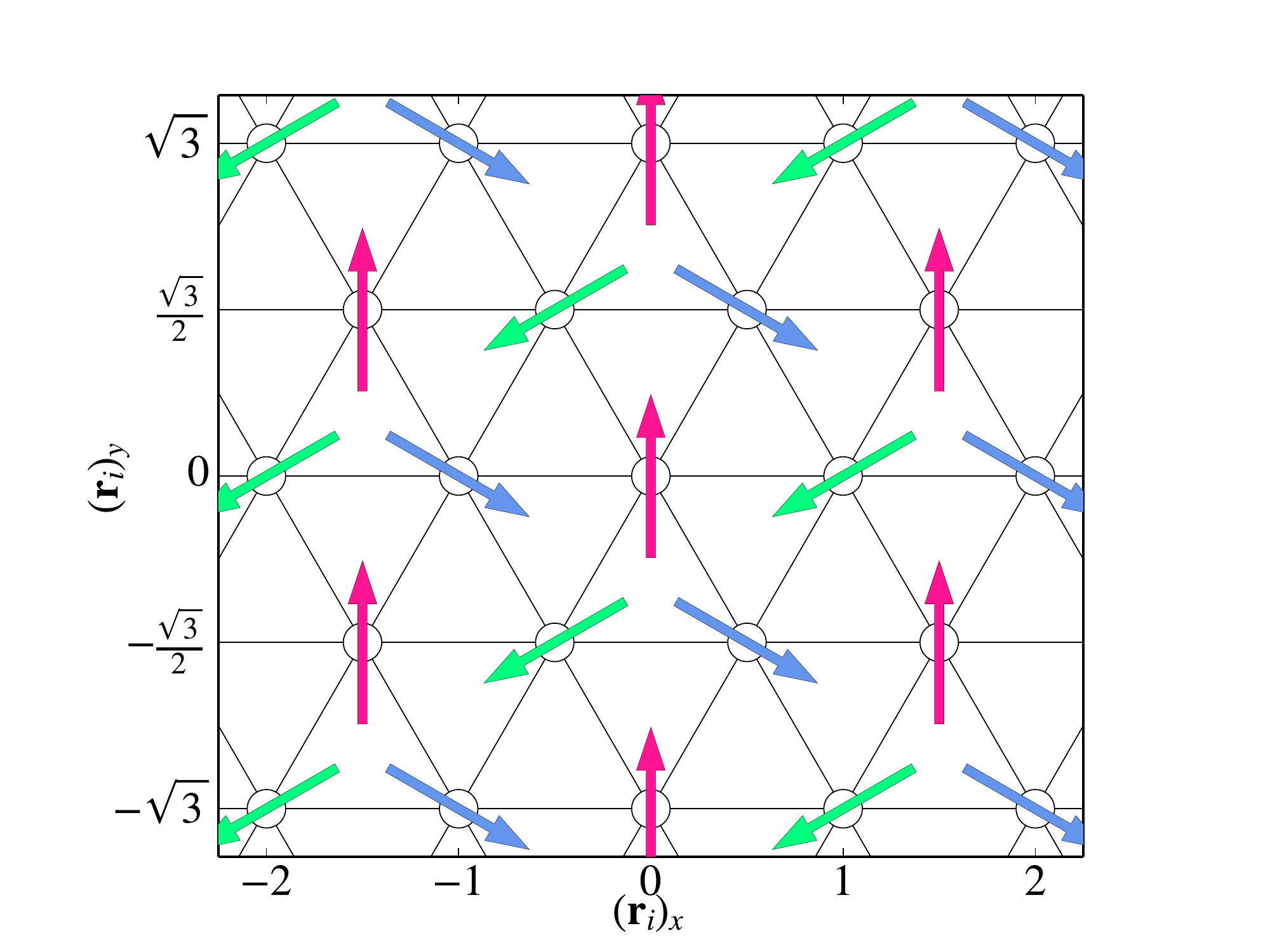} \\
    \caption{
      Schematic figure of the 120$^{\circ}$ N\'{e}el ordered state on the
      triangular lattice.
      $(\mb{r}_i)_{x(y)}$ denotes the $x$ $(y)$ 
      coordinate of $\mb{r}_i$.
      \label{fig.120}}
  \end{center}
\end{figure}

The $120^\circ$ N\'{e}el ordered state has a three-sublattice structure,
as shown in Fig.~\ref{fig.120}.
However, the introduction of a rotating 
frame~\cite{Oguchi1985,Miyake1985,Leung1993,Ohyama1993,Deutscher1993,Trumper2000,Chernyshev2009,Zhitomirsky2013}
allows us to develop a one-sublattice spin-wave theory for
the $120^\circ$ N\'{e}el ordered state.

In terms of the spin operators in the rotating frame ($X$-$Y$-$Z$), 
the spin operators in the original frame ($x$-$y$-$z$) can be written as
\begin{eqnarray}
  \begin{array}{lll}  
    \hat{S}_i^x &=& \cos \theta_i \hat{S}_i^X + \sin \theta_i \hat{S}_{i}^Z,  \\
    \hat{S}_i^y &=& \hat{S}_i^Y,  \\
    \hat{S}_i^z &=& \cos \theta_i \hat{S}_i^Z - \sin \theta_i \hat{S}_{i}^X, 
  \end{array}
  \label{eq.rotXYZ}
\end{eqnarray}
where $\theta_{i}=\mathbf{Q}\cdot\mathbf{r}_i$ with 
$\mathbf{Q}=(4\pi/3,0)$ being a wave vector corresponding
to the 120$^\circ$ order 
and $\mathbf{r}_i$ the position of site $i$.  
The inner product of spin operators is thus given by 
\begin{eqnarray}
  \hat{\mb{S}}_i \cdot \hat{\mb{S}}_j
  &=&
  \hat{S}_i^Y \hat{S}_j^Y 
  +\left(\hat{S}_i^Z \hat{S}_j^Z + \hat{S}_i^X \hat{S}_j^X\right) \cos\theta_{ij} \notag \\
  &+&\left(\hat{S}_i^Z \hat{S}_j^X - \hat{S}_i^X \hat{S}_j^Z\right) \sin\theta_{ij},
  \label{eq.SSrotate}
\end{eqnarray}
where $\theta_{ij}=\theta_i-\theta_j$.  

We assume that the spins are pointing along the $Z$ axis
of the rotating ($X$-$Y$-$Z$) frame.  
The Holstein-Primakoff transformation~\cite{Holstein1940} 
for the spin operators in the rotating frame results in 
\begin{eqnarray}
  \begin{array}{lll}  
    \hat{S}^Z_{i} &=& S - \hat{a}_{i}^\dag \hat{a}_{i}, \label{eq.HP1} \\
    \hat{S}^-_{i} &=& \sqrt{2S} \hat{a}_{i}^\dag \left(1-\frac{\hat{a}_{i}^\dag \hat{a}_{i}}{2S} \right)^{\frac{1}{2}}, \label{eq.HP2} \\
    \hat{S}^+_{i} &=& \sqrt{2S}       \left(1-\frac{\hat{a}_{i}^\dag \hat{a}_{i}}{2S} \right)^{\frac{1}{2}} \hat{a}_{i}, \label{eq.HP3}
  \end{array}
\end{eqnarray}
where
$\hat{S}^{-}_i=\hat{S}_i^X - \imag \hat{S}_{i}^Y$,
$\hat{S}^{+}_{i}=(\hat{S}^{-}_{i})^\dag$, and 
$\hat{a}_{i}$ and $\hat{a}_{i}^\dag$ 
are bosonic annihilation and creation operators, respectively, satisfying
the canonical commutation relations 
$[\hat{a}_{i},\hat{a}_{j}]=0$ and 
$[\hat{a}_{i},\hat{a}_{j}^\dag]=\delta_{ij}$.

\subsection{Linear spin-wave approximation}
Up to the quadratic terms of the bosonic operators,
the inner product of the spin operators is approximated as
\begin{eqnarray}
\hat{\mb{S}}_i \cdot \hat{\mb{S}}_j 
&\approx&
S^2+
S\cos{\theta_{ij}}\left(\hat{a}_{i}^\dag \hat{a}_{i}+\hat{a}_{j}^\dag \hat{a}_{j}\right)\notag \\
&+&
\frac{S}{2}(\cos{\theta_{ij}}+1)\left(\hat{a}_{i}^\dag \hat{a}_{j}+\hat{a}_{j}^\dag \hat{a}_{i}\right)\notag \\
&+&
\frac{S}{2}(\cos{\theta_{ij}}-1)\left(\hat{a}_{i}^\dag \hat{a}_{j}^\dag+\hat{a}_{j} \hat{a}_{i}\right).
\label{eq.SS}
\end{eqnarray}
Notice that 
$\cos \theta_{ij}=-1/2$ for the nearest neighbors and 
$\cos \theta_{ij}=1$ for next-nearest neighbors. 

Similarly, $\hat{Q}_{ijkl}$ is approximated as 
\begin{eqnarray}
  \hat{Q}_{ijkl}
  &\approx&\hat{S}_i^Z \hat{S}_j^Z \hat{S}_k^Z \hat{S}_l^Z \cos{\theta_{ij}} \cos{\theta_{kl}} \notag \\
  &+&\hat{S}_i^Z \hat{S}_j^Z \hat{S}_k^X \hat{S}_l^X \cos{\theta_{ij}} \cos{\theta_{kl}} \notag \\
  &+&\hat{S}_k^Z \hat{S}_l^Z \hat{S}_i^X \hat{S}_j^X \cos{\theta_{ij}} \cos{\theta_{kl}} \notag \\
  &+&\hat{S}_i^Z \hat{S}_j^Z \hat{S}_k^Y \hat{S}_l^Y \cos{\theta_{ij}}  \notag \\
  &+&\hat{S}_k^Z \hat{S}_l^Z \hat{S}_i^Y \hat{S}_j^Y \cos{\theta_{kl}}  \notag \\
  &+&
  \left(\hat{S}_i^Z \hat{S}_j^X - \hat{S}_i^X \hat{S}_j^Z \right)
  \left(\hat{S}_k^Z \hat{S}_l^X - \hat{S}_k^X \hat{S}_l^Z \right)
  \sin{\theta_{ij}} \sin{\theta_{kl}} \notag \\
  &\approx&
  \left[
  S^4 - S^3 \left(\hat{a}_i^\dag \hat{a}_{i}
  +\hat{a}_{j}^\dag \hat{a}_{j}
  +\hat{a}_{k}^\dag \hat{a}_{k}
  +\hat{a}_{l}^\dag \hat{a}_{l}
  \right)\right] \cos{\theta_{ij}} \cos{\theta_{kl}} \notag \\
  &+&\frac{S^3}{2} \left(
  \hat{a}_{i}^\dag \hat{a}_{j}
  +{\rm H.c.}
  \right) \cos{\theta_{kl}} \left(\cos{\theta_{ij}}+1\right) \notag \\
  &+&\frac{S^3}{2} \left(
  \hat{a}_{k}^\dag \hat{a}_{l}
  +{\rm H.c.}
  \right) \cos{\theta_{ij}} \left(\cos{\theta_{kl}}+1\right) \notag \\
  &+&\frac{S^3}{2} \left(
  \hat{a}_{i}^\dag \hat{a}_{j}^\dag
  +{\rm H.c.}
  \right) \cos{\theta_{kl}} \left(\cos{\theta_{ij}}-1\right) \notag \\
  &+&\frac{S^3}{2} \left(
  \hat{a}_{k}^\dag \hat{a}_{l}^\dag
  +{\rm H.c.}
  \right) \cos{\theta_{ij}} \left(\cos{\theta_{kl}}-1\right) \notag \\
  &+&\frac{S^3}{2} \sin{\theta_{ij}} \sin{\theta_{kl}} \notag \\
&\times&  \left( 
    \hat{a}_{j}^\dag \hat{a}_{l}
   -\hat{a}_{j}^\dag \hat{a}_{k}
   -\hat{a}_{i}^\dag \hat{a}_{l}
   +\hat{a}_{i}^\dag \hat{a}_{k}
   +
   \right. \notag \\ 
&&   \left.
   \hat{a}_{j}^\dag \hat{a}_{l}^\dag
   -\hat{a}_{j}^\dag \hat{a}_{k}^\dag
   -\hat{a}_{i}^\dag \hat{a}_{l}^\dag
   +\hat{a}_{i}^\dag \hat{a}_{k}^\dag
   + {\rm H.c.}
   \right).
\end{eqnarray}
By substituting 
$ \cos\theta_{ij}
= \cos\theta_{kl}
= \cos\theta_{il}
= \cos\theta_{jk}
= \cos\theta_{jl}
=-1/2$,   
$\cos\theta_{ik}=1$, 
$\sin\theta_{ij} \sin\theta_{kl} = -3/4$, 
$\sin\theta_{il} \sin\theta_{jk} = 3/4$, and 
$\sin\theta_{ik} \sin\theta_{jl} = 0$
for
$\hat{Q}_{ijkl}$,
$\hat{Q}_{iljk}$, and
$\hat{Q}_{ikjl}$ 
in the last term of Eq.~(\ref{eq:P4}), 
we find 
\begin{eqnarray}
  &&
  \hat{Q}_{ijkl}+\hat{Q}_{iljk}-\hat{Q}_{ikjl} 
  \notag \\
  &\approx&
  S^4 - S^3 \left(\hat{a}_i^\dag \hat{a}_{i}
  +\hat{a}_{j}^\dag \hat{a}_{j}
  +\hat{a}_{k}^\dag \hat{a}_{k}
  +\hat{a}_{l}^\dag \hat{a}_{l}
  \right)  \notag \\
  &+&\frac{S^3}{4} \left(
  \hat{a}_{i}^\dag \hat{a}_{j}
  +\hat{a}_{k}^\dag \hat{a}_{l}
  +\hat{a}_{i}^\dag \hat{a}_{l}
  +\hat{a}_{j}^\dag \hat{a}_{k}
  -\hat{a}_{i}^\dag \hat{a}_{k}
  -4\hat{a}_{j}^\dag \hat{a}_{l}
  +{\rm H.c.}
  \right) \notag \\
  &+&\frac{3S^3}{4} \left(
    \hat{a}_{i}^\dag \hat{a}_{j}^\dag
  + \hat{a}_{k}^\dag \hat{a}_{l}^\dag
  + \hat{a}_{i}^\dag \hat{a}_{l}^\dag
  + \hat{a}_{j}^\dag \hat{a}_{k}^\dag
  - \hat{a}_{i}^\dag \hat{a}_{k}^\dag
  +{\rm H.c.}
  \right).
  \label{eq.SSSS}
\end{eqnarray}
Notice in Eq.~(\ref{eq.SSSS}) that 
the subscript pair $(i,k)$ contributes
to the next-nearest-neighbor terms, 
while the others to the nearest-neighbor terms.

\subsection{Spin-wave Hamiltonian}
By substituting the approximations in Eqs.~(\ref{eq.SS})
and (\ref{eq.SSSS}) into the Hamiltonian in Eq.~(\ref{eq.Hfull}), 
and carefully evaluating the sum over all plaquettes,
similarly in Eq.~(\ref{eq.JcSS}), 
we obtain the spin-wave Hamiltonian 
\begin{eqnarray}
  &&\hat{H}\approx\hat{H}_{\rm sw} 
  =
  E_{\rm sw} + 3SA_0 \sum_{i}\hat{a}_i^\dag \hat{a}_i \notag \\
  &+&\frac{S}{4} \left[
    A_1 \sum_{\langle ij \rangle}               \left(\hat{a}_i^\dag \hat{a}_j + \hat{a}_j^\dag \hat{a}_j \right)
    + A_2 \sum_{\langle\langle ij \rangle\rangle} \left(\hat{a}_i^\dag \hat{a}_j + \hat{a}_j^\dag \hat{a}_j \right) \right]\notag \\
  &-&\frac{3S}{4} \left[
    B_1 \sum_{\langle ij \rangle}               \left(\hat{a}_i^\dag \hat{a}_j^\dag + \hat{a}_i \hat{a}_j \right)
    +B_2 \sum_{\langle\langle ij \rangle\rangle} \left(\hat{a}_i^\dag \hat{a}_j^\dag + \hat{a}_i \hat{a}_j \right)\right],
  \label{eq.Hsw_real}
\end{eqnarray}
where
\begin{eqnarray}
  \begin{array}{lll}  
    A_0 &=& J+3J_{\rm c} - 16S^2J_{\rm c}, \\
    A_1 &=& J+5J_{\rm c}, \\
    A_2 &=& 4(1-S^2)J_{\rm c}, \\
    B_1 &=& J+5J_{\rm c} - 16S^2J_{\rm c}, \\
    B_2 &=& 4S^2J_{\rm c}, 
    \end{array}
\end{eqnarray}
and 
\begin{eqnarray}
  E_{\rm sw} 
  = -\frac{3}{2}\left[\left(J+3J_{\rm c}-8S^2J_{\rm c}\right) S^2 - \frac{J_{\rm c}}{2}\right]L.
\end{eqnarray}

With the Fourier transformation of the bosonic operators 
$
\hat{a}_i = \frac{1}{\sqrt{L}} \sum_{\mb{q}} \hat{a}_{\mb{q}}\e^{\imag \mb{q}\cdot \mb{r}_i}
$,
$\hat{H}_{\rm sw}$ in the momentum space is given by
\begin{eqnarray}
  \hat{H}_{\rm sw} &=& E_{\rm sw}
  + \sum_{\mb{q}} \left[
    A(\mb{q}) \hat{a}_{\mb{q}}^\dag \hat{a}_{\mb{q}}
    - \frac{1}{2} B(\mb{q})
    \left(
    \hat{a}_{\mb{q}}^\dag \hat{a}_{-\mb{q}}^\dag +
    \hat{a}_{-\mb{q}} \hat{a}_{\mb{q}}
    \right)
    \right]\notag \\
  &=&
  E_{\rm sw} - \frac{1}{2} \sum_{\mb q} A(\mb{q})
  \notag \\
  &+&
  \frac{1}{2}
  \sum_{\mb{q}}
  \left(
  \begin{array}{cc}
    \hat{a}_{\mb{q}}^\dag & \hat{a}_{-\mb{q}}
  \end{array}
  \right)
  \left(
  \begin{array}{cc}
     A(\mb{q}) & -B(\mb{q}) \\
    -B(\mb{q}) &  A(\mb{q}) 
  \end{array}
  \right)
  \left(
  \begin{array}{c}
    \hat{a}_{\mb{q}} \\
    \hat{a}^\dag_{-\mb{q}}  
  \end{array}
  \right),
  \label{eq.Hsw_momentum}
\end{eqnarray}
  where 
\begin{eqnarray}
  A(\mb{q}) &=&       3S     \left[ A_0 + \frac{A_1}{2} \gamma(\mb{q}) + \frac{A_2}{2} \gamma'(\mb{q}) \right],\label{eq.Aq} \\
  B(\mb{q}) &=& \frac{9S}{2} \left[             B_1      \gamma(\mb{q})+  B_2          \gamma'(\mb{q}) \right],\label{eq.Bq} 
\end{eqnarray}
$\gamma(\mb{q})= \frac{1}{6}\sum_{i=1}^{6} \e^{\imag \mb{q} \cdot \bs{\delta}_i}$, 
and
$\gamma'(\mb{q}) = \frac{1}{6}\sum_{i=1}^{6} \e^{\imag \mb{q} \cdot \bs{\delta}'_i}$ 
with
$\bs{\delta}_i$ ($\bs{\delta}'_i$) being the vectors connecting
the nearest (next-nearest) neighbors.

\subsection{Spin-wave dispersion}\label{sec.swdispersion}
We now introduce a Bogoliubov transformation
\begin{equation}
\left(
\begin{array}{c}
  \hat{a}_{\mb{q}} \\
  \hat{a}^\dag_{-\mb{q}}
\end{array}
\right)
=
\left(
  \begin{array}{cc}
     u_{\mb{q}} & v_{\mb{q}} \\
     v_{\mb{q}} & u_{\mb{q}} 
  \end{array}
  \right)
  \left(
  \begin{array}{c}
    \hat{b}_{\mb{q}} \\
    \hat{b}^\dag_{-\mb{q}}  
  \end{array}
  \right)  
\end{equation}
under the condition $u_{\mb{q}}^2-v_{\mb{q}}^2=1$ and thus 
the new operators $\hat{b}_{\mb{q}}$ and $\hat{b}_{\mb{q}}^\dag$ 
obey the canonical bosonic commutation relations. 
If $u_{\mb{q}}$ and $v_{\mb{q}}$ are chosen to satisfy
$u_{\mb{q}}^2+v_{\mb{q}}^2=A(\mb{q})/\Omega(\mb{q})$ and
$2u_{\mb{q}}v_{\mb{q}}=B(\mb{q})/\Omega(\mb{q})$ with 
\begin{equation}
  \Omega(\mb{q})=\sqrt{A(\mb{q})^2-B(\mb{q})^2}, 
\end{equation}
then the spin-wave Hamiltonian is given by  
\begin{eqnarray}
  \hat{H}_{\rm sw} &=& E_{\rm sw}
  -\frac{1}{2} \sum_{\mb{q}} A(\mb{q}) 
  + \sum_{\mb{q}} \Omega(\mb{q}) \left(\hat{b}_{\mb{q}}^\dag \hat{b}_{\mb{q}} + \frac{1}{2} \right), 
\end{eqnarray}
where $\Omega(\mb{q})$ is the spin-wave dispersion.

\begin{figure}
  \begin{center}
    \includegraphics[width=0.95\columnwidth]{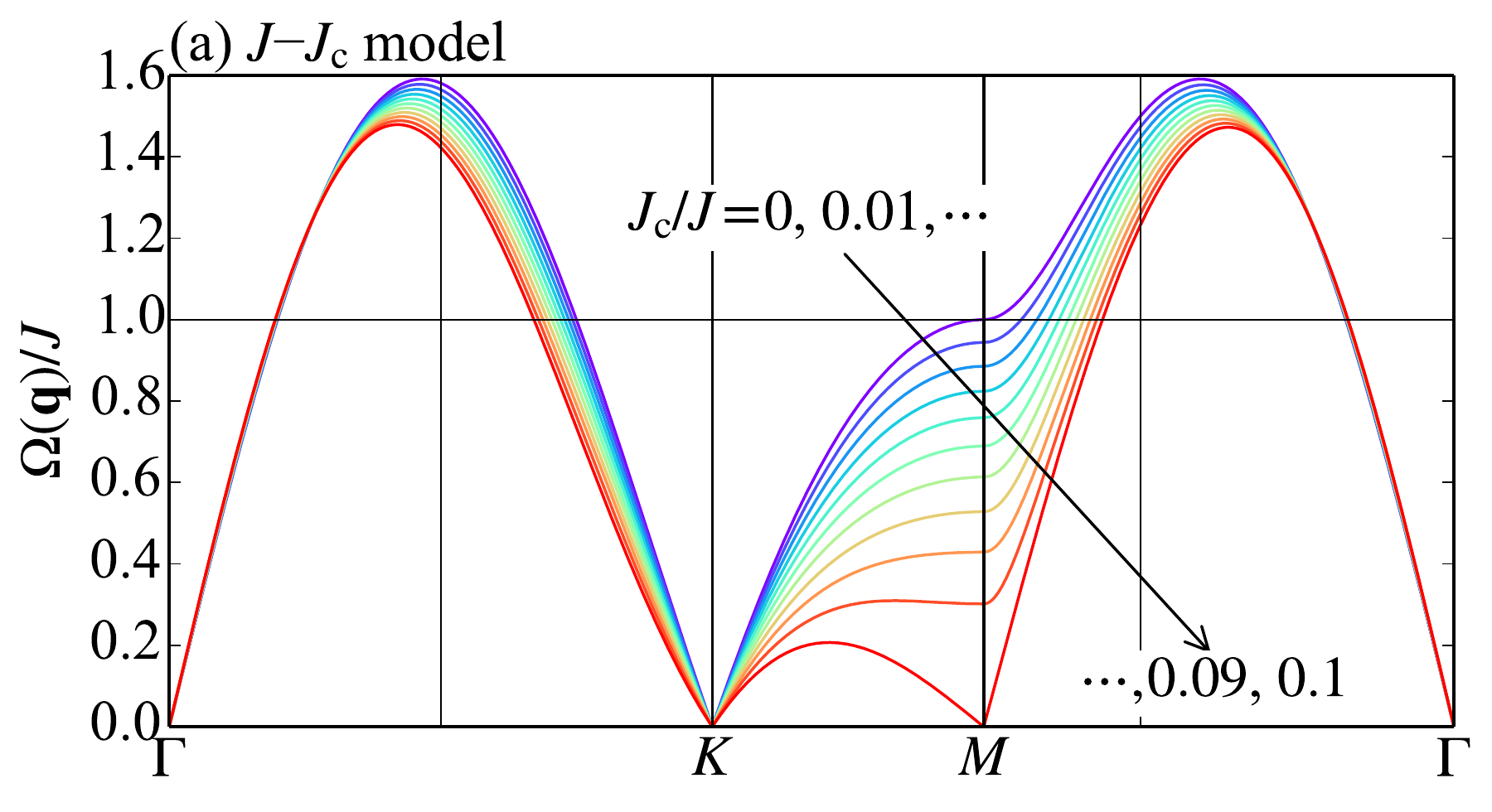} \\
    \includegraphics[width=0.95\columnwidth]{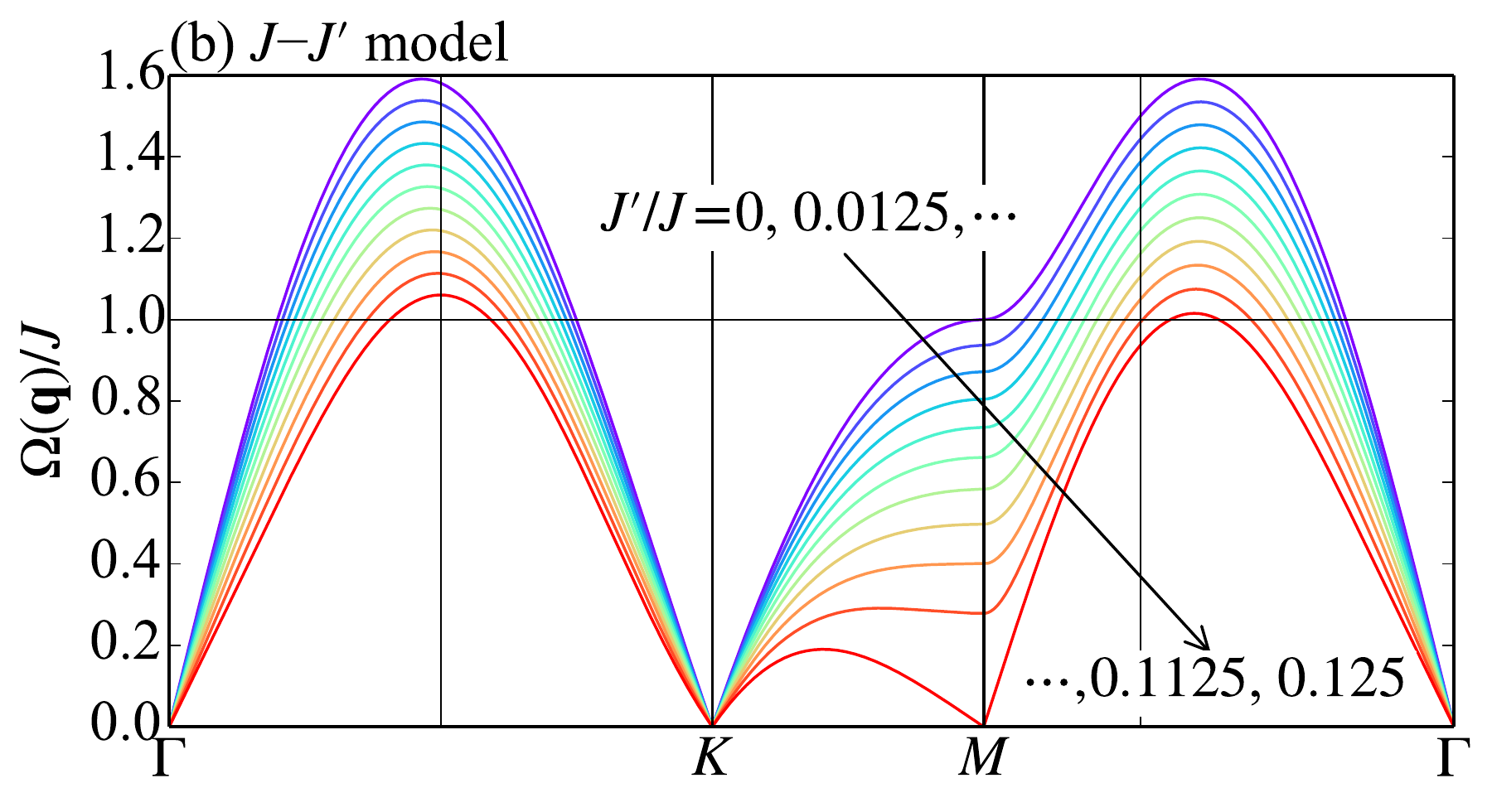}
    \caption{
      Linear spin-wave dispersions for
      (a) the $J$-$J_{\rm c}$ model and (b) the $J$-$J'$ model 
      with several $J_{\rm c}/J$ and $J'/J$ values indicated respectively 
      in the figures.
      The horizontal axis is momentum $\mb{q}$ 
      along the $\Gamma$--$K$--$M$--$\Gamma$ points in the (nonmagnetic) Brillouin zone, 
      where $\Gamma=(0,0)$, $K=(4\pi/3,0)$, and $M=(\pi,\pi/\sqrt{3})$. 
      Thin vertical lines indicate the magnetic Brillouin-zone boundaries corresponding      
      to the $120^\circ$ N\'{e}el order.
      The horizontal line at $\Omega(\mb{q})/J=1$ indicates 
      the spin-wave excitation energy at $M$ point in the purely triangular case with $J_{\rm c}=J'=0$. 
      \label{lsw}}
  \end{center}
\end{figure}

Figure~\ref{lsw}(a) 
shows the $S=1/2$ spin-wave dispersion $\Omega(\mb{q})$
for several values of $J_{\rm c}/J$
along the high symmetric momentum direction $\Gamma$--$K$--$M$--$\Gamma$,
where $\Gamma=(0,0)$, $K=(4\pi/3,0)$,
and $M=(\pi,\pi/\sqrt{3})$ (also see Fig.~\ref{fig.kpoints}). 
The zero modes at the $\Gamma$, $K$, and $K'$ points 
are preserved because 
$A(\Gamma)=B(\Gamma)=\frac{9S}{2}\left(J+5J_{\rm c}-12S^2J_{\rm c}\right)$ and
$A(\pm K)=-B(\pm K)=\frac{9S}{4}\left(J+5J_{\rm c}-24S^2J_{\rm c}\right)$.
The excitation energy at the $M$ point is given by
\begin{eqnarray}
  \Omega(M)
  &=& 2S\sqrt{
    \left[J-(3+28S^2)J_{\rm c}\right]
    \left[J+(3-16S^2)J_{\rm c}\right]
  }\notag \\
  &=& \sqrt{
    \left(J-10J_{\rm c}\right)
    \left(J-J_{\rm c}\right)
  }, 
\end{eqnarray}
where the second line is for $S=1/2$. 
It is found that the spin-wave excitation energy along
the $K$--$M$ line, especially at the $M$ point,  
reduces drastically with increasing $J_{\rm c}/J$,
and eventually becomes zero when $J_{\rm c}/J=0.1$, 
implying instability of the 120$^{\circ}$ N\'{e}el order.
On the other hand,
the spin-wave velocity around the $\Gamma$ point remains the same and  
the highest spin-wave excitation energy is kept 
around $1.5J$ as $J_{\rm c}/J$ is increased.

For a comparison, Fig.~\ref{lsw}(b) shows
the $S=1/2$ linear spin-wave dispersion $\Omega(\mb{q})$
for the $J$-$J'$ model defined as
\begin{equation}
\hat{H}_{JJ'} =
J \sum_{\langle ij\rangle} \hat{\mathbf{S}}_i \cdot \hat{\mathbf{S}}_j
+J' \sum_{\langle\langle ij\rangle\rangle} \hat{\mathbf{S}}_i \cdot \hat{\mathbf{S}}_j
\label{eq.J1J2}
\end{equation}
with $J'$ being the next-nearest-neighbor exchange interaction. 
The linear spin-wave dispersion for this model can be obtained
by replacing
$A_0$, $A_1$, $A_2$, $B_1$, and $B_2$ in Eqs.~(\ref{eq.Aq}) and (\ref{eq.Bq}) with
$\tilde{A}_0$, $\tilde{A}_1$, $\tilde{A}_2$, $\tilde{B}_1$, and $\tilde{B}_2$, where
\begin{eqnarray}
  \begin{array}{lll}  
    \tilde{A}_0 &=& J-2J',\\
    \tilde{A}_1 &=& J, \\
    \tilde{A}_2 &=& 4J', \\
    \tilde{B}_1 &=& J,\\
    \tilde{B}_2 &=& 0.
  \end{array}
\end{eqnarray}
Again the zero modes at the $\Gamma$, $K$, and $K'$ points are preserved with increasing $J'/J$. 
The excitation energy at the $M$ point is given by 
\begin{eqnarray}
  \Omega(M)
  &=& 2S\sqrt{
    \left(J-8J'\right)
    \left(J-2J'\right)
  }\notag \\
  &=& \sqrt{
    \left(J-8J'\right)
    \left(J-2J'\right)
  }. 
\end{eqnarray}
Similarly to the $J$-$J_{\rm c}$ model, 
the spin-wave excitation energy 
at the $M$ point reduces most significantly with increasing $J'/J$,  
and eventually becomes zero when $J'/J=1/8$, 
However, differently from the $J$-$J_{\rm c}$ model,
the spin-wave velocity around the $\Gamma$ point reduces and 
the highest spin-wave excitation energy is also
reduced from $\sim 1.6J$ to $\sim 1.05J$  
as $J'/J$ is increased. 
A similar dependence of the excitation energy
on the interaction parameter $J'$ has been found
also in the square lattice with the linear spin-wave theory~\cite{Rutonjski2016}.

The spin-wave excitation has two characteristic energy scales. 
One is the maxima of $\Omega(\mb{q})$ 
and the other is the saddle points, minima,
and nearly flat dispersion of $\Omega(\mb{q})$
at and around the $M$ and equivalent points. 
The comparison of the spin-wave dispersions 
suggests that, although both $J_{\rm c}$ and $J'$ can 
increase the separation of the two energy scales,
the more significant separation may appear in 
the $J$-$J_{\rm c}$ model rather than in the $J$-$J'$ model. 
Note however that analytical and numerical studies 
beyond the linear spin-wave theory~\cite{ZhengSeriese2006,Ghioldi2015,Ferrari2019} 
have shown a strong renormalization of the magnon excitation energy  
as compared to the spin-wave theory for the pure triangular-lattice case with $J_{\rm c}=J'=0$.

Finally, we note that the spin-wave analysis 
captures the magnon excitations but not nonmagnetic ones. 
Indeed, we were not able to find
the double-peak structure of the specific heat within the
spin-wave analysis. 
This implies that the nonmagnetic excitations beyond the simple magnon excitations 
might be essential to understand the characteristic double-peak structure of the specific heat  
found here in the finite-temperature Lanczos calculations.

\bibliography{biball}

\begin{thebibliography}{124}%
\makeatletter
\providecommand \@ifxundefined [1]{%
 \@ifx{#1\undefined}
}%
\providecommand \@ifnum [1]{%
 \ifnum #1\expandafter \@firstoftwo
 \else \expandafter \@secondoftwo
 \fi
}%
\providecommand \@ifx [1]{%
 \ifx #1\expandafter \@firstoftwo
 \else \expandafter \@secondoftwo
 \fi
}%
\providecommand \natexlab [1]{#1}%
\providecommand \enquote  [1]{``#1''}%
\providecommand \bibnamefont  [1]{#1}%
\providecommand \bibfnamefont [1]{#1}%
\providecommand \citenamefont [1]{#1}%
\providecommand \href@noop [0]{\@secondoftwo}%
\providecommand \href [0]{\begingroup \@sanitize@url \@href}%
\providecommand \@href[1]{\@@startlink{#1}\@@href}%
\providecommand \@@href[1]{\endgroup#1\@@endlink}%
\providecommand \@sanitize@url [0]{\catcode `\\12\catcode `\$12\catcode
  `\&12\catcode `\#12\catcode `\^12\catcode `\_12\catcode `\%12\relax}%
\providecommand \@@startlink[1]{}%
\providecommand \@@endlink[0]{}%
\providecommand \url  [0]{\begingroup\@sanitize@url \@url }%
\providecommand \@url [1]{\endgroup\@href {#1}{\urlprefix }}%
\providecommand \urlprefix  [0]{URL }%
\providecommand \Eprint [0]{\href }%
\providecommand \doibase [0]{http://dx.doi.org/}%
\providecommand \selectlanguage [0]{\@gobble}%
\providecommand \bibinfo  [0]{\@secondoftwo}%
\providecommand \bibfield  [0]{\@secondoftwo}%
\providecommand \translation [1]{[#1]}%
\providecommand \BibitemOpen [0]{}%
\providecommand \bibitemStop [0]{}%
\providecommand \bibitemNoStop [0]{.\EOS\space}%
\providecommand \EOS [0]{\spacefactor3000\relax}%
\providecommand \BibitemShut  [1]{\csname bibitem#1\endcsname}%
\let\auto@bib@innerbib\@empty
\bibitem [{\citenamefont {Anderson}(1973)}]{Anderson1973}%
  \BibitemOpen
  \bibfield  {author} {\bibinfo {author} {\bibfnamefont {P.}~\bibnamefont
  {Anderson}},\ }\href {\doibase https://doi.org/10.1016/0025-5408(73)90167-0}
  {\bibfield  {journal} {\bibinfo  {journal} {Materials Research Bulletin}\
  }\textbf {\bibinfo {volume} {8}},\ \bibinfo {pages} {153 } (\bibinfo {year}
  {1973})}\BibitemShut {NoStop}%
\bibitem [{\citenamefont {Fazekas}\ and\ \citenamefont
  {Anderson}(1974)}]{Fazekas1974}%
  \BibitemOpen
  \bibfield  {author} {\bibinfo {author} {\bibfnamefont {P.}~\bibnamefont
  {Fazekas}}\ and\ \bibinfo {author} {\bibfnamefont {P.~W.}\ \bibnamefont
  {Anderson}},\ }\href {\doibase 10.1080/14786439808206568} {\bibfield
  {journal} {\bibinfo  {journal} {The Philosophical Magazine: A Journal of
  Theoretical Experimental and Applied Physics}\ }\textbf {\bibinfo {volume}
  {30}},\ \bibinfo {pages} {423} (\bibinfo {year} {1974})}\BibitemShut
  {NoStop}%
\bibitem [{\citenamefont {Bernu}\ \emph {et~al.}(1992)\citenamefont {Bernu},
  \citenamefont {Lhuillier},\ and\ \citenamefont {Pierre}}]{Bernu1992}%
  \BibitemOpen
  \bibfield  {author} {\bibinfo {author} {\bibfnamefont {B.}~\bibnamefont
  {Bernu}}, \bibinfo {author} {\bibfnamefont {C.}~\bibnamefont {Lhuillier}}, \
  and\ \bibinfo {author} {\bibfnamefont {L.}~\bibnamefont {Pierre}},\ }\href
  {\doibase 10.1103/PhysRevLett.69.2590} {\bibfield  {journal} {\bibinfo
  {journal} {Phys. Rev. Lett.}\ }\textbf {\bibinfo {volume} {69}},\ \bibinfo
  {pages} {2590} (\bibinfo {year} {1992})}\BibitemShut {NoStop}%
\bibitem [{\citenamefont {Bernu}\ \emph {et~al.}(1994)\citenamefont {Bernu},
  \citenamefont {Lecheminant}, \citenamefont {Lhuillier},\ and\ \citenamefont
  {Pierre}}]{Bernu1994}%
  \BibitemOpen
  \bibfield  {author} {\bibinfo {author} {\bibfnamefont {B.}~\bibnamefont
  {Bernu}}, \bibinfo {author} {\bibfnamefont {P.}~\bibnamefont {Lecheminant}},
  \bibinfo {author} {\bibfnamefont {C.}~\bibnamefont {Lhuillier}}, \ and\
  \bibinfo {author} {\bibfnamefont {L.}~\bibnamefont {Pierre}},\ }\href
  {\doibase 10.1103/PhysRevB.50.10048} {\bibfield  {journal} {\bibinfo
  {journal} {Phys. Rev. B}\ }\textbf {\bibinfo {volume} {50}},\ \bibinfo
  {pages} {10048} (\bibinfo {year} {1994})}\BibitemShut {NoStop}%
\bibitem [{\citenamefont {Misguich}\ \emph {et~al.}(1999)\citenamefont
  {Misguich}, \citenamefont {Lhuillier}, \citenamefont {Bernu},\ and\
  \citenamefont {Waldtmann}}]{Misguich1999}%
  \BibitemOpen
  \bibfield  {author} {\bibinfo {author} {\bibfnamefont {G.}~\bibnamefont
  {Misguich}}, \bibinfo {author} {\bibfnamefont {C.}~\bibnamefont {Lhuillier}},
  \bibinfo {author} {\bibfnamefont {B.}~\bibnamefont {Bernu}}, \ and\ \bibinfo
  {author} {\bibfnamefont {C.}~\bibnamefont {Waldtmann}},\ }\href {\doibase
  10.1103/PhysRevB.60.1064} {\bibfield  {journal} {\bibinfo  {journal} {Phys.
  Rev. B}\ }\textbf {\bibinfo {volume} {60}},\ \bibinfo {pages} {1064}
  (\bibinfo {year} {1999})}\BibitemShut {NoStop}%
\bibitem [{\citenamefont {Capriotti}\ \emph {et~al.}(1999)\citenamefont
  {Capriotti}, \citenamefont {Trumper},\ and\ \citenamefont
  {Sorella}}]{Capriotti1999}%
  \BibitemOpen
  \bibfield  {author} {\bibinfo {author} {\bibfnamefont {L.}~\bibnamefont
  {Capriotti}}, \bibinfo {author} {\bibfnamefont {A.~E.}\ \bibnamefont
  {Trumper}}, \ and\ \bibinfo {author} {\bibfnamefont {S.}~\bibnamefont
  {Sorella}},\ }\href {\doibase 10.1103/PhysRevLett.82.3899} {\bibfield
  {journal} {\bibinfo  {journal} {Phys. Rev. Lett.}\ }\textbf {\bibinfo
  {volume} {82}},\ \bibinfo {pages} {3899} (\bibinfo {year}
  {1999})}\BibitemShut {NoStop}%
\bibitem [{\citenamefont {Yunoki}\ and\ \citenamefont
  {Sorella}(2006)}]{Yunoki2006}%
  \BibitemOpen
  \bibfield  {author} {\bibinfo {author} {\bibfnamefont {S.}~\bibnamefont
  {Yunoki}}\ and\ \bibinfo {author} {\bibfnamefont {S.}~\bibnamefont
  {Sorella}},\ }\href {\doibase 10.1103/PhysRevB.74.014408} {\bibfield
  {journal} {\bibinfo  {journal} {Phys. Rev. B}\ }\textbf {\bibinfo {volume}
  {74}},\ \bibinfo {pages} {014408} (\bibinfo {year} {2006})}\BibitemShut
  {NoStop}%
\bibitem [{\citenamefont {White}\ and\ \citenamefont
  {Chernyshev}(2007)}]{White2007}%
  \BibitemOpen
  \bibfield  {author} {\bibinfo {author} {\bibfnamefont {S.~R.}\ \bibnamefont
  {White}}\ and\ \bibinfo {author} {\bibfnamefont {A.~L.}\ \bibnamefont
  {Chernyshev}},\ }\href {\doibase 10.1103/PhysRevLett.99.127004} {\bibfield
  {journal} {\bibinfo  {journal} {Phys. Rev. Lett.}\ }\textbf {\bibinfo
  {volume} {99}},\ \bibinfo {pages} {127004} (\bibinfo {year}
  {2007})}\BibitemShut {NoStop}%
\bibitem [{\citenamefont {Lecheminant}\ \emph {et~al.}(1995)\citenamefont
  {Lecheminant}, \citenamefont {Bernu}, \citenamefont {Lhuillier},\ and\
  \citenamefont {Pierre}}]{Lecheminant1995}%
  \BibitemOpen
  \bibfield  {author} {\bibinfo {author} {\bibfnamefont {P.}~\bibnamefont
  {Lecheminant}}, \bibinfo {author} {\bibfnamefont {B.}~\bibnamefont {Bernu}},
  \bibinfo {author} {\bibfnamefont {C.}~\bibnamefont {Lhuillier}}, \ and\
  \bibinfo {author} {\bibfnamefont {L.}~\bibnamefont {Pierre}},\ }\href
  {\doibase 10.1103/PhysRevB.52.6647} {\bibfield  {journal} {\bibinfo
  {journal} {Phys. Rev. B}\ }\textbf {\bibinfo {volume} {52}},\ \bibinfo
  {pages} {6647} (\bibinfo {year} {1995})}\BibitemShut {NoStop}%
\bibitem [{\citenamefont {Manuel}\ and\ \citenamefont
  {Ceccatto}(1999)}]{Manuel1999}%
  \BibitemOpen
  \bibfield  {author} {\bibinfo {author} {\bibfnamefont {L.~O.}\ \bibnamefont
  {Manuel}}\ and\ \bibinfo {author} {\bibfnamefont {H.~A.}\ \bibnamefont
  {Ceccatto}},\ }\href {\doibase 10.1103/PhysRevB.60.9489} {\bibfield
  {journal} {\bibinfo  {journal} {Phys. Rev. B}\ }\textbf {\bibinfo {volume}
  {60}},\ \bibinfo {pages} {9489} (\bibinfo {year} {1999})}\BibitemShut
  {NoStop}%
\bibitem [{\citenamefont {Kaneko}\ \emph {et~al.}(2014)\citenamefont {Kaneko},
  \citenamefont {Morita},\ and\ \citenamefont {Imada}}]{Kaneko2014}%
  \BibitemOpen
  \bibfield  {author} {\bibinfo {author} {\bibfnamefont {R.}~\bibnamefont
  {Kaneko}}, \bibinfo {author} {\bibfnamefont {S.}~\bibnamefont {Morita}}, \
  and\ \bibinfo {author} {\bibfnamefont {M.}~\bibnamefont {Imada}},\ }\href
  {\doibase 10.7566/JPSJ.83.093707} {\bibfield  {journal} {\bibinfo  {journal}
  {J. Phys. Soc. Jpn.}\ }\textbf {\bibinfo {volume} {83}},\ \bibinfo {pages}
  {093707} (\bibinfo {year} {2014})}\BibitemShut {NoStop}%
\bibitem [{\citenamefont {Iqbal}\ \emph {et~al.}(2016)\citenamefont {Iqbal},
  \citenamefont {Hu}, \citenamefont {Thomale}, \citenamefont {Poilblanc},\ and\
  \citenamefont {Becca}}]{Iqbal2016}%
  \BibitemOpen
  \bibfield  {author} {\bibinfo {author} {\bibfnamefont {Y.}~\bibnamefont
  {Iqbal}}, \bibinfo {author} {\bibfnamefont {W.-J.}\ \bibnamefont {Hu}},
  \bibinfo {author} {\bibfnamefont {R.}~\bibnamefont {Thomale}}, \bibinfo
  {author} {\bibfnamefont {D.}~\bibnamefont {Poilblanc}}, \ and\ \bibinfo
  {author} {\bibfnamefont {F.}~\bibnamefont {Becca}},\ }\href {\doibase
  10.1103/PhysRevB.93.144411} {\bibfield  {journal} {\bibinfo  {journal} {Phys.
  Rev. B}\ }\textbf {\bibinfo {volume} {93}},\ \bibinfo {pages} {144411}
  (\bibinfo {year} {2016})}\BibitemShut {NoStop}%
\bibitem [{\citenamefont {Ferrari}\ and\ \citenamefont
  {Becca}(2019)}]{Ferrari2019}%
  \BibitemOpen
  \bibfield  {author} {\bibinfo {author} {\bibfnamefont {F.}~\bibnamefont
  {Ferrari}}\ and\ \bibinfo {author} {\bibfnamefont {F.}~\bibnamefont
  {Becca}},\ }\href {\doibase 10.1103/PhysRevX.9.031026} {\bibfield  {journal}
  {\bibinfo  {journal} {Phys. Rev. X}\ }\textbf {\bibinfo {volume} {9}},\
  \bibinfo {pages} {031026} (\bibinfo {year} {2019})}\BibitemShut {NoStop}%
\bibitem [{\citenamefont {Misguich}\ \emph {et~al.}(1998)\citenamefont
  {Misguich}, \citenamefont {Bernu}, \citenamefont {Lhuillier},\ and\
  \citenamefont {Waldtmann}}]{Misguich1998}%
  \BibitemOpen
  \bibfield  {author} {\bibinfo {author} {\bibfnamefont {G.}~\bibnamefont
  {Misguich}}, \bibinfo {author} {\bibfnamefont {B.}~\bibnamefont {Bernu}},
  \bibinfo {author} {\bibfnamefont {C.}~\bibnamefont {Lhuillier}}, \ and\
  \bibinfo {author} {\bibfnamefont {C.}~\bibnamefont {Waldtmann}},\ }\href
  {\doibase 10.1103/PhysRevLett.81.1098} {\bibfield  {journal} {\bibinfo
  {journal} {Phys. Rev. Lett.}\ }\textbf {\bibinfo {volume} {81}},\ \bibinfo
  {pages} {1098} (\bibinfo {year} {1998})}\BibitemShut {NoStop}%
\bibitem [{\citenamefont {Motrunich}(2005)}]{Motrunich2005}%
  \BibitemOpen
  \bibfield  {author} {\bibinfo {author} {\bibfnamefont {O.~I.}\ \bibnamefont
  {Motrunich}},\ }\href {\doibase 10.1103/PhysRevB.72.045105} {\bibfield
  {journal} {\bibinfo  {journal} {Phys. Rev. B}\ }\textbf {\bibinfo {volume}
  {72}},\ \bibinfo {pages} {045105} (\bibinfo {year} {2005})}\BibitemShut
  {NoStop}%
\bibitem [{\citenamefont {Mishmash}\ \emph {et~al.}(2013)\citenamefont
  {Mishmash}, \citenamefont {Garrison}, \citenamefont {Bieri},\ and\
  \citenamefont {Xu}}]{Mishmash2013}%
  \BibitemOpen
  \bibfield  {author} {\bibinfo {author} {\bibfnamefont {R.~V.}\ \bibnamefont
  {Mishmash}}, \bibinfo {author} {\bibfnamefont {J.~R.}\ \bibnamefont
  {Garrison}}, \bibinfo {author} {\bibfnamefont {S.}~\bibnamefont {Bieri}}, \
  and\ \bibinfo {author} {\bibfnamefont {C.}~\bibnamefont {Xu}},\ }\href
  {\doibase 10.1103/PhysRevLett.111.157203} {\bibfield  {journal} {\bibinfo
  {journal} {Phys. Rev. Lett.}\ }\textbf {\bibinfo {volume} {111}},\ \bibinfo
  {pages} {157203} (\bibinfo {year} {2013})}\BibitemShut {NoStop}%
\bibitem [{\citenamefont {Calzado}\ and\ \citenamefont
  {Malrieu}(2004)}]{Calzado2004}%
  \BibitemOpen
  \bibfield  {author} {\bibinfo {author} {\bibfnamefont {C.~J.}\ \bibnamefont
  {Calzado}}\ and\ \bibinfo {author} {\bibfnamefont {J.-P.}\ \bibnamefont
  {Malrieu}},\ }\href {\doibase 10.1103/PhysRevB.69.094435} {\bibfield
  {journal} {\bibinfo  {journal} {Phys. Rev. B}\ }\textbf {\bibinfo {volume}
  {69}},\ \bibinfo {pages} {094435} (\bibinfo {year} {2004})}\BibitemShut
  {NoStop}%
\bibitem [{\citenamefont {Tanaka}\ \emph {et~al.}(2018)\citenamefont {Tanaka},
  \citenamefont {Yokoyama},\ and\ \citenamefont {Hotta}}]{Tanaka2018}%
  \BibitemOpen
  \bibfield  {author} {\bibinfo {author} {\bibfnamefont {K.}~\bibnamefont
  {Tanaka}}, \bibinfo {author} {\bibfnamefont {Y.}~\bibnamefont {Yokoyama}}, \
  and\ \bibinfo {author} {\bibfnamefont {C.}~\bibnamefont {Hotta}},\ }\href
  {\doibase 10.7566/JPSJ.87.023702} {\bibfield  {journal} {\bibinfo  {journal}
  {J. Phys. Soc. Jpn.}\ }\textbf {\bibinfo {volume} {87}},\ \bibinfo {pages}
  {023702} (\bibinfo {year} {2018})}\BibitemShut {NoStop}%
\bibitem [{\citenamefont {Morita}\ \emph {et~al.}(2002)\citenamefont {Morita},
  \citenamefont {Watanabe},\ and\ \citenamefont {Imada}}]{Morita2002}%
  \BibitemOpen
  \bibfield  {author} {\bibinfo {author} {\bibfnamefont {H.}~\bibnamefont
  {Morita}}, \bibinfo {author} {\bibfnamefont {S.}~\bibnamefont {Watanabe}}, \
  and\ \bibinfo {author} {\bibfnamefont {M.}~\bibnamefont {Imada}},\ }\href
  {\doibase 10.1143/JPSJ.71.2109} {\bibfield  {journal} {\bibinfo  {journal}
  {J. Phys. Soc. Jpn.}\ }\textbf {\bibinfo {volume} {71}},\ \bibinfo {pages}
  {2109} (\bibinfo {year} {2002})}\BibitemShut {NoStop}%
\bibitem [{\citenamefont {Koretsune}\ \emph {et~al.}(2007)\citenamefont
  {Koretsune}, \citenamefont {Motome},\ and\ \citenamefont
  {Furusaki}}]{Koretsune2007}%
  \BibitemOpen
  \bibfield  {author} {\bibinfo {author} {\bibfnamefont {T.}~\bibnamefont
  {Koretsune}}, \bibinfo {author} {\bibfnamefont {Y.}~\bibnamefont {Motome}}, \
  and\ \bibinfo {author} {\bibfnamefont {A.}~\bibnamefont {Furusaki}},\ }\href
  {\doibase 10.1143/JPSJ.76.074719} {\bibfield  {journal} {\bibinfo  {journal}
  {Journal of the Physical Society of Japan}\ }\textbf {\bibinfo {volume}
  {76}},\ \bibinfo {pages} {074719} (\bibinfo {year} {2007})}\BibitemShut
  {NoStop}%
\bibitem [{\citenamefont {Sahebsara}\ and\ \citenamefont
  {S\'en\'echal}(2008)}]{Sahebsara2008}%
  \BibitemOpen
  \bibfield  {author} {\bibinfo {author} {\bibfnamefont {P.}~\bibnamefont
  {Sahebsara}}\ and\ \bibinfo {author} {\bibfnamefont {D.}~\bibnamefont
  {S\'en\'echal}},\ }\href {\doibase 10.1103/PhysRevLett.100.136402} {\bibfield
   {journal} {\bibinfo  {journal} {Phys. Rev. Lett.}\ }\textbf {\bibinfo
  {volume} {100}},\ \bibinfo {pages} {136402} (\bibinfo {year}
  {2008})}\BibitemShut {NoStop}%
\bibitem [{\citenamefont {Yoshioka}\ \emph {et~al.}(2009)\citenamefont
  {Yoshioka}, \citenamefont {Koga},\ and\ \citenamefont
  {Kawakami}}]{Yoshioka2009}%
  \BibitemOpen
  \bibfield  {author} {\bibinfo {author} {\bibfnamefont {T.}~\bibnamefont
  {Yoshioka}}, \bibinfo {author} {\bibfnamefont {A.}~\bibnamefont {Koga}}, \
  and\ \bibinfo {author} {\bibfnamefont {N.}~\bibnamefont {Kawakami}},\ }\href
  {\doibase 10.1103/PhysRevLett.103.036401} {\bibfield  {journal} {\bibinfo
  {journal} {Phys. Rev. Lett.}\ }\textbf {\bibinfo {volume} {103}},\ \bibinfo
  {pages} {036401} (\bibinfo {year} {2009})}\BibitemShut {NoStop}%
\bibitem [{\citenamefont {Tocchio}\ \emph {et~al.}(2013)\citenamefont
  {Tocchio}, \citenamefont {Feldner}, \citenamefont {Becca}, \citenamefont
  {Valent\'{\i}},\ and\ \citenamefont {Gros}}]{Tocchio2013}%
  \BibitemOpen
  \bibfield  {author} {\bibinfo {author} {\bibfnamefont {L.~F.}\ \bibnamefont
  {Tocchio}}, \bibinfo {author} {\bibfnamefont {H.}~\bibnamefont {Feldner}},
  \bibinfo {author} {\bibfnamefont {F.}~\bibnamefont {Becca}}, \bibinfo
  {author} {\bibfnamefont {R.}~\bibnamefont {Valent\'{\i}}}, \ and\ \bibinfo
  {author} {\bibfnamefont {C.}~\bibnamefont {Gros}},\ }\href {\doibase
  10.1103/PhysRevB.87.035143} {\bibfield  {journal} {\bibinfo  {journal} {Phys.
  Rev. B}\ }\textbf {\bibinfo {volume} {87}},\ \bibinfo {pages} {035143}
  (\bibinfo {year} {2013})}\BibitemShut {NoStop}%
\bibitem [{\citenamefont {Yamada}(2014)}]{Yamada2014}%
  \BibitemOpen
  \bibfield  {author} {\bibinfo {author} {\bibfnamefont {A.}~\bibnamefont
  {Yamada}},\ }\href {\doibase 10.1103/PhysRevB.89.195108} {\bibfield
  {journal} {\bibinfo  {journal} {Phys. Rev. B}\ }\textbf {\bibinfo {volume}
  {89}},\ \bibinfo {pages} {195108} (\bibinfo {year} {2014})}\BibitemShut
  {NoStop}%
\bibitem [{\citenamefont {Laubach}\ \emph {et~al.}(2015)\citenamefont
  {Laubach}, \citenamefont {Thomale}, \citenamefont {Platt}, \citenamefont
  {Hanke},\ and\ \citenamefont {Li}}]{Laubach2015}%
  \BibitemOpen
  \bibfield  {author} {\bibinfo {author} {\bibfnamefont {M.}~\bibnamefont
  {Laubach}}, \bibinfo {author} {\bibfnamefont {R.}~\bibnamefont {Thomale}},
  \bibinfo {author} {\bibfnamefont {C.}~\bibnamefont {Platt}}, \bibinfo
  {author} {\bibfnamefont {W.}~\bibnamefont {Hanke}}, \ and\ \bibinfo {author}
  {\bibfnamefont {G.}~\bibnamefont {Li}},\ }\href {\doibase
  10.1103/PhysRevB.91.245125} {\bibfield  {journal} {\bibinfo  {journal} {Phys.
  Rev. B}\ }\textbf {\bibinfo {volume} {91}},\ \bibinfo {pages} {245125}
  (\bibinfo {year} {2015})}\BibitemShut {NoStop}%
\bibitem [{\citenamefont {Misumi}\ \emph {et~al.}(2017)\citenamefont {Misumi},
  \citenamefont {Kaneko},\ and\ \citenamefont {Ohta}}]{Misumi2017}%
  \BibitemOpen
  \bibfield  {author} {\bibinfo {author} {\bibfnamefont {K.}~\bibnamefont
  {Misumi}}, \bibinfo {author} {\bibfnamefont {T.}~\bibnamefont {Kaneko}}, \
  and\ \bibinfo {author} {\bibfnamefont {Y.}~\bibnamefont {Ohta}},\ }\href
  {\doibase 10.1103/PhysRevB.95.075124} {\bibfield  {journal} {\bibinfo
  {journal} {Phys. Rev. B}\ }\textbf {\bibinfo {volume} {95}},\ \bibinfo
  {pages} {075124} (\bibinfo {year} {2017})}\BibitemShut {NoStop}%
\bibitem [{\citenamefont {Shirakawa}\ \emph {et~al.}(2017)\citenamefont
  {Shirakawa}, \citenamefont {Tohyama}, \citenamefont {Kokalj}, \citenamefont
  {Sota},\ and\ \citenamefont {Yunoki}}]{Shirakawa2017}%
  \BibitemOpen
  \bibfield  {author} {\bibinfo {author} {\bibfnamefont {T.}~\bibnamefont
  {Shirakawa}}, \bibinfo {author} {\bibfnamefont {T.}~\bibnamefont {Tohyama}},
  \bibinfo {author} {\bibfnamefont {J.}~\bibnamefont {Kokalj}}, \bibinfo
  {author} {\bibfnamefont {S.}~\bibnamefont {Sota}}, \ and\ \bibinfo {author}
  {\bibfnamefont {S.}~\bibnamefont {Yunoki}},\ }\href {\doibase
  10.1103/PhysRevB.96.205130} {\bibfield  {journal} {\bibinfo  {journal} {Phys.
  Rev. B}\ }\textbf {\bibinfo {volume} {96}},\ \bibinfo {pages} {205130}
  (\bibinfo {year} {2017})}\BibitemShut {NoStop}%
\bibitem [{\citenamefont {Szasz}\ \emph {et~al.}(2020)\citenamefont {Szasz},
  \citenamefont {Motruk}, \citenamefont {Zaletel},\ and\ \citenamefont
  {Moore}}]{Szasz2018}%
  \BibitemOpen
  \bibfield  {author} {\bibinfo {author} {\bibfnamefont {A.}~\bibnamefont
  {Szasz}}, \bibinfo {author} {\bibfnamefont {J.}~\bibnamefont {Motruk}},
  \bibinfo {author} {\bibfnamefont {M.~P.}\ \bibnamefont {Zaletel}}, \ and\
  \bibinfo {author} {\bibfnamefont {J.~E.}\ \bibnamefont {Moore}},\ }\href
  {\doibase 10.1103/PhysRevX.10.021042} {\bibfield  {journal} {\bibinfo
  {journal} {Phys. Rev. X}\ }\textbf {\bibinfo {volume} {10}},\ \bibinfo
  {pages} {021042} (\bibinfo {year} {2020})}\BibitemShut {NoStop}%
\bibitem [{\citenamefont {Skolimowski}\ \emph {et~al.}(2019)\citenamefont
  {Skolimowski}, \citenamefont {Gerasimenko},\ and\ \citenamefont
  {\ifmmode~\check{Z}\else \v{Z}\fi{}itko}}]{Skolimowski2019}%
  \BibitemOpen
  \bibfield  {author} {\bibinfo {author} {\bibfnamefont {J.}~\bibnamefont
  {Skolimowski}}, \bibinfo {author} {\bibfnamefont {Y.}~\bibnamefont
  {Gerasimenko}}, \ and\ \bibinfo {author} {\bibfnamefont {R.}~\bibnamefont
  {\ifmmode~\check{Z}\else \v{Z}\fi{}itko}},\ }\href {\doibase
  10.1103/PhysRevLett.122.036802} {\bibfield  {journal} {\bibinfo  {journal}
  {Phys. Rev. Lett.}\ }\textbf {\bibinfo {volume} {122}},\ \bibinfo {pages}
  {036802} (\bibinfo {year} {2019})}\BibitemShut {NoStop}%
\bibitem [{\citenamefont {McMahan}\ and\ \citenamefont
  {Wilkins}(1975)}]{McMahan1975}%
  \BibitemOpen
  \bibfield  {author} {\bibinfo {author} {\bibfnamefont {A.~K.}\ \bibnamefont
  {McMahan}}\ and\ \bibinfo {author} {\bibfnamefont {J.~W.}\ \bibnamefont
  {Wilkins}},\ }\href {\doibase 10.1103/PhysRevLett.35.376} {\bibfield
  {journal} {\bibinfo  {journal} {Phys. Rev. Lett.}\ }\textbf {\bibinfo
  {volume} {35}},\ \bibinfo {pages} {376} (\bibinfo {year} {1975})}\BibitemShut
  {NoStop}%
\bibitem [{\citenamefont {Hetherington}\ and\ \citenamefont
  {Willard}(1975)}]{Hetherington1975}%
  \BibitemOpen
  \bibfield  {author} {\bibinfo {author} {\bibfnamefont {J.~H.}\ \bibnamefont
  {Hetherington}}\ and\ \bibinfo {author} {\bibfnamefont {F.~D.~C.}\
  \bibnamefont {Willard}},\ }\href {\doibase 10.1103/PhysRevLett.35.1442}
  {\bibfield  {journal} {\bibinfo  {journal} {Phys. Rev. Lett.}\ }\textbf
  {\bibinfo {volume} {35}},\ \bibinfo {pages} {1442} (\bibinfo {year}
  {1975})}\BibitemShut {NoStop}%
\bibitem [{\citenamefont {Roger}\ \emph {et~al.}(1980)\citenamefont {Roger},
  \citenamefont {Delrieu},\ and\ \citenamefont {Hetherington}}]{Roger1980}%
  \BibitemOpen
  \bibfield  {author} {\bibinfo {author} {\bibfnamefont {M.}~\bibnamefont
  {Roger}}, \bibinfo {author} {\bibfnamefont {J.~M.}\ \bibnamefont {Delrieu}},
  \ and\ \bibinfo {author} {\bibfnamefont {J.~H.}\ \bibnamefont
  {Hetherington}},\ }\href {\doibase 10.1103/PhysRevLett.45.137} {\bibfield
  {journal} {\bibinfo  {journal} {Phys. Rev. Lett.}\ }\textbf {\bibinfo
  {volume} {45}},\ \bibinfo {pages} {137} (\bibinfo {year} {1980})}\BibitemShut
  {NoStop}%
\bibitem [{\citenamefont {Yosida}(1980)}]{Yosida1980}%
  \BibitemOpen
  \bibfield  {author} {\bibinfo {author} {\bibfnamefont {K.}~\bibnamefont
  {Yosida}},\ }\href {\doibase 10.1143/PTP.69.475} {\bibfield  {journal}
  {\bibinfo  {journal} {Progress of Theoretical Physics Supplement}\ }\textbf
  {\bibinfo {volume} {69}},\ \bibinfo {pages} {475} (\bibinfo {year}
  {1980})}\BibitemShut {NoStop}%
\bibitem [{\citenamefont {Ceperley}\ and\ \citenamefont
  {Jacucci}(1987)}]{Ceperley1987}%
  \BibitemOpen
  \bibfield  {author} {\bibinfo {author} {\bibfnamefont {D.~M.}\ \bibnamefont
  {Ceperley}}\ and\ \bibinfo {author} {\bibfnamefont {G.}~\bibnamefont
  {Jacucci}},\ }\href {\doibase 10.1103/PhysRevLett.58.1648} {\bibfield
  {journal} {\bibinfo  {journal} {Phys. Rev. Lett.}\ }\textbf {\bibinfo
  {volume} {58}},\ \bibinfo {pages} {1648} (\bibinfo {year}
  {1987})}\BibitemShut {NoStop}%
\bibitem [{\citenamefont {Roger}(2011)}]{Roger2011}%
  \BibitemOpen
  \bibfield  {author} {\bibinfo {author} {\bibfnamefont {M.}~\bibnamefont
  {Roger}},\ }\href {\doibase 10.1007/s10909-010-0293-1} {\bibfield  {journal}
  {\bibinfo  {journal} {Journal of Low Temperature Physics}\ }\textbf {\bibinfo
  {volume} {162}},\ \bibinfo {pages} {625} (\bibinfo {year}
  {2011})}\BibitemShut {NoStop}%
\bibitem [{\citenamefont {C\^andido}\ \emph {et~al.}(2011)\citenamefont
  {C\^andido}, \citenamefont {Hai},\ and\ \citenamefont
  {Ceperley}}]{Candido2011}%
  \BibitemOpen
  \bibfield  {author} {\bibinfo {author} {\bibfnamefont {L.}~\bibnamefont
  {C\^andido}}, \bibinfo {author} {\bibfnamefont {G.-Q.}\ \bibnamefont {Hai}},
  \ and\ \bibinfo {author} {\bibfnamefont {D.~M.}\ \bibnamefont {Ceperley}},\
  }\href {\doibase 10.1103/PhysRevB.84.064515} {\bibfield  {journal} {\bibinfo
  {journal} {Phys. Rev. B}\ }\textbf {\bibinfo {volume} {84}},\ \bibinfo
  {pages} {064515} (\bibinfo {year} {2011})}\BibitemShut {NoStop}%
\bibitem [{\citenamefont {Yosida}\ and\ \citenamefont
  {Inagaki}(1981)}]{Yosida1981}%
  \BibitemOpen
  \bibfield  {author} {\bibinfo {author} {\bibfnamefont {K.}~\bibnamefont
  {Yosida}}\ and\ \bibinfo {author} {\bibfnamefont {S.}~\bibnamefont
  {Inagaki}},\ }\href {\doibase 10.1143/JPSJ.50.3268} {\bibfield  {journal}
  {\bibinfo  {journal} {J. Phys. Soc. Jpn.}\ }\textbf {\bibinfo {volume}
  {50}},\ \bibinfo {pages} {3268} (\bibinfo {year} {1981})}\BibitemShut
  {NoStop}%
\bibitem [{\citenamefont {Lorenzana}\ \emph {et~al.}(1999)\citenamefont
  {Lorenzana}, \citenamefont {Eroles},\ and\ \citenamefont
  {Sorella}}]{Lorenzana1999}%
  \BibitemOpen
  \bibfield  {author} {\bibinfo {author} {\bibfnamefont {J.}~\bibnamefont
  {Lorenzana}}, \bibinfo {author} {\bibfnamefont {J.}~\bibnamefont {Eroles}}, \
  and\ \bibinfo {author} {\bibfnamefont {S.}~\bibnamefont {Sorella}},\ }\href
  {\doibase 10.1103/PhysRevLett.83.5122} {\bibfield  {journal} {\bibinfo
  {journal} {Phys. Rev. Lett.}\ }\textbf {\bibinfo {volume} {83}},\ \bibinfo
  {pages} {5122} (\bibinfo {year} {1999})}\BibitemShut {NoStop}%
\bibitem [{\citenamefont {Coldea}\ \emph {et~al.}(2001)\citenamefont {Coldea},
  \citenamefont {Hayden}, \citenamefont {Aeppli}, \citenamefont {Perring},
  \citenamefont {Frost}, \citenamefont {Mason}, \citenamefont {Cheong},\ and\
  \citenamefont {Fisk}}]{Coldea2001}%
  \BibitemOpen
  \bibfield  {author} {\bibinfo {author} {\bibfnamefont {R.}~\bibnamefont
  {Coldea}}, \bibinfo {author} {\bibfnamefont {S.~M.}\ \bibnamefont {Hayden}},
  \bibinfo {author} {\bibfnamefont {G.}~\bibnamefont {Aeppli}}, \bibinfo
  {author} {\bibfnamefont {T.~G.}\ \bibnamefont {Perring}}, \bibinfo {author}
  {\bibfnamefont {C.~D.}\ \bibnamefont {Frost}}, \bibinfo {author}
  {\bibfnamefont {T.~E.}\ \bibnamefont {Mason}}, \bibinfo {author}
  {\bibfnamefont {S.-W.}\ \bibnamefont {Cheong}}, \ and\ \bibinfo {author}
  {\bibfnamefont {Z.}~\bibnamefont {Fisk}},\ }\href {\doibase
  10.1103/PhysRevLett.86.5377} {\bibfield  {journal} {\bibinfo  {journal}
  {Phys. Rev. Lett.}\ }\textbf {\bibinfo {volume} {86}},\ \bibinfo {pages}
  {5377} (\bibinfo {year} {2001})}\BibitemShut {NoStop}%
\bibitem [{\citenamefont {Katanin}\ and\ \citenamefont
  {Kampf}(2002)}]{Katanin2002}%
  \BibitemOpen
  \bibfield  {author} {\bibinfo {author} {\bibfnamefont {A.~A.}\ \bibnamefont
  {Katanin}}\ and\ \bibinfo {author} {\bibfnamefont {A.~P.}\ \bibnamefont
  {Kampf}},\ }\href {\doibase 10.1103/PhysRevB.66.100403} {\bibfield  {journal}
  {\bibinfo  {journal} {Phys. Rev. B}\ }\textbf {\bibinfo {volume} {66}},\
  \bibinfo {pages} {100403} (\bibinfo {year} {2002})}\BibitemShut {NoStop}%
\bibitem [{\citenamefont {Headings}\ \emph {et~al.}(2010)\citenamefont
  {Headings}, \citenamefont {Hayden}, \citenamefont {Coldea},\ and\
  \citenamefont {Perring}}]{Headings2010}%
  \BibitemOpen
  \bibfield  {author} {\bibinfo {author} {\bibfnamefont {N.~S.}\ \bibnamefont
  {Headings}}, \bibinfo {author} {\bibfnamefont {S.~M.}\ \bibnamefont
  {Hayden}}, \bibinfo {author} {\bibfnamefont {R.}~\bibnamefont {Coldea}}, \
  and\ \bibinfo {author} {\bibfnamefont {T.~G.}\ \bibnamefont {Perring}},\
  }\href {\doibase 10.1103/PhysRevLett.105.247001} {\bibfield  {journal}
  {\bibinfo  {journal} {Phys. Rev. Lett.}\ }\textbf {\bibinfo {volume} {105}},\
  \bibinfo {pages} {247001} (\bibinfo {year} {2010})}\BibitemShut {NoStop}%
\bibitem [{\citenamefont {Rutonjski}\ \emph {et~al.}(2016)\citenamefont
  {Rutonjski}, \citenamefont {Pavkov-Hrvojevi\'{c}},\ and\ \citenamefont
  {Berovi\'{c}}}]{Rutonjski2016}%
  \BibitemOpen
  \bibfield  {author} {\bibinfo {author} {\bibfnamefont {M.~S.}\ \bibnamefont
  {Rutonjski}}, \bibinfo {author} {\bibfnamefont {M.~V.}\ \bibnamefont
  {Pavkov-Hrvojevi\'{c}}}, \ and\ \bibinfo {author} {\bibfnamefont {M.~B.}\
  \bibnamefont {Berovi\'{c}}},\ }\href {\doibase 10.1142/S0217979215502513}
  {\bibfield  {journal} {\bibinfo  {journal} {International Journal of Modern
  Physics B}\ }\textbf {\bibinfo {volume} {30}},\ \bibinfo {pages} {1550251}
  (\bibinfo {year} {2016})}\BibitemShut {NoStop}%
\bibitem [{\citenamefont {Yamamoto}\ and\ \citenamefont
  {Noriki}(2019)}]{Yamamoto2019}%
  \BibitemOpen
  \bibfield  {author} {\bibinfo {author} {\bibfnamefont {S.}~\bibnamefont
  {Yamamoto}}\ and\ \bibinfo {author} {\bibfnamefont {Y.}~\bibnamefont
  {Noriki}},\ }\href {\doibase 10.1103/PhysRevB.99.094412} {\bibfield
  {journal} {\bibinfo  {journal} {Phys. Rev. B}\ }\textbf {\bibinfo {volume}
  {99}},\ \bibinfo {pages} {094412} (\bibinfo {year} {2019})}\BibitemShut
  {NoStop}%
\bibitem [{\citenamefont {Law}\ and\ \citenamefont {Lee}(2017)}]{Law2017}%
  \BibitemOpen
  \bibfield  {author} {\bibinfo {author} {\bibfnamefont {K.~T.}\ \bibnamefont
  {Law}}\ and\ \bibinfo {author} {\bibfnamefont {P.~A.}\ \bibnamefont {Lee}},\
  }\href {\doibase 10.1073/pnas.1706769114} {\bibfield  {journal} {\bibinfo
  {journal} {Proceedings of the National Academy of Sciences}\ }\textbf
  {\bibinfo {volume} {114}},\ \bibinfo {pages} {6996} (\bibinfo {year}
  {2017})}\BibitemShut {NoStop}%
\bibitem [{\citenamefont {He}\ \emph {et~al.}(2018)\citenamefont {He},
  \citenamefont {Xu}, \citenamefont {Chen}, \citenamefont {Law},\ and\
  \citenamefont {Lee}}]{He2018}%
  \BibitemOpen
  \bibfield  {author} {\bibinfo {author} {\bibfnamefont {W.-Y.}\ \bibnamefont
  {He}}, \bibinfo {author} {\bibfnamefont {X.~Y.}\ \bibnamefont {Xu}}, \bibinfo
  {author} {\bibfnamefont {G.}~\bibnamefont {Chen}}, \bibinfo {author}
  {\bibfnamefont {K.~T.}\ \bibnamefont {Law}}, \ and\ \bibinfo {author}
  {\bibfnamefont {P.~A.}\ \bibnamefont {Lee}},\ }\href {\doibase
  10.1103/PhysRevLett.121.046401} {\bibfield  {journal} {\bibinfo  {journal}
  {Phys. Rev. Lett.}\ }\textbf {\bibinfo {volume} {121}},\ \bibinfo {pages}
  {046401} (\bibinfo {year} {2018})}\BibitemShut {NoStop}%
\bibitem [{\citenamefont {Shimizu}\ \emph {et~al.}(2003)\citenamefont
  {Shimizu}, \citenamefont {Miyagawa}, \citenamefont {Kanoda}, \citenamefont
  {Maesato},\ and\ \citenamefont {Saito}}]{Shimizu2003}%
  \BibitemOpen
  \bibfield  {author} {\bibinfo {author} {\bibfnamefont {Y.}~\bibnamefont
  {Shimizu}}, \bibinfo {author} {\bibfnamefont {K.}~\bibnamefont {Miyagawa}},
  \bibinfo {author} {\bibfnamefont {K.}~\bibnamefont {Kanoda}}, \bibinfo
  {author} {\bibfnamefont {M.}~\bibnamefont {Maesato}}, \ and\ \bibinfo
  {author} {\bibfnamefont {G.}~\bibnamefont {Saito}},\ }\href {\doibase
  10.1103/PhysRevLett.91.107001} {\bibfield  {journal} {\bibinfo  {journal}
  {Phys. Rev. Lett.}\ }\textbf {\bibinfo {volume} {91}},\ \bibinfo {pages}
  {107001} (\bibinfo {year} {2003})}\BibitemShut {NoStop}%
\bibitem [{\citenamefont {Kurosaki}\ \emph {et~al.}(2005)\citenamefont
  {Kurosaki}, \citenamefont {Shimizu}, \citenamefont {Miyagawa}, \citenamefont
  {Kanoda},\ and\ \citenamefont {Saito}}]{Kurosaki2005}%
  \BibitemOpen
  \bibfield  {author} {\bibinfo {author} {\bibfnamefont {Y.}~\bibnamefont
  {Kurosaki}}, \bibinfo {author} {\bibfnamefont {Y.}~\bibnamefont {Shimizu}},
  \bibinfo {author} {\bibfnamefont {K.}~\bibnamefont {Miyagawa}}, \bibinfo
  {author} {\bibfnamefont {K.}~\bibnamefont {Kanoda}}, \ and\ \bibinfo {author}
  {\bibfnamefont {G.}~\bibnamefont {Saito}},\ }\href {\doibase
  10.1103/PhysRevLett.95.177001} {\bibfield  {journal} {\bibinfo  {journal}
  {Phys. Rev. Lett.}\ }\textbf {\bibinfo {volume} {95}},\ \bibinfo {pages}
  {177001} (\bibinfo {year} {2005})}\BibitemShut {NoStop}%
\bibitem [{\citenamefont {Manna}\ \emph {et~al.}(2010)\citenamefont {Manna},
  \citenamefont {de~Souza}, \citenamefont {Br\"uhl}, \citenamefont
  {Schlueter},\ and\ \citenamefont {Lang}}]{Manna2010}%
  \BibitemOpen
  \bibfield  {author} {\bibinfo {author} {\bibfnamefont {R.~S.}\ \bibnamefont
  {Manna}}, \bibinfo {author} {\bibfnamefont {M.}~\bibnamefont {de~Souza}},
  \bibinfo {author} {\bibfnamefont {A.}~\bibnamefont {Br\"uhl}}, \bibinfo
  {author} {\bibfnamefont {J.~A.}\ \bibnamefont {Schlueter}}, \ and\ \bibinfo
  {author} {\bibfnamefont {M.}~\bibnamefont {Lang}},\ }\href {\doibase
  10.1103/PhysRevLett.104.016403} {\bibfield  {journal} {\bibinfo  {journal}
  {Phys. Rev. Lett.}\ }\textbf {\bibinfo {volume} {104}},\ \bibinfo {pages}
  {016403} (\bibinfo {year} {2010})}\BibitemShut {NoStop}%
\bibitem [{\citenamefont {Itou}\ \emph {et~al.}(2008)\citenamefont {Itou},
  \citenamefont {Oyamada}, \citenamefont {Maegawa}, \citenamefont {Tamura},\
  and\ \citenamefont {Kato}}]{Itou2008}%
  \BibitemOpen
  \bibfield  {author} {\bibinfo {author} {\bibfnamefont {T.}~\bibnamefont
  {Itou}}, \bibinfo {author} {\bibfnamefont {A.}~\bibnamefont {Oyamada}},
  \bibinfo {author} {\bibfnamefont {S.}~\bibnamefont {Maegawa}}, \bibinfo
  {author} {\bibfnamefont {M.}~\bibnamefont {Tamura}}, \ and\ \bibinfo {author}
  {\bibfnamefont {R.}~\bibnamefont {Kato}},\ }\href {\doibase
  10.1103/PhysRevB.77.104413} {\bibfield  {journal} {\bibinfo  {journal} {Phys.
  Rev. B}\ }\textbf {\bibinfo {volume} {77}},\ \bibinfo {pages} {104413}
  (\bibinfo {year} {2008})}\BibitemShut {NoStop}%
\bibitem [{\citenamefont {Yamashita}\ \emph {et~al.}(2010)\citenamefont
  {Yamashita}, \citenamefont {Nakata}, \citenamefont {Senshu}, \citenamefont
  {Nagata}, \citenamefont {Yamamoto}, \citenamefont {Kato}, \citenamefont
  {Shibauchi},\ and\ \citenamefont {Matsuda}}]{Yamashita2010}%
  \BibitemOpen
  \bibfield  {author} {\bibinfo {author} {\bibfnamefont {M.}~\bibnamefont
  {Yamashita}}, \bibinfo {author} {\bibfnamefont {N.}~\bibnamefont {Nakata}},
  \bibinfo {author} {\bibfnamefont {Y.}~\bibnamefont {Senshu}}, \bibinfo
  {author} {\bibfnamefont {M.}~\bibnamefont {Nagata}}, \bibinfo {author}
  {\bibfnamefont {H.~M.}\ \bibnamefont {Yamamoto}}, \bibinfo {author}
  {\bibfnamefont {R.}~\bibnamefont {Kato}}, \bibinfo {author} {\bibfnamefont
  {T.}~\bibnamefont {Shibauchi}}, \ and\ \bibinfo {author} {\bibfnamefont
  {Y.}~\bibnamefont {Matsuda}},\ }\href {\doibase 10.1126/science.1188200}
  {\bibfield  {journal} {\bibinfo  {journal} {Science}\ }\textbf {\bibinfo
  {volume} {328}},\ \bibinfo {pages} {1246} (\bibinfo {year}
  {2010})}\BibitemShut {NoStop}%
\bibitem [{\citenamefont {Fazekas}\ and\ \citenamefont
  {Tosatti}(1979)}]{FazekasTosatti1979}%
  \BibitemOpen
  \bibfield  {author} {\bibinfo {author} {\bibfnamefont {P.}~\bibnamefont
  {Fazekas}}\ and\ \bibinfo {author} {\bibfnamefont {E.}~\bibnamefont
  {Tosatti}},\ }\href {\doibase 10.1080/13642817908245359} {\bibfield
  {journal} {\bibinfo  {journal} {Philosophical Magazine B}\ }\textbf {\bibinfo
  {volume} {39}},\ \bibinfo {pages} {229} (\bibinfo {year} {1979})}\BibitemShut
  {NoStop}%
\bibitem [{\citenamefont {Klanjsek}\ \emph {et~al.}(2017)\citenamefont
  {Klanjsek}, \citenamefont {Zorko}, \citenamefont {Zitko}, \citenamefont
  {Mravlje}, \citenamefont {Jaglicic}, \citenamefont {Biswas}, \citenamefont
  {Prelovsek}, \citenamefont {Mihailovic},\ and\ \citenamefont
  {Arcon}}]{Klanjsek2017}%
  \BibitemOpen
  \bibfield  {author} {\bibinfo {author} {\bibfnamefont {M.}~\bibnamefont
  {Klanjsek}}, \bibinfo {author} {\bibfnamefont {A.}~\bibnamefont {Zorko}},
  \bibinfo {author} {\bibfnamefont {R.}~\bibnamefont {Zitko}}, \bibinfo
  {author} {\bibfnamefont {J.}~\bibnamefont {Mravlje}}, \bibinfo {author}
  {\bibfnamefont {Z.}~\bibnamefont {Jaglicic}}, \bibinfo {author}
  {\bibfnamefont {P.}~\bibnamefont {Biswas}}, \bibinfo {author} {\bibfnamefont
  {P.}~\bibnamefont {Prelovsek}}, \bibinfo {author} {\bibfnamefont
  {D.}~\bibnamefont {Mihailovic}}, \ and\ \bibinfo {author} {\bibfnamefont
  {D.}~\bibnamefont {Arcon}},\ }\href {https://doi.org/10.1038/nphys4212}
  {\bibfield  {journal} {\bibinfo  {journal} {Nature Physics}\ }\textbf
  {\bibinfo {volume} {13}},\ \bibinfo {pages} {1130} (\bibinfo {year}
  {2017})}\BibitemShut {NoStop}%
\bibitem [{\citenamefont {Roger}(1984)}]{Roger1984}%
  \BibitemOpen
  \bibfield  {author} {\bibinfo {author} {\bibfnamefont {M.}~\bibnamefont
  {Roger}},\ }\href {\doibase 10.1103/PhysRevB.30.6432} {\bibfield  {journal}
  {\bibinfo  {journal} {Phys. Rev. B}\ }\textbf {\bibinfo {volume} {30}},\
  \bibinfo {pages} {6432} (\bibinfo {year} {1984})}\BibitemShut {NoStop}%
\bibitem [{\citenamefont {Roger}(1990)}]{Roger1990}%
  \BibitemOpen
  \bibfield  {author} {\bibinfo {author} {\bibfnamefont {M.}~\bibnamefont
  {Roger}},\ }\href {\doibase 10.1103/PhysRevLett.64.297} {\bibfield  {journal}
  {\bibinfo  {journal} {Phys. Rev. Lett.}\ }\textbf {\bibinfo {volume} {64}},\
  \bibinfo {pages} {297} (\bibinfo {year} {1990})}\BibitemShut {NoStop}%
\bibitem [{\citenamefont {Ishida}\ \emph {et~al.}(1997)\citenamefont {Ishida},
  \citenamefont {Morishita}, \citenamefont {Yawata},\ and\ \citenamefont
  {Fukuyama}}]{Ishida1997}%
  \BibitemOpen
  \bibfield  {author} {\bibinfo {author} {\bibfnamefont {K.}~\bibnamefont
  {Ishida}}, \bibinfo {author} {\bibfnamefont {M.}~\bibnamefont {Morishita}},
  \bibinfo {author} {\bibfnamefont {K.}~\bibnamefont {Yawata}}, \ and\ \bibinfo
  {author} {\bibfnamefont {H.}~\bibnamefont {Fukuyama}},\ }\href {\doibase
  10.1103/PhysRevLett.79.3451} {\bibfield  {journal} {\bibinfo  {journal}
  {Phys. Rev. Lett.}\ }\textbf {\bibinfo {volume} {79}},\ \bibinfo {pages}
  {3451} (\bibinfo {year} {1997})}\BibitemShut {NoStop}%
\bibitem [{\citenamefont {Roger}\ \emph {et~al.}(1998)\citenamefont {Roger},
  \citenamefont {B\"auerle}, \citenamefont {Bunkov}, \citenamefont {Chen},\
  and\ \citenamefont {Godfrin}}]{Roger1998}%
  \BibitemOpen
  \bibfield  {author} {\bibinfo {author} {\bibfnamefont {M.}~\bibnamefont
  {Roger}}, \bibinfo {author} {\bibfnamefont {C.}~\bibnamefont {B\"auerle}},
  \bibinfo {author} {\bibfnamefont {Y.~M.}\ \bibnamefont {Bunkov}}, \bibinfo
  {author} {\bibfnamefont {A.-S.}\ \bibnamefont {Chen}}, \ and\ \bibinfo
  {author} {\bibfnamefont {H.}~\bibnamefont {Godfrin}},\ }\href {\doibase
  10.1103/PhysRevLett.80.1308} {\bibfield  {journal} {\bibinfo  {journal}
  {Phys. Rev. Lett.}\ }\textbf {\bibinfo {volume} {80}},\ \bibinfo {pages}
  {1308} (\bibinfo {year} {1998})}\BibitemShut {NoStop}%
\bibitem [{\citenamefont {Momoi}\ \emph {et~al.}(1999)\citenamefont {Momoi},
  \citenamefont {Sakamoto},\ and\ \citenamefont {Kubo}}]{Momoi1999}%
  \BibitemOpen
  \bibfield  {author} {\bibinfo {author} {\bibfnamefont {T.}~\bibnamefont
  {Momoi}}, \bibinfo {author} {\bibfnamefont {H.}~\bibnamefont {Sakamoto}}, \
  and\ \bibinfo {author} {\bibfnamefont {K.}~\bibnamefont {Kubo}},\ }\href
  {\doibase 10.1103/PhysRevB.59.9491} {\bibfield  {journal} {\bibinfo
  {journal} {Phys. Rev. B}\ }\textbf {\bibinfo {volume} {59}},\ \bibinfo
  {pages} {9491} (\bibinfo {year} {1999})}\BibitemShut {NoStop}%
\bibitem [{\citenamefont {Momoi}\ \emph {et~al.}(2006)\citenamefont {Momoi},
  \citenamefont {Sindzingre},\ and\ \citenamefont {Shannon}}]{Momoi2006}%
  \BibitemOpen
  \bibfield  {author} {\bibinfo {author} {\bibfnamefont {T.}~\bibnamefont
  {Momoi}}, \bibinfo {author} {\bibfnamefont {P.}~\bibnamefont {Sindzingre}}, \
  and\ \bibinfo {author} {\bibfnamefont {N.}~\bibnamefont {Shannon}},\ }\href
  {\doibase 10.1103/PhysRevLett.97.257204} {\bibfield  {journal} {\bibinfo
  {journal} {Phys. Rev. Lett.}\ }\textbf {\bibinfo {volume} {97}},\ \bibinfo
  {pages} {257204} (\bibinfo {year} {2006})}\BibitemShut {NoStop}%
\bibitem [{\citenamefont {Fukuyama}(2008)}]{Fukuyama2008}%
  \BibitemOpen
  \bibfield  {author} {\bibinfo {author} {\bibfnamefont {H.}~\bibnamefont
  {Fukuyama}},\ }\href {http://dx.doi.org/10.1143/JPSJ.77.111013} {\bibfield
  {journal} {\bibinfo  {journal} {J. Phys. Soc. Jpn.}\ }\textbf {\bibinfo
  {volume} {77}},\ \bibinfo {pages} {111013} (\bibinfo {year}
  {2008})}\BibitemShut {NoStop}%
\bibitem [{\citenamefont {Fuseya}\ and\ \citenamefont
  {Ogata}(2009)}]{Fuseya2009}%
  \BibitemOpen
  \bibfield  {author} {\bibinfo {author} {\bibfnamefont {Y.}~\bibnamefont
  {Fuseya}}\ and\ \bibinfo {author} {\bibfnamefont {M.}~\bibnamefont {Ogata}},\
  }\href {\doibase 10.1143/JPSJ.78.013601} {\bibfield  {journal} {\bibinfo
  {journal} {J. Phys. Soc. Jpn.}\ }\textbf {\bibinfo {volume} {78}},\ \bibinfo
  {pages} {013601} (\bibinfo {year} {2009})}\BibitemShut {NoStop}%
\bibitem [{\citenamefont {Seki}\ \emph {et~al.}(2009)\citenamefont {Seki},
  \citenamefont {Shirakawa},\ and\ \citenamefont {Ohta}}]{Seki2009}%
  \BibitemOpen
  \bibfield  {author} {\bibinfo {author} {\bibfnamefont {K.}~\bibnamefont
  {Seki}}, \bibinfo {author} {\bibfnamefont {T.}~\bibnamefont {Shirakawa}}, \
  and\ \bibinfo {author} {\bibfnamefont {Y.}~\bibnamefont {Ohta}},\ }\href
  {\doibase 10.1103/PhysRevB.79.024303} {\bibfield  {journal} {\bibinfo
  {journal} {Phys. Rev. B}\ }\textbf {\bibinfo {volume} {79}},\ \bibinfo
  {pages} {024303} (\bibinfo {year} {2009})}\BibitemShut {NoStop}%
\bibitem [{\citenamefont {Momoi}\ \emph {et~al.}(2012)\citenamefont {Momoi},
  \citenamefont {Sindzingre},\ and\ \citenamefont {Kubo}}]{Momoi2012}%
  \BibitemOpen
  \bibfield  {author} {\bibinfo {author} {\bibfnamefont {T.}~\bibnamefont
  {Momoi}}, \bibinfo {author} {\bibfnamefont {P.}~\bibnamefont {Sindzingre}}, \
  and\ \bibinfo {author} {\bibfnamefont {K.}~\bibnamefont {Kubo}},\ }\href
  {\doibase 10.1103/PhysRevLett.108.057206} {\bibfield  {journal} {\bibinfo
  {journal} {Phys. Rev. Lett.}\ }\textbf {\bibinfo {volume} {108}},\ \bibinfo
  {pages} {057206} (\bibinfo {year} {2012})}\BibitemShut {NoStop}%
\bibitem [{\citenamefont {Moroni}\ and\ \citenamefont
  {Boninsegni}(2019)}]{Moroni2019}%
  \BibitemOpen
  \bibfield  {author} {\bibinfo {author} {\bibfnamefont {S.}~\bibnamefont
  {Moroni}}\ and\ \bibinfo {author} {\bibfnamefont {M.}~\bibnamefont
  {Boninsegni}},\ }\href {\doibase 10.1103/PhysRevB.99.195441} {\bibfield
  {journal} {\bibinfo  {journal} {Phys. Rev. B}\ }\textbf {\bibinfo {volume}
  {99}},\ \bibinfo {pages} {195441} (\bibinfo {year} {2019})}\BibitemShut
  {NoStop}%
\bibitem [{\citenamefont {LiMing}\ \emph {et~al.}(2000)\citenamefont {LiMing},
  \citenamefont {Misguich}, \citenamefont {Sindzingre},\ and\ \citenamefont
  {Lhuillier}}]{LiMing2000}%
  \BibitemOpen
  \bibfield  {author} {\bibinfo {author} {\bibfnamefont {W.}~\bibnamefont
  {LiMing}}, \bibinfo {author} {\bibfnamefont {G.}~\bibnamefont {Misguich}},
  \bibinfo {author} {\bibfnamefont {P.}~\bibnamefont {Sindzingre}}, \ and\
  \bibinfo {author} {\bibfnamefont {C.}~\bibnamefont {Lhuillier}},\ }\href
  {\doibase 10.1103/PhysRevB.62.6372} {\bibfield  {journal} {\bibinfo
  {journal} {Phys. Rev. B}\ }\textbf {\bibinfo {volume} {62}},\ \bibinfo
  {pages} {6372} (\bibinfo {year} {2000})}\BibitemShut {NoStop}%
\bibitem [{\citenamefont {Kato}(2014)}]{Kato2014}%
  \BibitemOpen
  \bibfield  {author} {\bibinfo {author} {\bibfnamefont {R.}~\bibnamefont
  {Kato}},\ }\href {\doibase 10.1246/bcsj.20130290} {\bibfield  {journal}
  {\bibinfo  {journal} {Bulletin of the Chemical Society of Japan}\ }\textbf
  {\bibinfo {volume} {87}},\ \bibinfo {pages} {355} (\bibinfo {year}
  {2014})}\BibitemShut {NoStop}%
\bibitem [{\citenamefont {Rawl}\ \emph {et~al.}(2017)\citenamefont {Rawl},
  \citenamefont {Ge}, \citenamefont {Agrawal}, \citenamefont {Kamiya},
  \citenamefont {Dela~Cruz}, \citenamefont {Butch}, \citenamefont {Sun},
  \citenamefont {Lee}, \citenamefont {Choi}, \citenamefont {Oitmaa},
  \citenamefont {Batista}, \citenamefont {Mourigal}, \citenamefont {Zhou},\
  and\ \citenamefont {Ma}}]{Rawl2017}%
  \BibitemOpen
  \bibfield  {author} {\bibinfo {author} {\bibfnamefont {R.}~\bibnamefont
  {Rawl}}, \bibinfo {author} {\bibfnamefont {L.}~\bibnamefont {Ge}}, \bibinfo
  {author} {\bibfnamefont {H.}~\bibnamefont {Agrawal}}, \bibinfo {author}
  {\bibfnamefont {Y.}~\bibnamefont {Kamiya}}, \bibinfo {author} {\bibfnamefont
  {C.~R.}\ \bibnamefont {Dela~Cruz}}, \bibinfo {author} {\bibfnamefont {N.~P.}\
  \bibnamefont {Butch}}, \bibinfo {author} {\bibfnamefont {X.~F.}\ \bibnamefont
  {Sun}}, \bibinfo {author} {\bibfnamefont {M.}~\bibnamefont {Lee}}, \bibinfo
  {author} {\bibfnamefont {E.~S.}\ \bibnamefont {Choi}}, \bibinfo {author}
  {\bibfnamefont {J.}~\bibnamefont {Oitmaa}}, \bibinfo {author} {\bibfnamefont
  {C.~D.}\ \bibnamefont {Batista}}, \bibinfo {author} {\bibfnamefont
  {M.}~\bibnamefont {Mourigal}}, \bibinfo {author} {\bibfnamefont {H.~D.}\
  \bibnamefont {Zhou}}, \ and\ \bibinfo {author} {\bibfnamefont
  {J.}~\bibnamefont {Ma}},\ }\href {\doibase 10.1103/PhysRevB.95.060412}
  {\bibfield  {journal} {\bibinfo  {journal} {Phys. Rev. B}\ }\textbf {\bibinfo
  {volume} {95}},\ \bibinfo {pages} {060412} (\bibinfo {year}
  {2017})}\BibitemShut {NoStop}%
\bibitem [{\citenamefont {Cui}\ \emph {et~al.}(2018)\citenamefont {Cui},
  \citenamefont {Dai}, \citenamefont {Zhou}, \citenamefont {Wang},
  \citenamefont {Li}, \citenamefont {Song}, \citenamefont {Wang}, \citenamefont
  {Ma}, \citenamefont {Zhang}, \citenamefont {Li}, \citenamefont {Luke},
  \citenamefont {Normand}, \citenamefont {Xiang},\ and\ \citenamefont
  {Yu}}]{Cui2018}%
  \BibitemOpen
  \bibfield  {author} {\bibinfo {author} {\bibfnamefont {Y.}~\bibnamefont
  {Cui}}, \bibinfo {author} {\bibfnamefont {J.}~\bibnamefont {Dai}}, \bibinfo
  {author} {\bibfnamefont {P.}~\bibnamefont {Zhou}}, \bibinfo {author}
  {\bibfnamefont {P.~S.}\ \bibnamefont {Wang}}, \bibinfo {author}
  {\bibfnamefont {T.~R.}\ \bibnamefont {Li}}, \bibinfo {author} {\bibfnamefont
  {W.~H.}\ \bibnamefont {Song}}, \bibinfo {author} {\bibfnamefont {J.~C.}\
  \bibnamefont {Wang}}, \bibinfo {author} {\bibfnamefont {L.}~\bibnamefont
  {Ma}}, \bibinfo {author} {\bibfnamefont {Z.}~\bibnamefont {Zhang}}, \bibinfo
  {author} {\bibfnamefont {S.~Y.}\ \bibnamefont {Li}}, \bibinfo {author}
  {\bibfnamefont {G.~M.}\ \bibnamefont {Luke}}, \bibinfo {author}
  {\bibfnamefont {B.}~\bibnamefont {Normand}}, \bibinfo {author} {\bibfnamefont
  {T.}~\bibnamefont {Xiang}}, \ and\ \bibinfo {author} {\bibfnamefont
  {W.}~\bibnamefont {Yu}},\ }\href {\doibase 10.1103/PhysRevMaterials.2.044403}
  {\bibfield  {journal} {\bibinfo  {journal} {Phys. Rev. Materials}\ }\textbf
  {\bibinfo {volume} {2}},\ \bibinfo {pages} {044403} (\bibinfo {year}
  {2018})}\BibitemShut {NoStop}%
\bibitem [{\citenamefont {Prelov\ifmmode~\check{s}\else \v{s}\fi{}ek}\ \emph
  {et~al.}(2020)\citenamefont {Prelov\ifmmode~\check{s}\else \v{s}\fi{}ek},
  \citenamefont {Morita}, \citenamefont {Tohyama},\ and\ \citenamefont
  {Herbrych}}]{prelovsek2019vanishing}%
  \BibitemOpen
  \bibfield  {author} {\bibinfo {author} {\bibfnamefont {P.}~\bibnamefont
  {Prelov\ifmmode~\check{s}\else \v{s}\fi{}ek}}, \bibinfo {author}
  {\bibfnamefont {K.}~\bibnamefont {Morita}}, \bibinfo {author} {\bibfnamefont
  {T.}~\bibnamefont {Tohyama}}, \ and\ \bibinfo {author} {\bibfnamefont
  {J.}~\bibnamefont {Herbrych}},\ }\href {\doibase
  10.1103/PhysRevResearch.2.023024} {\bibfield  {journal} {\bibinfo  {journal}
  {Phys. Rev. Research}\ }\textbf {\bibinfo {volume} {2}},\ \bibinfo {pages}
  {023024} (\bibinfo {year} {2020})}\BibitemShut {NoStop}%
\bibitem [{\citenamefont {Morita}\ and\ \citenamefont
  {Tohyama}(2020)}]{Morita2019}%
  \BibitemOpen
  \bibfield  {author} {\bibinfo {author} {\bibfnamefont {K.}~\bibnamefont
  {Morita}}\ and\ \bibinfo {author} {\bibfnamefont {T.}~\bibnamefont
  {Tohyama}},\ }\href {\doibase 10.1103/PhysRevResearch.2.013205} {\bibfield
  {journal} {\bibinfo  {journal} {Phys. Rev. Research}\ }\textbf {\bibinfo
  {volume} {2}},\ \bibinfo {pages} {013205} (\bibinfo {year}
  {2020})}\BibitemShut {NoStop}%
\bibitem [{\citenamefont {Jakli\ifmmode~\check{c}\else \v{c}\fi{}}\ and\
  \citenamefont {Prelov\ifmmode~\check{s}\else
  \v{s}\fi{}ek}(1994)}]{Jaklic1994}%
  \BibitemOpen
  \bibfield  {author} {\bibinfo {author} {\bibfnamefont {J.}~\bibnamefont
  {Jakli\ifmmode~\check{c}\else \v{c}\fi{}}}\ and\ \bibinfo {author}
  {\bibfnamefont {P.}~\bibnamefont {Prelov\ifmmode~\check{s}\else
  \v{s}\fi{}ek}},\ }\href {\doibase 10.1103/PhysRevB.49.5065} {\bibfield
  {journal} {\bibinfo  {journal} {Phys. Rev. B}\ }\textbf {\bibinfo {volume}
  {49}},\ \bibinfo {pages} {5065} (\bibinfo {year} {1994})}\BibitemShut
  {NoStop}%
\bibitem [{\citenamefont {Jakli\v{c}}\ and\ \citenamefont
  {Prelov\v{s}ek}(2000)}]{Jaklic2000}%
  \BibitemOpen
  \bibfield  {author} {\bibinfo {author} {\bibfnamefont {J.}~\bibnamefont
  {Jakli\v{c}}}\ and\ \bibinfo {author} {\bibfnamefont {P.}~\bibnamefont
  {Prelov\v{s}ek}},\ }\href {http://dx.doi.org/10.1080/000187300243381}
  {\bibfield  {journal} {\bibinfo  {journal} {Adv. Phys.}\ }\textbf {\bibinfo
  {volume} {49}},\ \bibinfo {pages} {1} (\bibinfo {year} {2000})}\BibitemShut
  {NoStop}%
\bibitem [{\citenamefont {Prelov{\v{s}}ek}\ and\ \citenamefont
  {Bon{\v{c}}a}(2013)}]{Prelovsek}%
  \BibitemOpen
  \bibfield  {author} {\bibinfo {author} {\bibfnamefont {P.}~\bibnamefont
  {Prelov{\v{s}}ek}}\ and\ \bibinfo {author} {\bibfnamefont {J.}~\bibnamefont
  {Bon{\v{c}}a}},\ }\enquote {\bibinfo {title} {Ground state and finite
  temperature lanczos methods},}\ in\ \href {\doibase
  10.1007/978-3-642-35106-8_1} {\emph {\bibinfo {booktitle} {Strongly
  Correlated Systems: Numerical Methods}}},\ \bibinfo {editor} {edited by\
  \bibinfo {editor} {\bibfnamefont {A.}~\bibnamefont {Avella}}\ and\ \bibinfo
  {editor} {\bibfnamefont {F.}~\bibnamefont {Mancini}}}\ (\bibinfo  {publisher}
  {Springer Berlin Heidelberg},\ \bibinfo {address} {Berlin, Heidelberg},\
  \bibinfo {year} {2013})\ pp.\ \bibinfo {pages} {1--30}\BibitemShut {NoStop}%
\bibitem [{\citenamefont {Iitaka}\ and\ \citenamefont
  {Ebisuzaki}(2004)}]{Iitaka2004}%
  \BibitemOpen
  \bibfield  {author} {\bibinfo {author} {\bibfnamefont {T.}~\bibnamefont
  {Iitaka}}\ and\ \bibinfo {author} {\bibfnamefont {T.}~\bibnamefont
  {Ebisuzaki}},\ }\href {\doibase 10.1103/PhysRevE.69.057701} {\bibfield
  {journal} {\bibinfo  {journal} {Phys. Rev. E}\ }\textbf {\bibinfo {volume}
  {69}},\ \bibinfo {pages} {057701} (\bibinfo {year} {2004})}\BibitemShut
  {NoStop}%
\bibitem [{\citenamefont {Wei\ss{}e}\ \emph {et~al.}(2006)\citenamefont
  {Wei\ss{}e}, \citenamefont {Wellein}, \citenamefont {Alvermann},\ and\
  \citenamefont {Fehske}}]{Weisse2006}%
  \BibitemOpen
  \bibfield  {author} {\bibinfo {author} {\bibfnamefont {A.}~\bibnamefont
  {Wei\ss{}e}}, \bibinfo {author} {\bibfnamefont {G.}~\bibnamefont {Wellein}},
  \bibinfo {author} {\bibfnamefont {A.}~\bibnamefont {Alvermann}}, \ and\
  \bibinfo {author} {\bibfnamefont {H.}~\bibnamefont {Fehske}},\ }\href
  {\doibase 10.1103/RevModPhys.78.275} {\bibfield  {journal} {\bibinfo
  {journal} {Rev. Mod. Phys.}\ }\textbf {\bibinfo {volume} {78}},\ \bibinfo
  {pages} {275} (\bibinfo {year} {2006})}\BibitemShut {NoStop}%
\bibitem [{\citenamefont {Drabold}\ and\ \citenamefont
  {Sankey}(1993)}]{Drabold1993}%
  \BibitemOpen
  \bibfield  {author} {\bibinfo {author} {\bibfnamefont {D.~A.}\ \bibnamefont
  {Drabold}}\ and\ \bibinfo {author} {\bibfnamefont {O.~F.}\ \bibnamefont
  {Sankey}},\ }\href {\doibase 10.1103/PhysRevLett.70.3631} {\bibfield
  {journal} {\bibinfo  {journal} {Phys. Rev. Lett.}\ }\textbf {\bibinfo
  {volume} {70}},\ \bibinfo {pages} {3631} (\bibinfo {year}
  {1993})}\BibitemShut {NoStop}%
\bibitem [{\citenamefont {Chatelin}(1988)}]{Chatelin}%
  \BibitemOpen
  \bibfield  {author} {\bibinfo {author} {\bibfnamefont {F.}~\bibnamefont
  {Chatelin}},\ }\href@noop {} {\emph {\bibinfo {title} {{Valeurs propres de
  matrices}}}}\ (\bibinfo  {publisher} {Masson},\ \bibinfo {address} {Paris},\
  \bibinfo {year} {1988})\ \bibinfo {note} {{translation by M. Iri and Y. Iri
  (Springer, Tokyo, 2003)}}\BibitemShut {NoStop}%
\bibitem [{\citenamefont {Shirakawa}\ and\ \citenamefont
  {Yunoki}(2014)}]{Shirakawa2014}%
  \BibitemOpen
  \bibfield  {author} {\bibinfo {author} {\bibfnamefont {T.}~\bibnamefont
  {Shirakawa}}\ and\ \bibinfo {author} {\bibfnamefont {S.}~\bibnamefont
  {Yunoki}},\ }\href {\doibase 10.1103/PhysRevB.90.195109} {\bibfield
  {journal} {\bibinfo  {journal} {Phys. Rev. B}\ }\textbf {\bibinfo {volume}
  {90}},\ \bibinfo {pages} {195109} (\bibinfo {year} {2014})}\BibitemShut
  {NoStop}%
\bibitem [{\citenamefont {Allerdt}\ \emph {et~al.}(2015)\citenamefont
  {Allerdt}, \citenamefont {B\"usser}, \citenamefont {Martins},\ and\
  \citenamefont {Feiguin}}]{Allerdt2015}%
  \BibitemOpen
  \bibfield  {author} {\bibinfo {author} {\bibfnamefont {A.}~\bibnamefont
  {Allerdt}}, \bibinfo {author} {\bibfnamefont {C.~A.}\ \bibnamefont
  {B\"usser}}, \bibinfo {author} {\bibfnamefont {G.~B.}\ \bibnamefont
  {Martins}}, \ and\ \bibinfo {author} {\bibfnamefont {A.~E.}\ \bibnamefont
  {Feiguin}},\ }\href {\doibase 10.1103/PhysRevB.91.085101} {\bibfield
  {journal} {\bibinfo  {journal} {Phys. Rev. B}\ }\textbf {\bibinfo {volume}
  {91}},\ \bibinfo {pages} {085101} (\bibinfo {year} {2015})}\BibitemShut
  {NoStop}%
\bibitem [{\citenamefont {Seki}\ \emph {et~al.}(2018)\citenamefont {Seki},
  \citenamefont {Shirakawa},\ and\ \citenamefont {Yunoki}}]{Seki2018}%
  \BibitemOpen
  \bibfield  {author} {\bibinfo {author} {\bibfnamefont {K.}~\bibnamefont
  {Seki}}, \bibinfo {author} {\bibfnamefont {T.}~\bibnamefont {Shirakawa}}, \
  and\ \bibinfo {author} {\bibfnamefont {S.}~\bibnamefont {Yunoki}},\ }\href
  {\doibase 10.1103/PhysRevB.98.205114} {\bibfield  {journal} {\bibinfo
  {journal} {Phys. Rev. B}\ }\textbf {\bibinfo {volume} {98}},\ \bibinfo
  {pages} {205114} (\bibinfo {year} {2018})}\BibitemShut {NoStop}%
\bibitem [{\citenamefont {Aichhorn}\ \emph {et~al.}(2003)\citenamefont
  {Aichhorn}, \citenamefont {Daghofer}, \citenamefont {Evertz},\ and\
  \citenamefont {von~der Linden}}]{Aichhorn2003}%
  \BibitemOpen
  \bibfield  {author} {\bibinfo {author} {\bibfnamefont {M.}~\bibnamefont
  {Aichhorn}}, \bibinfo {author} {\bibfnamefont {M.}~\bibnamefont {Daghofer}},
  \bibinfo {author} {\bibfnamefont {H.~G.}\ \bibnamefont {Evertz}}, \ and\
  \bibinfo {author} {\bibfnamefont {W.}~\bibnamefont {von~der Linden}},\ }\href
  {\doibase 10.1103/PhysRevB.67.161103} {\bibfield  {journal} {\bibinfo
  {journal} {Phys. Rev. B}\ }\textbf {\bibinfo {volume} {67}},\ \bibinfo
  {pages} {161103} (\bibinfo {year} {2003})}\BibitemShut {NoStop}%
\bibitem [{\citenamefont {Schnack}\ \emph {et~al.}(2018)\citenamefont
  {Schnack}, \citenamefont {Schulenburg},\ and\ \citenamefont
  {Richter}}]{Schnack2018}%
  \BibitemOpen
  \bibfield  {author} {\bibinfo {author} {\bibfnamefont {J.}~\bibnamefont
  {Schnack}}, \bibinfo {author} {\bibfnamefont {J.}~\bibnamefont
  {Schulenburg}}, \ and\ \bibinfo {author} {\bibfnamefont {J.}~\bibnamefont
  {Richter}},\ }\href {\doibase 10.1103/PhysRevB.98.094423} {\bibfield
  {journal} {\bibinfo  {journal} {Phys. Rev. B}\ }\textbf {\bibinfo {volume}
  {98}},\ \bibinfo {pages} {094423} (\bibinfo {year} {2018})}\BibitemShut
  {NoStop}%
\bibitem [{\citenamefont {Schnack}\ \emph {et~al.}(2020)\citenamefont
  {Schnack}, \citenamefont {Richter},\ and\ \citenamefont
  {Steinigeweg}}]{Schnack2020}%
  \BibitemOpen
  \bibfield  {author} {\bibinfo {author} {\bibfnamefont {J.}~\bibnamefont
  {Schnack}}, \bibinfo {author} {\bibfnamefont {J.}~\bibnamefont {Richter}}, \
  and\ \bibinfo {author} {\bibfnamefont {R.}~\bibnamefont {Steinigeweg}},\
  }\href {\doibase 10.1103/PhysRevResearch.2.013186} {\bibfield  {journal}
  {\bibinfo  {journal} {Phys. Rev. Research}\ }\textbf {\bibinfo {volume}
  {2}},\ \bibinfo {pages} {013186} (\bibinfo {year} {2020})}\BibitemShut
  {NoStop}%
\bibitem [{\citenamefont {Sugiura}\ and\ \citenamefont
  {Shimizu}(2013)}]{Sugiura2013}%
  \BibitemOpen
  \bibfield  {author} {\bibinfo {author} {\bibfnamefont {S.}~\bibnamefont
  {Sugiura}}\ and\ \bibinfo {author} {\bibfnamefont {A.}~\bibnamefont
  {Shimizu}},\ }\href {\doibase 10.1103/PhysRevLett.111.010401} {\bibfield
  {journal} {\bibinfo  {journal} {Phys. Rev. Lett.}\ }\textbf {\bibinfo
  {volume} {111}},\ \bibinfo {pages} {010401} (\bibinfo {year}
  {2013})}\BibitemShut {NoStop}%
\bibitem [{\citenamefont {Nishida}\ \emph {et~al.}(2020)\citenamefont
  {Nishida}, \citenamefont {Fujiuchi}, \citenamefont {Sugimoto},\ and\
  \citenamefont {Ohta}}]{nishida2019typicalitybased}%
  \BibitemOpen
  \bibfield  {author} {\bibinfo {author} {\bibfnamefont {H.}~\bibnamefont
  {Nishida}}, \bibinfo {author} {\bibfnamefont {R.}~\bibnamefont {Fujiuchi}},
  \bibinfo {author} {\bibfnamefont {K.}~\bibnamefont {Sugimoto}}, \ and\
  \bibinfo {author} {\bibfnamefont {Y.}~\bibnamefont {Ohta}},\ }\href {\doibase
  10.7566/JPSJ.89.023702} {\bibfield  {journal} {\bibinfo  {journal} {J. Phys.
  Soc. Jpn.}\ }\textbf {\bibinfo {volume} {89}},\ \bibinfo {pages} {023702}
  (\bibinfo {year} {2020})}\BibitemShut {NoStop}%
\bibitem [{\citenamefont {Tal-Ezer}\ and\ \citenamefont
  {Kosloff}(1984)}]{TalEzer1984}%
  \BibitemOpen
  \bibfield  {author} {\bibinfo {author} {\bibfnamefont {H.}~\bibnamefont
  {Tal-Ezer}}\ and\ \bibinfo {author} {\bibfnamefont {R.}~\bibnamefont
  {Kosloff}},\ }\href {\doibase 10.1063/1.448136} {\bibfield  {journal}
  {\bibinfo  {journal} {The Journal of Chemical Physics}\ }\textbf {\bibinfo
  {volume} {81}},\ \bibinfo {pages} {3967} (\bibinfo {year}
  {1984})}\BibitemShut {NoStop}%
\bibitem [{\citenamefont {Wang}(1994)}]{Wang1994}%
  \BibitemOpen
  \bibfield  {author} {\bibinfo {author} {\bibfnamefont {L.-W.}\ \bibnamefont
  {Wang}},\ }\href {\doibase 10.1103/PhysRevB.49.10154} {\bibfield  {journal}
  {\bibinfo  {journal} {Phys. Rev. B}\ }\textbf {\bibinfo {volume} {49}},\
  \bibinfo {pages} {10154} (\bibinfo {year} {1994})}\BibitemShut {NoStop}%
\bibitem [{\citenamefont {Wang}\ and\ \citenamefont
  {Zunger}(1994)}]{Wang1994PRL}%
  \BibitemOpen
  \bibfield  {author} {\bibinfo {author} {\bibfnamefont {L.-W.}\ \bibnamefont
  {Wang}}\ and\ \bibinfo {author} {\bibfnamefont {A.}~\bibnamefont {Zunger}},\
  }\href {\doibase 10.1103/PhysRevLett.73.1039} {\bibfield  {journal} {\bibinfo
   {journal} {Phys. Rev. Lett.}\ }\textbf {\bibinfo {volume} {73}},\ \bibinfo
  {pages} {1039} (\bibinfo {year} {1994})}\BibitemShut {NoStop}%
\bibitem [{\citenamefont {Iitaka}(1994)}]{Iitaka1994}%
  \BibitemOpen
  \bibfield  {author} {\bibinfo {author} {\bibfnamefont {T.}~\bibnamefont
  {Iitaka}},\ }\href {\doibase 10.1103/PhysRevE.49.4684} {\bibfield  {journal}
  {\bibinfo  {journal} {Phys. Rev. E}\ }\textbf {\bibinfo {volume} {49}},\
  \bibinfo {pages} {4684} (\bibinfo {year} {1994})}\BibitemShut {NoStop}%
\bibitem [{\citenamefont {Vijay}\ and\ \citenamefont
  {Metiu}(2002)}]{Vijay2002}%
  \BibitemOpen
  \bibfield  {author} {\bibinfo {author} {\bibfnamefont {A.}~\bibnamefont
  {Vijay}}\ and\ \bibinfo {author} {\bibfnamefont {H.}~\bibnamefont {Metiu}},\
  }\href {\doibase 10.1063/1.1425824} {\bibfield  {journal} {\bibinfo
  {journal} {The Journal of Chemical Physics}\ }\textbf {\bibinfo {volume}
  {116}},\ \bibinfo {pages} {60} (\bibinfo {year} {2002})}\BibitemShut
  {NoStop}%
\bibitem [{\citenamefont {Iitaka}\ and\ \citenamefont
  {Ebisuzaki}(2003)}]{Iitaka2003}%
  \BibitemOpen
  \bibfield  {author} {\bibinfo {author} {\bibfnamefont {T.}~\bibnamefont
  {Iitaka}}\ and\ \bibinfo {author} {\bibfnamefont {T.}~\bibnamefont
  {Ebisuzaki}},\ }\href {\doibase 10.1103/PhysRevLett.90.047203} {\bibfield
  {journal} {\bibinfo  {journal} {Phys. Rev. Lett.}\ }\textbf {\bibinfo
  {volume} {90}},\ \bibinfo {pages} {047203} (\bibinfo {year}
  {2003})}\BibitemShut {NoStop}%
\bibitem [{\citenamefont {Machida}\ \emph {et~al.}(2005)\citenamefont
  {Machida}, \citenamefont {Iitaka},\ and\ \citenamefont
  {Miyashita}}]{Machida2005}%
  \BibitemOpen
  \bibfield  {author} {\bibinfo {author} {\bibfnamefont {M.}~\bibnamefont
  {Machida}}, \bibinfo {author} {\bibfnamefont {T.}~\bibnamefont {Iitaka}}, \
  and\ \bibinfo {author} {\bibfnamefont {S.}~\bibnamefont {Miyashita}},\ }\href
  {\doibase 10.1143/JPSJS.74S.107} {\bibfield  {journal} {\bibinfo  {journal}
  {J. Phys. Soc. Jpn.}\ }\textbf {\bibinfo {volume} {74}},\ \bibinfo {pages}
  {107} (\bibinfo {year} {2005})}\BibitemShut {NoStop}%
\bibitem [{\citenamefont {Wei{\ss}e}\ and\ \citenamefont
  {Fehske}(2008)}]{Weisse_book}%
  \BibitemOpen
  \bibfield  {author} {\bibinfo {author} {\bibfnamefont {A.}~\bibnamefont
  {Wei{\ss}e}}\ and\ \bibinfo {author} {\bibfnamefont {H.}~\bibnamefont
  {Fehske}},\ }\href@noop {} {\emph {\bibinfo {title} {Computational
  Many-Particle Physics}}},\ edited by\ \bibinfo {editor} {\bibfnamefont
  {H.}~\bibnamefont {Fehske}}, \bibinfo {editor} {\bibfnamefont
  {R.}~\bibnamefont {Schneider}}, \ and\ \bibinfo {editor} {\bibfnamefont
  {A.}~\bibnamefont {Wei{\ss}e}},\ Lect. Notes Phys. 739\ (\bibinfo
  {publisher} {Springer},\ \bibinfo {address} {Berlin Heidelberg},\ \bibinfo
  {year} {2008})\ Chap.~\bibinfo {chapter} {19}, pp.\ \bibinfo {pages}
  {545--577}\BibitemShut {NoStop}%
\bibitem [{\citenamefont {Seki}\ \emph {et~al.}(2019)\citenamefont {Seki},
  \citenamefont {Otsuka}, \citenamefont {Yunoki},\ and\ \citenamefont
  {Sorella}}]{Seki2019Dirac}%
  \BibitemOpen
  \bibfield  {author} {\bibinfo {author} {\bibfnamefont {K.}~\bibnamefont
  {Seki}}, \bibinfo {author} {\bibfnamefont {Y.}~\bibnamefont {Otsuka}},
  \bibinfo {author} {\bibfnamefont {S.}~\bibnamefont {Yunoki}}, \ and\ \bibinfo
  {author} {\bibfnamefont {S.}~\bibnamefont {Sorella}},\ }\href {\doibase
  10.1103/PhysRevB.99.125145} {\bibfield  {journal} {\bibinfo  {journal} {Phys.
  Rev. B}\ }\textbf {\bibinfo {volume} {99}},\ \bibinfo {pages} {125145}
  (\bibinfo {year} {2019})}\BibitemShut {NoStop}%
\bibitem [{\citenamefont {Deutsch}(1991)}]{Deutsch1991}%
  \BibitemOpen
  \bibfield  {author} {\bibinfo {author} {\bibfnamefont {J.~M.}\ \bibnamefont
  {Deutsch}},\ }\href {\doibase 10.1103/PhysRevA.43.2046} {\bibfield  {journal}
  {\bibinfo  {journal} {Phys. Rev. A}\ }\textbf {\bibinfo {volume} {43}},\
  \bibinfo {pages} {2046} (\bibinfo {year} {1991})}\BibitemShut {NoStop}%
\bibitem [{\citenamefont {Srednicki}(1994)}]{Srednicki1994}%
  \BibitemOpen
  \bibfield  {author} {\bibinfo {author} {\bibfnamefont {M.}~\bibnamefont
  {Srednicki}},\ }\href {\doibase 10.1103/PhysRevE.50.888} {\bibfield
  {journal} {\bibinfo  {journal} {Phys. Rev. E}\ }\textbf {\bibinfo {volume}
  {50}},\ \bibinfo {pages} {888} (\bibinfo {year} {1994})}\BibitemShut
  {NoStop}%
\bibitem [{\citenamefont {Rousochatzakis}\ \emph {et~al.}(2019)\citenamefont
  {Rousochatzakis}, \citenamefont {Kourtis}, \citenamefont {Knolle},
  \citenamefont {Moessner},\ and\ \citenamefont
  {Perkins}}]{Rousochatzakis2019}%
  \BibitemOpen
  \bibfield  {author} {\bibinfo {author} {\bibfnamefont {I.}~\bibnamefont
  {Rousochatzakis}}, \bibinfo {author} {\bibfnamefont {S.}~\bibnamefont
  {Kourtis}}, \bibinfo {author} {\bibfnamefont {J.}~\bibnamefont {Knolle}},
  \bibinfo {author} {\bibfnamefont {R.}~\bibnamefont {Moessner}}, \ and\
  \bibinfo {author} {\bibfnamefont {N.~B.}\ \bibnamefont {Perkins}},\ }\href
  {\doibase 10.1103/PhysRevB.100.045117} {\bibfield  {journal} {\bibinfo
  {journal} {Phys. Rev. B}\ }\textbf {\bibinfo {volume} {100}},\ \bibinfo
  {pages} {045117} (\bibinfo {year} {2019})}\BibitemShut {NoStop}%
\bibitem [{\citenamefont {Prelov\ifmmode~\check{s}\else \v{s}\fi{}ek}\ and\
  \citenamefont {Kokalj}(2020)}]{Prelovsek2019}%
  \BibitemOpen
  \bibfield  {author} {\bibinfo {author} {\bibfnamefont {P.}~\bibnamefont
  {Prelov\ifmmode~\check{s}\else \v{s}\fi{}ek}}\ and\ \bibinfo {author}
  {\bibfnamefont {J.}~\bibnamefont {Kokalj}},\ }\href {\doibase
  10.1103/PhysRevB.101.075105} {\bibfield  {journal} {\bibinfo  {journal}
  {Phys. Rev. B}\ }\textbf {\bibinfo {volume} {101}},\ \bibinfo {pages}
  {075105} (\bibinfo {year} {2020})}\BibitemShut {NoStop}%
\bibitem [{SM()}]{SM}%
  \BibitemOpen
  \href@noop {} {}\bibinfo {note} {See~\hyperlink{stitle}{Supplemental
  Material} for details on the finite-size effects.}\BibitemShut {Stop}%
\bibitem [{\citenamefont {Prelov\ifmmode~\check{s}\else \v{s}\fi{}ek}\ and\
  \citenamefont {Kokalj}(2018)}]{Prelovsek2018}%
  \BibitemOpen
  \bibfield  {author} {\bibinfo {author} {\bibfnamefont {P.}~\bibnamefont
  {Prelov\ifmmode~\check{s}\else \v{s}\fi{}ek}}\ and\ \bibinfo {author}
  {\bibfnamefont {J.}~\bibnamefont {Kokalj}},\ }\href {\doibase
  10.1103/PhysRevB.98.035107} {\bibfield  {journal} {\bibinfo  {journal} {Phys.
  Rev. B}\ }\textbf {\bibinfo {volume} {98}},\ \bibinfo {pages} {035107}
  (\bibinfo {year} {2018})}\BibitemShut {NoStop}%
\bibitem [{\citenamefont {Chen}\ \emph {et~al.}(2018)\citenamefont {Chen},
  \citenamefont {Chen}, \citenamefont {Chen}, \citenamefont {Li},\ and\
  \citenamefont {Weichselbaum}}]{Chen2018}%
  \BibitemOpen
  \bibfield  {author} {\bibinfo {author} {\bibfnamefont {B.-B.}\ \bibnamefont
  {Chen}}, \bibinfo {author} {\bibfnamefont {L.}~\bibnamefont {Chen}}, \bibinfo
  {author} {\bibfnamefont {Z.}~\bibnamefont {Chen}}, \bibinfo {author}
  {\bibfnamefont {W.}~\bibnamefont {Li}}, \ and\ \bibinfo {author}
  {\bibfnamefont {A.}~\bibnamefont {Weichselbaum}},\ }\href {\doibase
  10.1103/PhysRevX.8.031082} {\bibfield  {journal} {\bibinfo  {journal} {Phys.
  Rev. X}\ }\textbf {\bibinfo {volume} {8}},\ \bibinfo {pages} {031082}
  (\bibinfo {year} {2018})}\BibitemShut {NoStop}%
\bibitem [{\citenamefont {Chen}\ \emph {et~al.}(2019)\citenamefont {Chen},
  \citenamefont {Qu}, \citenamefont {Li}, \citenamefont {Chen}, \citenamefont
  {Gong}, \citenamefont {von Delft}, \citenamefont {Weichselbaum},\ and\
  \citenamefont {Li}}]{Chen2019TRG}%
  \BibitemOpen
  \bibfield  {author} {\bibinfo {author} {\bibfnamefont {L.}~\bibnamefont
  {Chen}}, \bibinfo {author} {\bibfnamefont {D.-W.}\ \bibnamefont {Qu}},
  \bibinfo {author} {\bibfnamefont {H.}~\bibnamefont {Li}}, \bibinfo {author}
  {\bibfnamefont {B.-B.}\ \bibnamefont {Chen}}, \bibinfo {author}
  {\bibfnamefont {S.-S.}\ \bibnamefont {Gong}}, \bibinfo {author}
  {\bibfnamefont {J.}~\bibnamefont {von Delft}}, \bibinfo {author}
  {\bibfnamefont {A.}~\bibnamefont {Weichselbaum}}, \ and\ \bibinfo {author}
  {\bibfnamefont {W.}~\bibnamefont {Li}},\ }\href {\doibase
  10.1103/PhysRevB.99.140404} {\bibfield  {journal} {\bibinfo  {journal} {Phys.
  Rev. B}\ }\textbf {\bibinfo {volume} {99}},\ \bibinfo {pages} {140404}
  (\bibinfo {year} {2019})}\BibitemShut {NoStop}%
\bibitem [{\citenamefont {Bernu}\ and\ \citenamefont
  {Misguich}(2001)}]{Bernu2001}%
  \BibitemOpen
  \bibfield  {author} {\bibinfo {author} {\bibfnamefont {B.}~\bibnamefont
  {Bernu}}\ and\ \bibinfo {author} {\bibfnamefont {G.}~\bibnamefont
  {Misguich}},\ }\href {\doibase 10.1103/PhysRevB.63.134409} {\bibfield
  {journal} {\bibinfo  {journal} {Phys. Rev. B}\ }\textbf {\bibinfo {volume}
  {63}},\ \bibinfo {pages} {134409} (\bibinfo {year} {2001})}\BibitemShut
  {NoStop}%
\bibitem [{\citenamefont {Yang}\ \emph {et~al.}(2010)\citenamefont {Yang},
  \citenamefont {L\"auchli}, \citenamefont {Mila},\ and\ \citenamefont
  {Schmidt}}]{Yang2010}%
  \BibitemOpen
  \bibfield  {author} {\bibinfo {author} {\bibfnamefont {H.-Y.}\ \bibnamefont
  {Yang}}, \bibinfo {author} {\bibfnamefont {A.~M.}\ \bibnamefont {L\"auchli}},
  \bibinfo {author} {\bibfnamefont {F.}~\bibnamefont {Mila}}, \ and\ \bibinfo
  {author} {\bibfnamefont {K.~P.}\ \bibnamefont {Schmidt}},\ }\href {\doibase
  10.1103/PhysRevLett.105.267204} {\bibfield  {journal} {\bibinfo  {journal}
  {Phys. Rev. Lett.}\ }\textbf {\bibinfo {volume} {105}},\ \bibinfo {pages}
  {267204} (\bibinfo {year} {2010})}\BibitemShut {NoStop}%
\bibitem [{\citenamefont {Lecheminant}\ \emph {et~al.}(1997)\citenamefont
  {Lecheminant}, \citenamefont {Bernu}, \citenamefont {Lhuillier},
  \citenamefont {Pierre},\ and\ \citenamefont {Sindzingre}}]{Lecheminant1997}%
  \BibitemOpen
  \bibfield  {author} {\bibinfo {author} {\bibfnamefont {P.}~\bibnamefont
  {Lecheminant}}, \bibinfo {author} {\bibfnamefont {B.}~\bibnamefont {Bernu}},
  \bibinfo {author} {\bibfnamefont {C.}~\bibnamefont {Lhuillier}}, \bibinfo
  {author} {\bibfnamefont {L.}~\bibnamefont {Pierre}}, \ and\ \bibinfo {author}
  {\bibfnamefont {P.}~\bibnamefont {Sindzingre}},\ }\href {\doibase
  10.1103/PhysRevB.56.2521} {\bibfield  {journal} {\bibinfo  {journal} {Phys.
  Rev. B}\ }\textbf {\bibinfo {volume} {56}},\ \bibinfo {pages} {2521}
  (\bibinfo {year} {1997})}\BibitemShut {NoStop}%
\bibitem [{\citenamefont {{Waldtmann, C.}}\ \emph {et~al.}(1998)\citenamefont
  {{Waldtmann, C.}}, \citenamefont {{Everts, H.-U.}}, \citenamefont {{Bernu,
  B.}}, \citenamefont {{Lhuillier, C.}}, \citenamefont {{Sindzingre, P.}},
  \citenamefont {{Lecheminant, P.}},\ and\ \citenamefont {{Pierre,
  L.}}}]{Waldtmann1998}%
  \BibitemOpen
  \bibfield  {author} {\bibinfo {author} {\bibnamefont {{Waldtmann, C.}}},
  \bibinfo {author} {\bibnamefont {{Everts, H.-U.}}}, \bibinfo {author}
  {\bibnamefont {{Bernu, B.}}}, \bibinfo {author} {\bibnamefont {{Lhuillier,
  C.}}}, \bibinfo {author} {\bibnamefont {{Sindzingre, P.}}}, \bibinfo {author}
  {\bibnamefont {{Lecheminant, P.}}}, \ and\ \bibinfo {author} {\bibnamefont
  {{Pierre, L.}}},\ }\href {\doibase 10.1007/s100510050274} {\bibfield
  {journal} {\bibinfo  {journal} {Eur. Phys. J. B}\ }\textbf {\bibinfo {volume}
  {2}},\ \bibinfo {pages} {501} (\bibinfo {year} {1998})}\BibitemShut {NoStop}%
\bibitem [{\citenamefont {Shimokawa}\ and\ \citenamefont
  {Kawamura}(2016)}]{Shimokawa2016}%
  \BibitemOpen
  \bibfield  {author} {\bibinfo {author} {\bibfnamefont {T.}~\bibnamefont
  {Shimokawa}}\ and\ \bibinfo {author} {\bibfnamefont {H.}~\bibnamefont
  {Kawamura}},\ }\href {\doibase 10.7566/JPSJ.85.113702} {\bibfield  {journal}
  {\bibinfo  {journal} {J. Phys. Soc. Jpn.}\ }\textbf {\bibinfo {volume}
  {85}},\ \bibinfo {pages} {113702} (\bibinfo {year} {2016})}\BibitemShut
  {NoStop}%
\bibitem [{\citenamefont {Yamaji}\ \emph {et~al.}(2016)\citenamefont {Yamaji},
  \citenamefont {Suzuki}, \citenamefont {Yamada}, \citenamefont {Suga},
  \citenamefont {Kawashima},\ and\ \citenamefont {Imada}}]{Yamaji2016}%
  \BibitemOpen
  \bibfield  {author} {\bibinfo {author} {\bibfnamefont {Y.}~\bibnamefont
  {Yamaji}}, \bibinfo {author} {\bibfnamefont {T.}~\bibnamefont {Suzuki}},
  \bibinfo {author} {\bibfnamefont {T.}~\bibnamefont {Yamada}}, \bibinfo
  {author} {\bibfnamefont {S.-i.}\ \bibnamefont {Suga}}, \bibinfo {author}
  {\bibfnamefont {N.}~\bibnamefont {Kawashima}}, \ and\ \bibinfo {author}
  {\bibfnamefont {M.}~\bibnamefont {Imada}},\ }\href {\doibase
  10.1103/PhysRevB.93.174425} {\bibfield  {journal} {\bibinfo  {journal} {Phys.
  Rev. B}\ }\textbf {\bibinfo {volume} {93}},\ \bibinfo {pages} {174425}
  (\bibinfo {year} {2016})}\BibitemShut {NoStop}%
\bibitem [{\citenamefont {Mermin}\ and\ \citenamefont
  {Wagner}(1966)}]{Mermin1966}%
  \BibitemOpen
  \bibfield  {author} {\bibinfo {author} {\bibfnamefont {N.~D.}\ \bibnamefont
  {Mermin}}\ and\ \bibinfo {author} {\bibfnamefont {H.}~\bibnamefont
  {Wagner}},\ }\href {\doibase 10.1103/PhysRevLett.17.1133} {\bibfield
  {journal} {\bibinfo  {journal} {Phys. Rev. Lett.}\ }\textbf {\bibinfo
  {volume} {17}},\ \bibinfo {pages} {1133} (\bibinfo {year}
  {1966})}\BibitemShut {NoStop}%
\bibitem [{\citenamefont {Kratochvilova}\ \emph {et~al.}(2017)\citenamefont
  {Kratochvilova}, \citenamefont {Hillier}, \citenamefont {Wildes},
  \citenamefont {Wang}, \citenamefont {Cheong},\ and\ \citenamefont
  {Park}}]{Kratochvilova2017}%
  \BibitemOpen
  \bibfield  {author} {\bibinfo {author} {\bibfnamefont {M.}~\bibnamefont
  {Kratochvilova}}, \bibinfo {author} {\bibfnamefont {A.~D.}\ \bibnamefont
  {Hillier}}, \bibinfo {author} {\bibfnamefont {A.~R.}\ \bibnamefont {Wildes}},
  \bibinfo {author} {\bibfnamefont {L.}~\bibnamefont {Wang}}, \bibinfo {author}
  {\bibfnamefont {S.-W.}\ \bibnamefont {Cheong}}, \ and\ \bibinfo {author}
  {\bibfnamefont {J.-G.}\ \bibnamefont {Park}},\ }\href {\doibase
  10.1038/s41535-017-0048-1} {\bibfield  {journal} {\bibinfo  {journal} {npj
  Quantum Materials}\ }\textbf {\bibinfo {volume} {2}},\ \bibinfo {pages} {42}
  (\bibinfo {year} {2017})}\BibitemShut {NoStop}%
\bibitem [{\citenamefont {Ribak}\ \emph {et~al.}(2017)\citenamefont {Ribak},
  \citenamefont {Silber}, \citenamefont {Baines}, \citenamefont {Chashka},
  \citenamefont {Salman}, \citenamefont {Dagan},\ and\ \citenamefont
  {Kanigel}}]{Ribak2017}%
  \BibitemOpen
  \bibfield  {author} {\bibinfo {author} {\bibfnamefont {A.}~\bibnamefont
  {Ribak}}, \bibinfo {author} {\bibfnamefont {I.}~\bibnamefont {Silber}},
  \bibinfo {author} {\bibfnamefont {C.}~\bibnamefont {Baines}}, \bibinfo
  {author} {\bibfnamefont {K.}~\bibnamefont {Chashka}}, \bibinfo {author}
  {\bibfnamefont {Z.}~\bibnamefont {Salman}}, \bibinfo {author} {\bibfnamefont
  {Y.}~\bibnamefont {Dagan}}, \ and\ \bibinfo {author} {\bibfnamefont
  {A.}~\bibnamefont {Kanigel}},\ }\href {\doibase 10.1103/PhysRevB.96.195131}
  {\bibfield  {journal} {\bibinfo  {journal} {Phys. Rev. B}\ }\textbf {\bibinfo
  {volume} {96}},\ \bibinfo {pages} {195131} (\bibinfo {year}
  {2017})}\BibitemShut {NoStop}%
\bibitem [{\citenamefont {Murayama}\ \emph {et~al.}(2020)\citenamefont
  {Murayama}, \citenamefont {Sato}, \citenamefont {Taniguchi}, \citenamefont
  {Kurihara}, \citenamefont {Xing}, \citenamefont {Huang}, \citenamefont
  {Kasahara}, \citenamefont {Kasahara}, \citenamefont {Kimchi}, \citenamefont
  {Yoshida}, \citenamefont {Iwasa}, \citenamefont {Mizukami}, \citenamefont
  {Shibauchi}, \citenamefont {Konczykowski},\ and\ \citenamefont
  {Matsuda}}]{Maruyama2020}%
  \BibitemOpen
  \bibfield  {author} {\bibinfo {author} {\bibfnamefont {H.}~\bibnamefont
  {Murayama}}, \bibinfo {author} {\bibfnamefont {Y.}~\bibnamefont {Sato}},
  \bibinfo {author} {\bibfnamefont {T.}~\bibnamefont {Taniguchi}}, \bibinfo
  {author} {\bibfnamefont {R.}~\bibnamefont {Kurihara}}, \bibinfo {author}
  {\bibfnamefont {X.~Z.}\ \bibnamefont {Xing}}, \bibinfo {author}
  {\bibfnamefont {W.}~\bibnamefont {Huang}}, \bibinfo {author} {\bibfnamefont
  {S.}~\bibnamefont {Kasahara}}, \bibinfo {author} {\bibfnamefont
  {Y.}~\bibnamefont {Kasahara}}, \bibinfo {author} {\bibfnamefont
  {I.}~\bibnamefont {Kimchi}}, \bibinfo {author} {\bibfnamefont
  {M.}~\bibnamefont {Yoshida}}, \bibinfo {author} {\bibfnamefont
  {Y.}~\bibnamefont {Iwasa}}, \bibinfo {author} {\bibfnamefont
  {Y.}~\bibnamefont {Mizukami}}, \bibinfo {author} {\bibfnamefont
  {T.}~\bibnamefont {Shibauchi}}, \bibinfo {author} {\bibfnamefont
  {M.}~\bibnamefont {Konczykowski}}, \ and\ \bibinfo {author} {\bibfnamefont
  {Y.}~\bibnamefont {Matsuda}},\ }\href {\doibase
  10.1103/PhysRevResearch.2.013099} {\bibfield  {journal} {\bibinfo  {journal}
  {Phys. Rev. Research}\ }\textbf {\bibinfo {volume} {2}},\ \bibinfo {pages}
  {013099} (\bibinfo {year} {2020})}\BibitemShut {NoStop}%
\bibitem [{\citenamefont {Greywall}\ and\ \citenamefont
  {Busch}(1989)}]{Greywall1989}%
  \BibitemOpen
  \bibfield  {author} {\bibinfo {author} {\bibfnamefont {D.~S.}\ \bibnamefont
  {Greywall}}\ and\ \bibinfo {author} {\bibfnamefont {P.~A.}\ \bibnamefont
  {Busch}},\ }\href {\doibase 10.1103/PhysRevLett.62.1868} {\bibfield
  {journal} {\bibinfo  {journal} {Phys. Rev. Lett.}\ }\textbf {\bibinfo
  {volume} {62}},\ \bibinfo {pages} {1868} (\bibinfo {year}
  {1989})}\BibitemShut {NoStop}%
\bibitem [{\citenamefont {Lin}(1990)}]{Lin1990}%
  \BibitemOpen
  \bibfield  {author} {\bibinfo {author} {\bibfnamefont {H.~Q.}\ \bibnamefont
  {Lin}},\ }\href {\doibase 10.1103/PhysRevB.42.6561} {\bibfield  {journal}
  {\bibinfo  {journal} {Phys. Rev. B}\ }\textbf {\bibinfo {volume} {42}},\
  \bibinfo {pages} {6561} (\bibinfo {year} {1990})}\BibitemShut {NoStop}%
\bibitem [{\citenamefont {Oguchi}\ \emph {et~al.}(1985)\citenamefont {Oguchi},
  \citenamefont {Nishimori},\ and\ \citenamefont {Taguchi}}]{Oguchi1985}%
  \BibitemOpen
  \bibfield  {author} {\bibinfo {author} {\bibfnamefont {T.}~\bibnamefont
  {Oguchi}}, \bibinfo {author} {\bibfnamefont {H.}~\bibnamefont {Nishimori}}, \
  and\ \bibinfo {author} {\bibfnamefont {Y.}~\bibnamefont {Taguchi}},\ }\href
  {\doibase 10.1143/JPSJ.54.4494} {\bibfield  {journal} {\bibinfo  {journal}
  {J. Phys. Soc. Jpn.}\ }\textbf {\bibinfo {volume} {54}},\ \bibinfo {pages}
  {4494} (\bibinfo {year} {1985})}\BibitemShut {NoStop}%
\bibitem [{\citenamefont {Miyake}(1985)}]{Miyake1985}%
  \BibitemOpen
  \bibfield  {author} {\bibinfo {author} {\bibfnamefont {S.~J.}\ \bibnamefont
  {Miyake}},\ }\href {\doibase 10.1143/PTP.74.468} {\bibfield  {journal}
  {\bibinfo  {journal} {Progress of Theoretical Physics}\ }\textbf {\bibinfo
  {volume} {74}},\ \bibinfo {pages} {468} (\bibinfo {year} {1985})}\BibitemShut
  {NoStop}%
\bibitem [{\citenamefont {Leung}\ and\ \citenamefont
  {Runge}(1993)}]{Leung1993}%
  \BibitemOpen
  \bibfield  {author} {\bibinfo {author} {\bibfnamefont {P.~W.}\ \bibnamefont
  {Leung}}\ and\ \bibinfo {author} {\bibfnamefont {K.~J.}\ \bibnamefont
  {Runge}},\ }\href {\doibase 10.1103/PhysRevB.47.5861} {\bibfield  {journal}
  {\bibinfo  {journal} {Phys. Rev. B}\ }\textbf {\bibinfo {volume} {47}},\
  \bibinfo {pages} {5861} (\bibinfo {year} {1993})}\BibitemShut {NoStop}%
\bibitem [{\citenamefont {Ohyama}\ and\ \citenamefont
  {Shiba}(1993)}]{Ohyama1993}%
  \BibitemOpen
  \bibfield  {author} {\bibinfo {author} {\bibfnamefont {T.}~\bibnamefont
  {Ohyama}}\ and\ \bibinfo {author} {\bibfnamefont {H.}~\bibnamefont {Shiba}},\
  }\href {\doibase 10.1143/JPSJ.62.3277} {\bibfield  {journal} {\bibinfo
  {journal} {J. Phys. Soc. Jpn.}\ }\textbf {\bibinfo {volume} {62}},\ \bibinfo
  {pages} {3277} (\bibinfo {year} {1993})}\BibitemShut {NoStop}%
\bibitem [{\citenamefont {Deutscher}\ and\ \citenamefont
  {Everts}(1993)}]{Deutscher1993}%
  \BibitemOpen
  \bibfield  {author} {\bibinfo {author} {\bibfnamefont {R.}~\bibnamefont
  {Deutscher}}\ and\ \bibinfo {author} {\bibfnamefont {H.~U.}\ \bibnamefont
  {Everts}},\ }\href {\doibase 10.1007/BF01308811} {\bibfield  {journal}
  {\bibinfo  {journal} {Zeitschrift f{\"u}r Physik B Condensed Matter}\
  }\textbf {\bibinfo {volume} {93}},\ \bibinfo {pages} {77} (\bibinfo {year}
  {1993})}\BibitemShut {NoStop}%
\bibitem [{\citenamefont {Trumper}\ \emph {et~al.}(2000)\citenamefont
  {Trumper}, \citenamefont {Capriotti},\ and\ \citenamefont
  {Sorella}}]{Trumper2000}%
  \BibitemOpen
  \bibfield  {author} {\bibinfo {author} {\bibfnamefont {A.~E.}\ \bibnamefont
  {Trumper}}, \bibinfo {author} {\bibfnamefont {L.}~\bibnamefont {Capriotti}},
  \ and\ \bibinfo {author} {\bibfnamefont {S.}~\bibnamefont {Sorella}},\ }\href
  {\doibase 10.1103/PhysRevB.61.11529} {\bibfield  {journal} {\bibinfo
  {journal} {Phys. Rev. B}\ }\textbf {\bibinfo {volume} {61}},\ \bibinfo
  {pages} {11529} (\bibinfo {year} {2000})}\BibitemShut {NoStop}%
\bibitem [{\citenamefont {Chernyshev}\ and\ \citenamefont
  {Zhitomirsky}(2009)}]{Chernyshev2009}%
  \BibitemOpen
  \bibfield  {author} {\bibinfo {author} {\bibfnamefont {A.~L.}\ \bibnamefont
  {Chernyshev}}\ and\ \bibinfo {author} {\bibfnamefont {M.~E.}\ \bibnamefont
  {Zhitomirsky}},\ }\href {\doibase 10.1103/PhysRevB.79.144416} {\bibfield
  {journal} {\bibinfo  {journal} {Phys. Rev. B}\ }\textbf {\bibinfo {volume}
  {79}},\ \bibinfo {pages} {144416} (\bibinfo {year} {2009})}\BibitemShut
  {NoStop}%
\bibitem [{\citenamefont {Zhitomirsky}\ and\ \citenamefont
  {Chernyshev}(2013)}]{Zhitomirsky2013}%
  \BibitemOpen
  \bibfield  {author} {\bibinfo {author} {\bibfnamefont {M.~E.}\ \bibnamefont
  {Zhitomirsky}}\ and\ \bibinfo {author} {\bibfnamefont {A.~L.}\ \bibnamefont
  {Chernyshev}},\ }\href {\doibase 10.1103/RevModPhys.85.219} {\bibfield
  {journal} {\bibinfo  {journal} {Rev. Mod. Phys.}\ }\textbf {\bibinfo {volume}
  {85}},\ \bibinfo {pages} {219} (\bibinfo {year} {2013})}\BibitemShut
  {NoStop}%
\bibitem [{\citenamefont {Holstein}\ and\ \citenamefont
  {Primakoff}(1940)}]{Holstein1940}%
  \BibitemOpen
  \bibfield  {author} {\bibinfo {author} {\bibfnamefont {T.}~\bibnamefont
  {Holstein}}\ and\ \bibinfo {author} {\bibfnamefont {H.}~\bibnamefont
  {Primakoff}},\ }\href {\doibase 10.1103/PhysRev.58.1098} {\bibfield
  {journal} {\bibinfo  {journal} {Phys. Rev.}\ }\textbf {\bibinfo {volume}
  {58}},\ \bibinfo {pages} {1098} (\bibinfo {year} {1940})}\BibitemShut
  {NoStop}%
\bibitem [{\citenamefont {Zheng}\ \emph {et~al.}(2006)\citenamefont {Zheng},
  \citenamefont {Fj\ae{}restad}, \citenamefont {Singh}, \citenamefont
  {McKenzie},\ and\ \citenamefont {Coldea}}]{ZhengSeriese2006}%
  \BibitemOpen
  \bibfield  {author} {\bibinfo {author} {\bibfnamefont {W.}~\bibnamefont
  {Zheng}}, \bibinfo {author} {\bibfnamefont {J.~O.}\ \bibnamefont
  {Fj\ae{}restad}}, \bibinfo {author} {\bibfnamefont {R.~R.~P.}\ \bibnamefont
  {Singh}}, \bibinfo {author} {\bibfnamefont {R.~H.}\ \bibnamefont {McKenzie}},
  \ and\ \bibinfo {author} {\bibfnamefont {R.}~\bibnamefont {Coldea}},\ }\href
  {\doibase 10.1103/PhysRevB.74.224420} {\bibfield  {journal} {\bibinfo
  {journal} {Phys. Rev. B}\ }\textbf {\bibinfo {volume} {74}},\ \bibinfo
  {pages} {224420} (\bibinfo {year} {2006})}\BibitemShut {NoStop}%
\bibitem [{\citenamefont {Ghioldi}\ \emph {et~al.}(2015)\citenamefont
  {Ghioldi}, \citenamefont {Mezio}, \citenamefont {Manuel}, \citenamefont
  {Singh}, \citenamefont {Oitmaa},\ and\ \citenamefont
  {Trumper}}]{Ghioldi2015}%
  \BibitemOpen
  \bibfield  {author} {\bibinfo {author} {\bibfnamefont {E.~A.}\ \bibnamefont
  {Ghioldi}}, \bibinfo {author} {\bibfnamefont {A.}~\bibnamefont {Mezio}},
  \bibinfo {author} {\bibfnamefont {L.~O.}\ \bibnamefont {Manuel}}, \bibinfo
  {author} {\bibfnamefont {R.~R.~P.}\ \bibnamefont {Singh}}, \bibinfo {author}
  {\bibfnamefont {J.}~\bibnamefont {Oitmaa}}, \ and\ \bibinfo {author}
  {\bibfnamefont {A.~E.}\ \bibnamefont {Trumper}},\ }\href {\doibase
  10.1103/PhysRevB.91.134423} {\bibfield  {journal} {\bibinfo  {journal} {Phys.
  Rev. B}\ }\textbf {\bibinfo {volume} {91}},\ \bibinfo {pages} {134423}
  (\bibinfo {year} {2015})}\BibitemShut {NoStop}%
\end{thebibliography}%

\clearpage
\clearpage
\setcounter{page}{1}
\newcommand{\beginsupplement}{%
        \setcounter{table}{0}
        \renewcommand{\thetable}{S\arabic{table}}%
        \setcounter{figure}{0}
        \renewcommand{\thefigure}{S\arabic{figure}}%
}

\beginsupplement

\onecolumngrid

\begin{center}
  \hypertarget{stitle}{}
  {\large{\textbf{        
        Supplemental Material:\\
        \vspace{0.1 cm}
        Thermodynamic properties of an $S=1/2$ ring-exchange model on the triangular lattice        
  }}}
  
  \vspace{0.2 cm}
  Kazuhiro Seki$^{1}$ and Seiji Yunoki$^{1,2,3}$  \\
  \vspace{0.2 cm}

  \centering{\small
  {\it
    $^1$ {Computational Quantum Matter Research Team, RIKEN, Center for Emergent Matter Science (CEMS), Saitama 351-0198, Japan} \\
    $^2$ {Computational Condensed Matter Physics Laboratory, RIKEN Cluster for Pioneering Research (CPR), Saitama 351-0198, Japan} \\
    $^3$ {Computational Materials Science Research Team, RIKEN Center for Computational Science (R-CCS),  Hyogo 650-0047,  Japan} 
  }
}
\vspace{0.2 cm}

\end{center}

This Supplemental Material contains calculated results of
the specific heat $c(T)$, entropy $s(T)$, susceptibility $\chi(T)$, and
Wilson ratio $R_{\rm W}(T)$ for $L=16,18,20,24,30$, and $36$.
The full diagonalization is employed for $L\leqslant 20$, while 
the block-extended finite-temperature Lanczos method is applied for $L\geqslant 24$
with the block-Lanczos parameters $(R_{\rm B},M_{\rm B},N_{\rm L})$
being the same as those reported in the main text.

\begin{center}
  \begin{figure}
    \includegraphics[width=0.95\columnwidth]{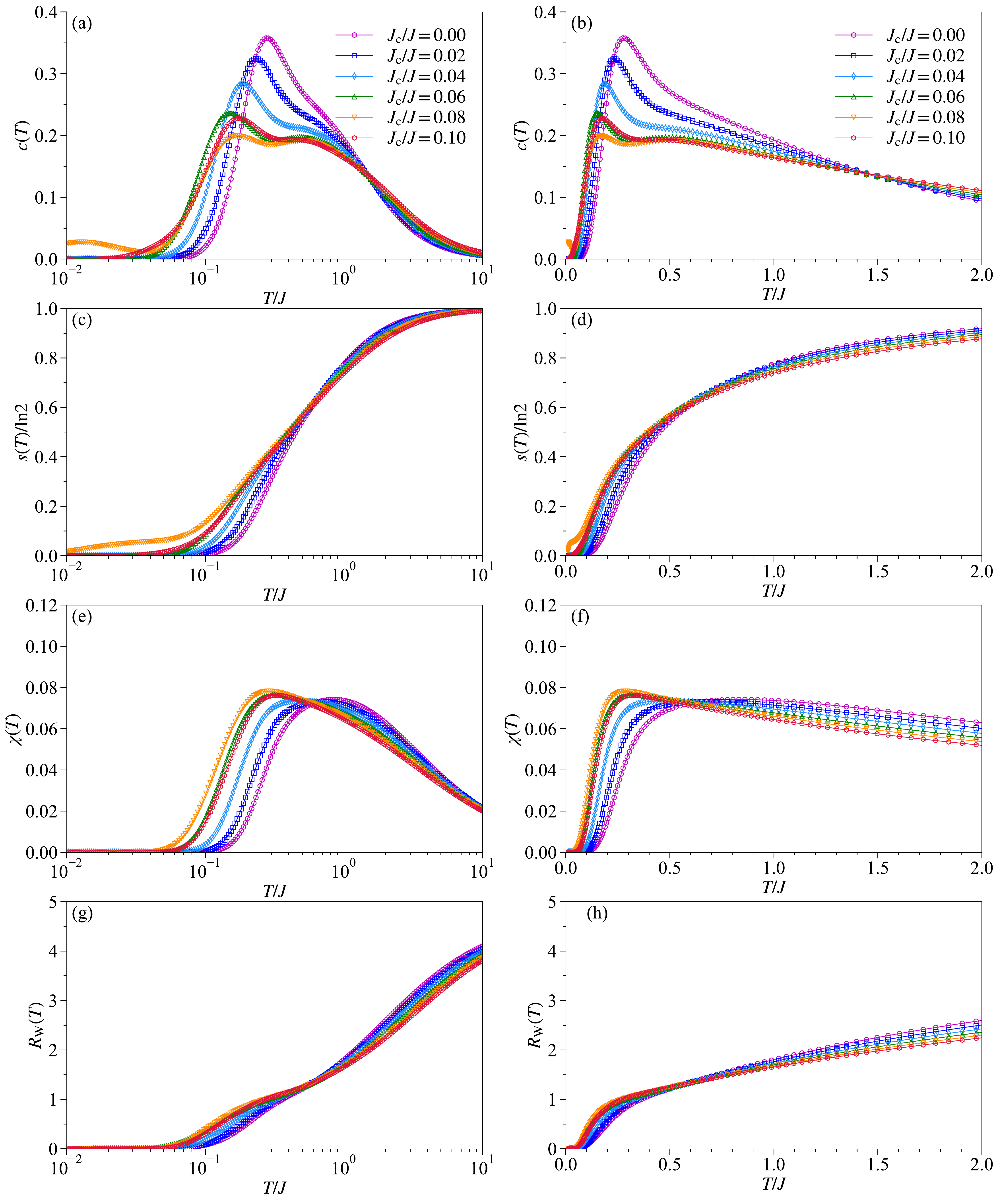}
    \caption{
      Same as Fig.~\ref{Kdep} of the maintext but for $L=16$. 
    }
  \end{figure}
  
  \begin{figure}
    \includegraphics[width=0.95\columnwidth]{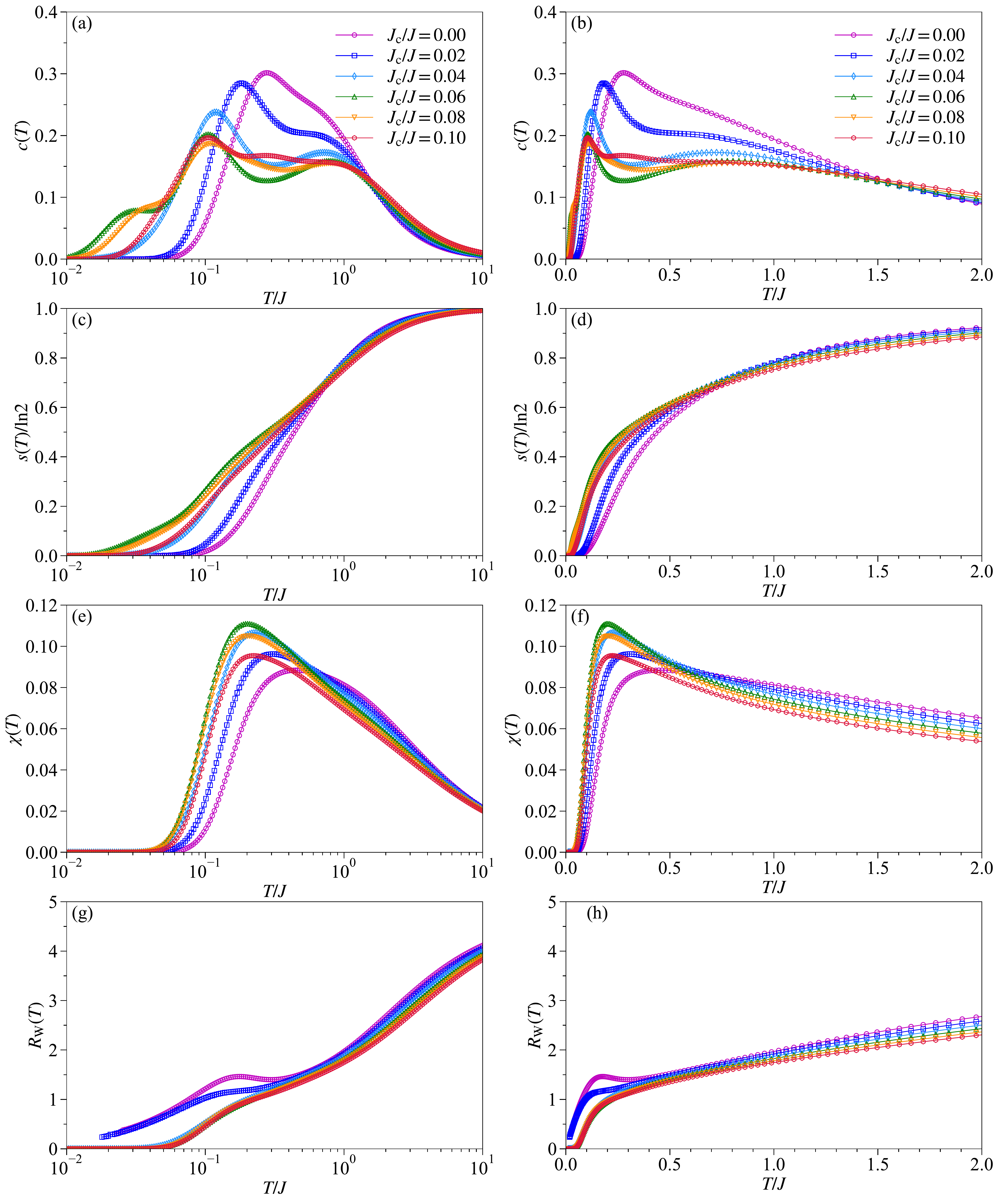}
    \caption{
      Same as Fig.~\ref{Kdep} of the maintext but for $L=18$. 
    }
  \end{figure}
    
  \begin{figure}
    \includegraphics[width=0.95\columnwidth]{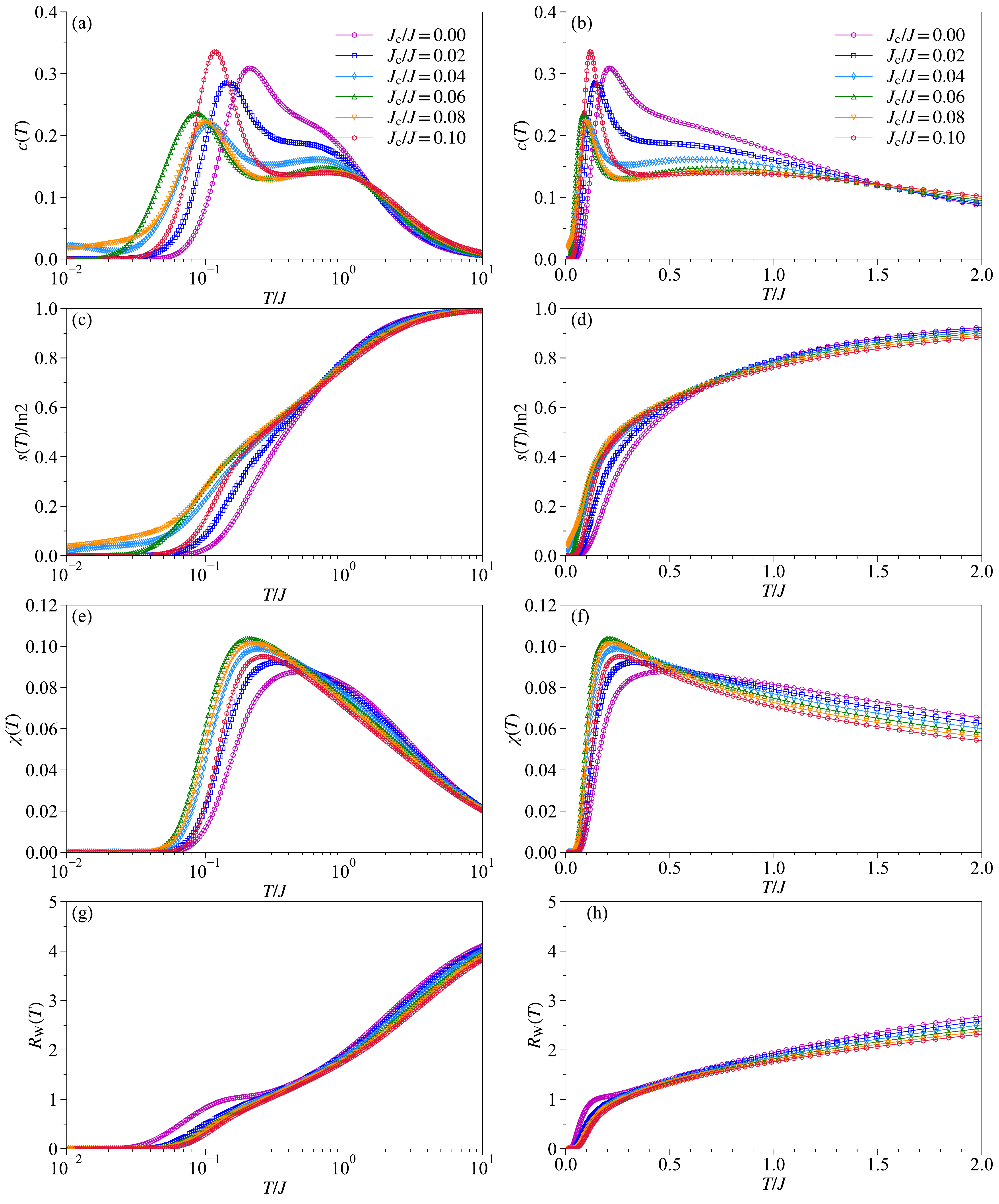}
    \caption{
      Same as Fig.~\ref{Kdep} of the maintext but for $L=20$. 
    }
  \end{figure}
  
  \begin{figure}
    \includegraphics[width=0.95\columnwidth]{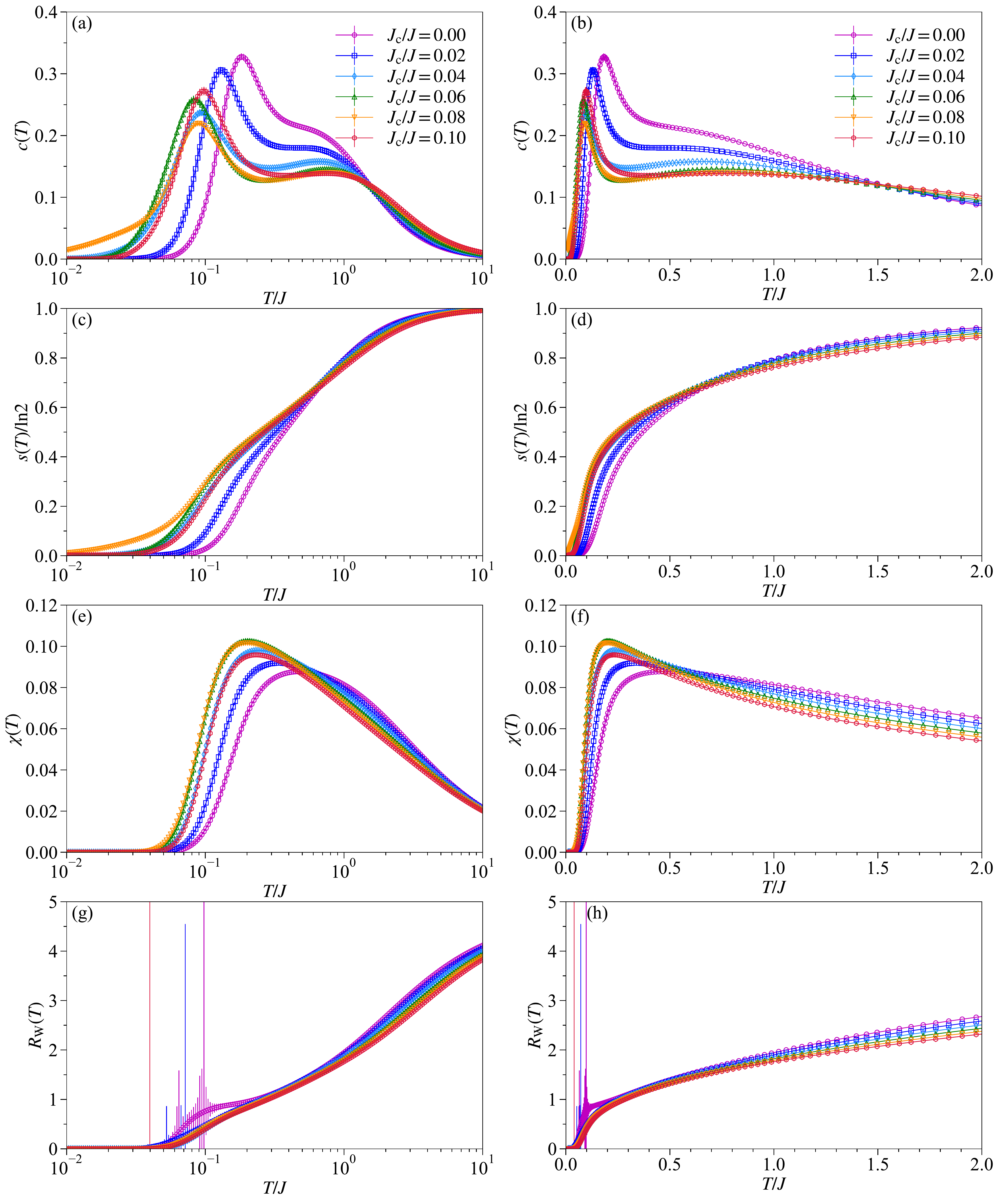}
    \caption{
      Same as Fig.~\ref{Kdep} of the maintext but for $L=24$. 
    }
  \end{figure}

  \begin{figure}
    \includegraphics[width=0.95\columnwidth]{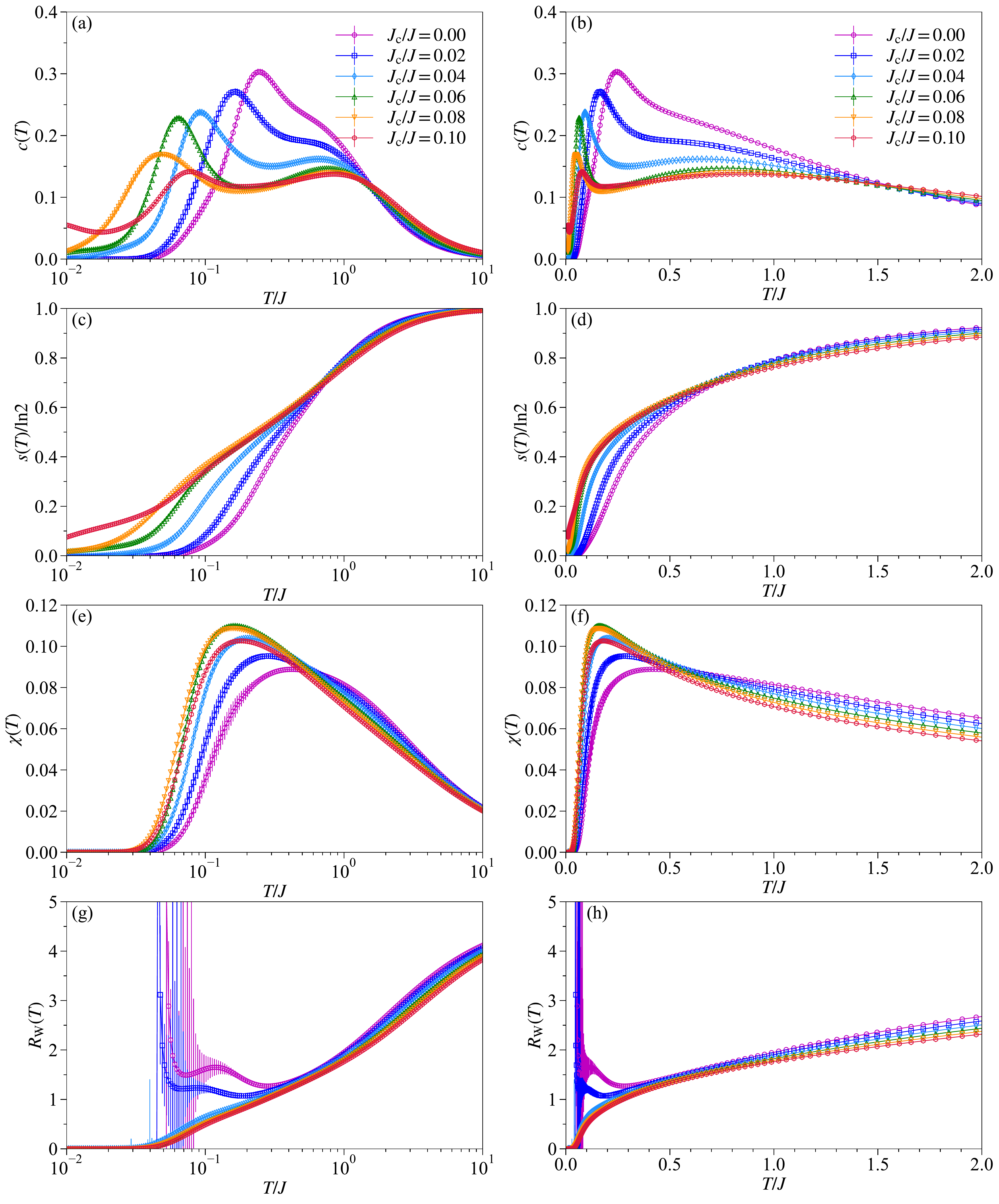}
    \caption{
      Same as Fig.~\ref{Kdep} of the maintext but for $L=30$. 
    }
  \end{figure}
  
  \begin{figure}
    \includegraphics[width=0.95\columnwidth]{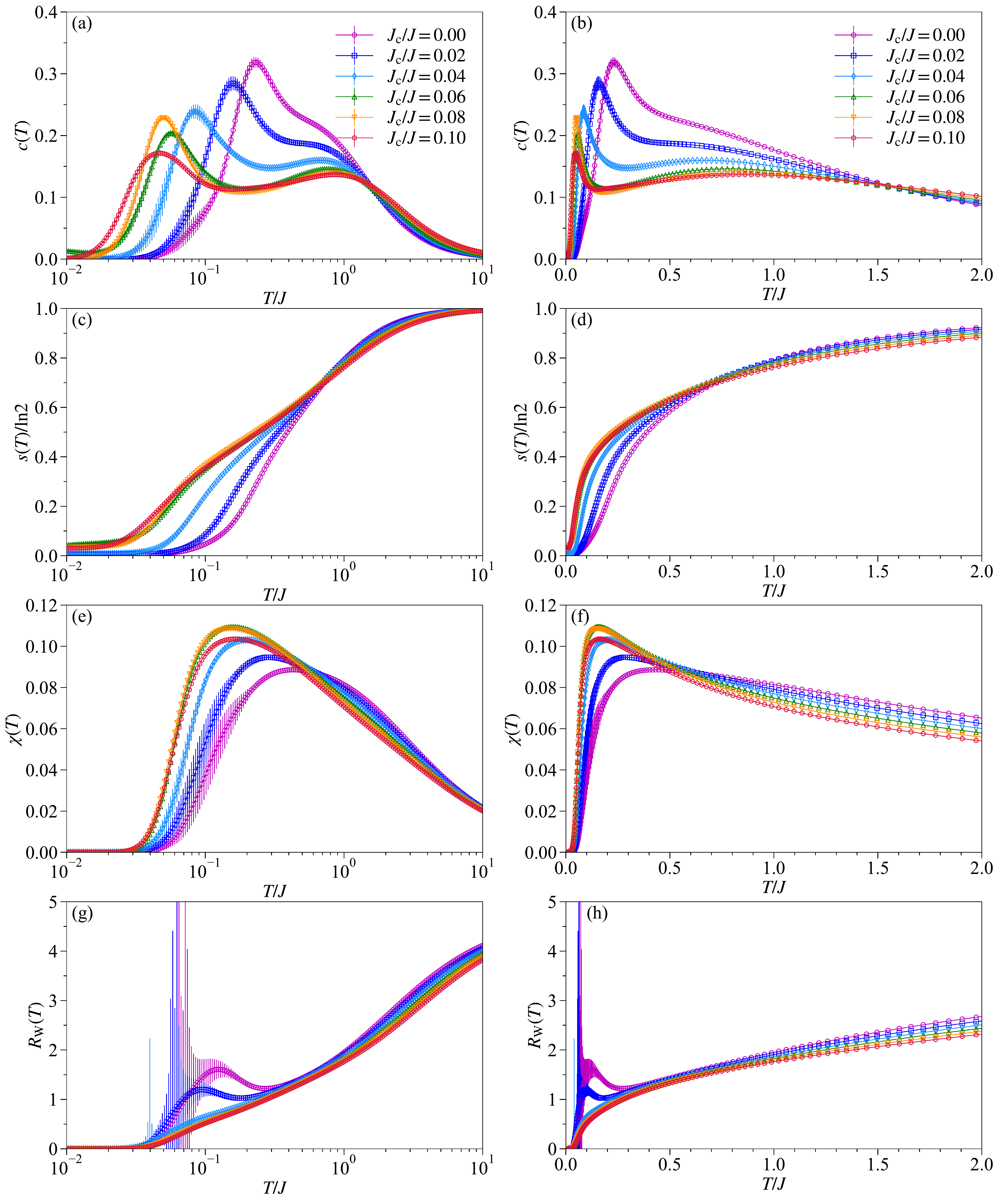}
    \caption{
      Same as Fig.~\ref{Kdep} of the maintext, i.e., for $L=36$
    }
  \end{figure}
\end{center}

\end{document}